\documentstyle[12pt,aaspp4,flushrt]{article}  

\def\epb#1{} 
\def\epb#1{\epsfbox{#1}} 

\def\capt{\small \baselineskip 12pt }
\def\be{\begin{equation}}
\def\ee{\end{equation}}

\def\eq#1{equation~(\ref{eq:#1})}
\def\equ#1{equation~(\ref{eq:#1})}
\def\Fig#1{Figure~\ref{fig:#1}}

\def\se#1{\S~\ref{sec:#1}}

\def\\{\hfill\break}

\def\ra{\rangle}
\def\la{\langle}
\def\av#1{\la #1 \ra}

\def\pot{{POTENT}}

\def\zoa{{ZoA}}
\def\dns{$D_n\!-\!\sigma$}

\def\sss{\scriptscriptstyle}

\def\bi{b_{\sss \rm IRAS}}
\def\betai{\beta_{\sss \rm IRAS}}
\def\betao{\beta_{\rm opt}}

\def\delp{\delta_{\sss \rm P}}
\def\delt{\delta_{\sss \rm T}}
\def\sigp{\sigma_{\sss \rm \bar P}}
\def\sigt{\sigma_{\sss \rm T}}

\def\po{P}
\def\tr{T}
\def\sigl{\sigma_{\rm loc}}
\def\epsl{\epsilon_{\rm loc}}
\def\sigsys{\sigma_{\rm sys}}
\def\epssys{\epsilon_{\rm sys}}
\def\sigran{\sigma_{\rm ran}}
\def\epsran{\epsilon_{\rm ran}}
\def\sigtot{\sigma_{\rm tot}}
\def\epstot{\epsilon_{\rm tot}}
\def\sw{s_{\rm w}}
\def\rw{r_{\rm w}}
\def\sigtw{\sigma_{{\rm  {\sss T}, w}}}
\def\epssysw{\epsilon_{\rm sys, w}}
\def\epsranw{\epsilon_{\rm ran, w}}
\def\epstotw{\epsilon_{\rm tot, w}}

\def\Rm{R_{\rm max}}
\def\Rs{R_{\rm s}}
\def\Re{R_{\rm e}}
\def\R4m{R_{4,{\rm max}}}
\def\sigd{\sigma_\delta}
\def\sigdm{\sigma_{\delta,{\rm max}}}
\def\sigv{\sigma_v}

\def\etal{{et al.\ }}

\def\eg{{e.g.}}
\def\ie{{i.e.}}

\def\ifm#1{\relax\ifmmode#1\else$\mathsurround=0pt #1$\fi}
\def\kms{\ifmmode\,{\rm km}\,{\rm s}^{-1}\else km$\,$s$^{-1}$\fi}
\def\hmpc{\,h\ifm{^{-1}}{\rm Mpc}}

\def\pa {\partial}

\def\ltsima{$\; \buildrel < \over \sim \;$}
\def\lsim{\lower.5ex\hbox{\ltsima}}
\def\gtsima{$\; \buildrel > \over \sim \;$}
\def\gsim{\lower.5ex\hbox{\gtsima}}

\def\pmb#1{\setbox0=\hbox{#1}%
 \kern-.025em\copy0\kern-\wd0
 \kern.05em\copy0\kern-\wd0
 \kern-.025em\raise.0433em\box0}
\def\vv{\pmb{$v$}}
\def\vx{\pmb{$x$}}

\def\vB{\pmb{$B$}}
\def\vL{\pmb{$L$}}
\def\v0{\pmb{$0$}}
\def\vnabla{\pmb{$\nabla$}}
\def\div{\vnabla  \cdot  }
\def\rot{\vnabla\!\times\!}
\def\divv{\div\vv}
\def\rotv{\rot\vv}

\def\omm{\Omega_{\rm m}}
\def\oml{\Omega_{\rm \Lambda}}

\def\today{\ifcase\month\or January\or February\or March\or April\or May\or
  June\or July\or August\or September\or October\or November\or December\fi
  \space\number\day, \number\year}

\begin{document}

\baselineskip 13pt \parskip 2pt 


\title{
POTENT RECONSTRUCTION FROM MARK III VELOCITIES}

\author{
A. Dekel\altaffilmark{1,2,3},
A. Eldar\altaffilmark{1},
T. Kolatt\altaffilmark{1,2},
A. Yahil\altaffilmark{4},\\
J.A. Willick\altaffilmark{5},
S.M. Faber\altaffilmark{3},
S. Courteau\altaffilmark{6},
D. Burstein\altaffilmark{7}
}

\altaffiltext{1}
{Racah Institute of Physics, The  Hebrew University, Jerusalem 91904, Israel}

\altaffiltext{2}
{Dept. of Physics, University of California, Santa Cruz, CA 95064}
		 
\altaffiltext{3}
{UCO/Lick Observatory, University of California, Santa Cruz, CA 95064}

\altaffiltext{4}
{Dept. of Physics \& Astronomy, State University of New York, Stony Brook, 
NY 11794-3800}

\altaffiltext{5}
{Dept. of Physics, Stanford University, Stanford, CA 94305-4060}

\altaffiltext{6}
{NRC/HIA, Dominion Astrophysical Observatory, 5071 W. Saanich Rd, 
Victoria, BC, V8X 4M6}

\altaffiltext{7}
{Dept. of Physics and Astronomy, Box 871504, Arizona State Univ.,
Tempe, AZ 85287}

\begin{abstract}

\rightskip=-0.5 true cm \leftskip=-0.5 true cm 

We present an improved version of the POTENT method for reconstructing
the velocity and mass density fields from radial peculiar velocities,
test it with mock catalogs, and apply it to the Mark III Catalog of 
Galaxy Peculiar Velocities.  The method is improved in several ways:
(a) the inhomogeneous Malmquist bias is reduced by grouping and 
corrected statistically in either forward or inverse analyses of 
inferred distances,
(b) the smoothing into a radial velocity field
is optimized such that window and sampling biases are reduced,
(c) the density field is derived from the velocity field using an improved
weakly non-linear approximation in Eulerian space, and 
(d) the computational 
errors are made negligible compared to the other errors.  
The method is carefully tested and optimized using realistic mock 
catalogs based on an N-body simulation that mimics our cosmological 
neighborhood, and the remaining systematic and random errors are 
evaluated quantitatively.
 
\medskip
The Mark III catalog, with $\sim\!3300$ grouped galaxies, 
allows a reliable reconstruction with fixed Gaussian smoothing of 
$10-12\hmpc$ out to $\sim\!60\hmpc$ and beyond in some directions. 
We present maps of the three-dimensional velocity and mass-density 
fields and the corresponding errors. 
The typical systematic and random errors in the density fluctuations inside 
$40\hmpc$ are $\pm 0.13$ and $\pm 0.18$.
The recovered mass distribution resembles in its gross features the
galaxy distribution in redshift surveys and the mass distribution
in a similar POTENT analysis of a complementary velocity catalog (SFI),
including the Great Attractor, Perseus-Pisces, and the large void in between.
The reconstruction inside $\sim 40\hmpc$ is not affected much by a 
revised calibration of the distance indicators (VM2, tailored to match 
the velocities from the IRAS 1.2 Jy redshift survey).
The bulk velocity within the sphere of radius $50\hmpc$ about the Local
Group is $V_{\rm 50}=370\pm 110\kms$ (including systematic errors), 
and is shown to be mostly generated by external mass fluctuations.  
With the VM2 calibration, $V_{\rm 50}$ is reduced to $305\pm 110\kms$. 

\end{abstract}


\rightskip=0 true cm \leftskip=0 true cm


{\baselineskip 12pt \parskip 2pt
\cleardoublepage
\setcounter{tocdepth}{2}  
\tableofcontents
\cleardoublepage
}

\section{INTRODUCTION}
\label{sec:intro}

The development of methods for estimating distances to galaxies
independently of their redshifts enables direct, quantitative study of
{\it large-scale dynamics} (reviews: Dekel 1994; Strauss \& Willick 1995;
Willick 1998; Dekel 1998a).  
The inferred peculiar velocities of thousands of galaxies
are interpreted as noisy tracers of an underlying peculiar velocity
field.  Under the assumption of structure evolution via
gravitational instability (GI), one can 
recover the velocity-potential field and the associated fields of
three-dimensional velocity and mass-density fluctuations.
These dynamical fields have important theoretical implications. For example,
they are related directly to the initial fluctuations
on scales ranging from $\sim\!10$ to $\sim\!100\hmpc$,
independent of galaxy-density ``biasing",
and they provide unique constraints on the value of the
cosmological density parameter, $\Omega$. 
Combined with galaxy redshift surveys, the dynamical fields
can be used to address the  ``biasing" relation between galaxies and mass
and help us better understand galaxy formation.
When compared to the fluctuations in the Cosmic Microwave Background (CMB), 
they allow a unique test of GI as the source for fluctuation growth,
and they provide constraints on the fluctuation power spectrum on 
intermediate scales.

There is growing evidence in support of the hypothesis that the
inferred large-scale peculiar velocity field is real, 
and of gravitational origin.
First, the $\delta T/T \sim 10^{-5}$ temperature fluctuations detected in
the CMB are consistent with gravity's generating $\sim
300$-$400\kms$ flows across regions of size $\sim 100\hmpc$, 
as inferred from the observed peculiar velocities 
(\eg, Bertschinger, Gorski \& Dekel 1990).  
Second, the
velocity fields inferred independently from spiral and from elliptical
galaxies, using different distance indicators, are consistent with 
being noisy versions of the same underlying velocity field 
(Kolatt \& Dekel 1994; re-confirmed for the Mark III data, unpublished;
Scodeggio 1997).  
Third, the gross features of the galaxy density field from redshift 
surveys are similar, within the errors, to the features of the 
mass-density field derived under GI from the observed peculiar 
velocities (Dekel \etal 1993; Hudson \etal 1995; Sigad \etal 1998), 
ruling out in particular certain alternative models in which the galaxy 
distribution could violate the continuity equation (Babul \etal 1994).

The important implications of the dynamical fields, and the
encouraging evidence for their reality and gravitational origin, motivate
an effort of analysis within the framework of GI.
This is the aim of the \pot\ program and related investigations.
Bertschinger and Dekel (1989, BD) proposed the original idea 
to recover the three-dimensional velocity field using the expected
irrotationality of gravitational flows in the weakly non-linear regime, 
and demonstrated the feasibility of such a method.  A first version of 
the method was developed and tested by Dekel, Bertschinger \& Faber 
(1990, DBF), and first maps of the recovered fields
in our near cosmological neighborhood were presented, based on the 
Mark II data that were available at that time 
(Bertschinger, Dekel, Faber, Dressler \& Burstein 1990, BDFDB).  
The present paper is the main paper of the second-generation
\pot\ analysis, describing and evaluating in detail the improved method
and presenting maps and simple statistics based on 
the extended Mark III data (Willick \etal 1997a).

The aim of the POTENT analysis is to recover with minimal systematic errors
the velocity and density fields that would be obtained if 
the true three-dimensional velocity field were sampled uniformly and 
with infinite density, and smoothed with a spherical 
Gaussian window of a fixed radius (hereafter G$\Rs$, where $\Rs$ is 
the smoothing radius in $\!\hmpc$, \eg, G10, G12, etc.).
The spatial statistical uniformity implied by the fixed smoothing scale, 
which is a special feature of POTENT, is useful both for 
pure cosmographic purposes and for simple, direct comparisons of the
recovered fields with theoretical models and other observations.
Note that, for certain specific purposes, such as a velocity-velocity
comparison with a redshift survey for determining
the parameter $\beta$ ($\equiv \Omega^{0.6}/b$, where $b$ is the
relevant biasing parameter), one also has the  
option of applying {\it variable\,} smoothing that can be optimized
to match the non-uniform sampling and errors in the data.
This is true in POTENT as well as in other methods
(\eg, DBF; BDFDB; Davis, Nusser \& Willick 1996, ITF; 
Willick \etal 1997b, VELMOD).
%
For example, a reconstruction method using a Wiener filter for a 
rigorous treatment of the random errors (Zaroubi, Hoffman \& Dekel 1998)
naturally applies such a variable smoothing that is determined by the data,
noise and a prior power spectrum. The Wiener fields are then 
forced to a fixed smoothing by generating constrained realizations.    

A few more introductory words about the key idea of POTENT are in 
order for the reader who is not familiar with DBF. 
If the large-scale structure evolved according to GI, the large-scale 
peculiar velocity field is expected to be {\it irrotational}, $\rotv=0$.  
Any vorticity mode would have decayed away during the linear phase 
of fluctuation growth as the universe expanded, and, based on Kelvin's 
circulation theorem, the flow remains vorticity-free in the 
weakly non-linear regime as long as it is laminar, \ie, with no orbit 
crossing (DBF).  
This has been shown to be a good approximation when collapsed regions 
are properly smoothed over, on scales of a few Mpc or more (DB; DBF).  
Irrotationality implies that the velocity field can be derived from a 
scalar potential, $\vv(\vx)\!=\!-\vnabla\Phi(\vx)$, and thus the radial 
velocity field $u(\vx)$, which also consists of one number at each point 
in space, should contain enough information for a 
full reconstruction.  In the standard \pot\ procedure, the velocity potential
is computed by integration along radial rays from the observer,
\be
\Phi(\vx) = -\int_0^r u (r',\theta,\phi) dr' \ .
\label{eq:pot}
\ee
The two missing transverse velocity components along $\theta$ and $\phi$
are then recovered 
by differentiation, and the underlying mass-density fluctuation 
field is computed from the partial derivatives of the velocity 
field using a mildly non-linear approximation (see \se{pot}).

The POTENT procedure thus recovers the underlying mass-density fluctuation
field from a whole-sky sample of observed radial peculiar velocities
via the following steps:
\newcounter{mynum}
\begin{list}{(\roman{mynum})}
            {\usecounter{mynum} \parsep 0in \itemsep .1cm \topsep -0pt} 
\item
Prepare the radial velocities for POTENT analysis, in particular 
correcting for Malmquist bias in different ways, including grouping.
\item
Smooth the peculiar velocities into a uniformly-smoothed radial
    velocity field that has minimum bias.
\item
Apply the ansatz of gravitating potential flow to recover
    the potential and three-dimensional velocity field. 
\item
Derive the underlying mass density field by an approximation to
GI in the mildly non-linear regime.
\item
Evaluate the remaining systematic and random errors using mock catalogs.
\end{list}
The most challenging part of the POTENT procedure is the
second step in the above list,
where one tries to obtain an unbiased radial-velocity field
$u(\vx)$ from the observed noisy and sparsely sampled radial velocities
of galaxies (\se{twf}).

The present analysis is superior to the original POTENT analysis of 1990
in several ways:  
\begin{list}{(\roman{mynum})}
            {\usecounter{mynum} \parsep 0in \itemsep .1cm \topsep -0pt} 
\item 
The new Mark III catalog contains $\sim 3300$ galaxies, mostly spirals 
with $\sim\! 17\!-\!21\%$ distance errors, compared with only $\sim\! 970$ 
galaxies in Mark II, which were dominated by ellipticals with 21\% errors. 
The Mark III catalog samples with higher resolution an extended volume 
of typical radius $\sim\! 60\!-\!70\hmpc$ about the Local Group (LG).  
This catalog, assembled from several different datasets, is  
carefully calibrated and merged into a self-consistent sample 
(\se{m3}). 
\item
New efforts have been made to minimize systematic errors in the data, 
such as Malmquist bias (\se{mb}).
\item
The derivation of the smoothed radial velocity field from the
discrete peculiar velocities is better designed and tested to minimize
systematic errors due to the tensor window imposed by the radial velocities
and due to the sparse and non-uniform sampling (\se{twf}).
\item 
The potential analysis is done with higher resolution and improved
accuracy such that the computational errors become negligible compared
to the other uncertainties (\se{pot}).
\item
The density field, which in DBF involved an elaborate iterative
procedure in Lagrangian space, is now recovered from the velocity field 
using a straightforward Eulerian prescription (\se{pot}).
\end{list}

This paper is organized as follows.
In \se{m3} we briefly describe the Mark III data. 
In the following six sections we describe the POTENT method and its testing
step by step using mock catalogs.
In \se{mock} we describe the mock catalogs that serve as
   our major tool for evaluating errors.
In \se{eval} we define the statistics to be used in the 
   evaluation of the reconstruction.
In \se{pot} we elaborate on the potential analysis and the 
  derivation of the density field.
In \se{twf} we discuss in detail the smoothing procedure  
  and the minimization of the associated systematic errors.
In \se{mb} we describe three different schemes for correcting Malmquist bias.
In \se{errors} we define our reference volumes and evaluate the remaining
errors in the POTENT analysis within these volumes.

The next two sections describe the results of applying POTENT 
to the actual Mark III catalog:
in \se{maps} we present maps of velocity and mass-density fields and  
in \se{bulk} we compute the bulk velocity in spheres and shells about the 
  Local Group and show preliminary results concerning a decomposition
  of the velocity field into its divergent and tidal components.

Finally, in \se{disc}, we discuss issues concerning the method and the data, 
summarize the cosmological implications and comment on alternative 
methods and other results. 
We conclude with a summary in \se{conc}.

\section{THE MARK III CATALOG}
\label{sec:m3}

The Mark III Catalog of Galaxy Peculiar Velocities 
(Willick \etal 1995; 1996; 1997a)
consists of roughly 3300 galaxies from several different datasets of 
spiral and elliptical/S0 galaxies.  The distances were inferred by 
the Tully-Fisher (TF) and $D_n\!-\!\sigma$ distance indicators,
respectively. 
These are based on empirical intrinsic linear correlations between a 
distance independent quantity (the log of the 
internal velocity in a galaxy, $\eta$, referring to rotation in spirals and
dispersion in ellipticals) and a distance-dependent quantity 
(the apparent magnitude in spirals and the log of the
apparent diameter in ellipticals). 
The ``forward" TF relation between the intrinsic quantities 
$\eta$ and absolute magnitude $M$ is
\be
M_{\sss \rm TF} (\eta)=a-b\eta \ . 
\label{eq:FTF}
\ee
It yields an inferred 
distance $d$ from observed values of $\eta$ and apparent magnitude $m$
via the standard distance-modulus relation
\be
5 \log d = m-M_{\sss \rm TF} (\eta) \ .
\ee
The CMB redshift $z$ of the galaxy is obtained from the helio-centric 
redshift using the standard transformation based on COBE's 4-year dipole
(Lineweaver \etal 1996),
and the inferred radial peculiar velocity at $d$, in the CMB frame, is 
\be
u=z-d \ ,
\ee
where all quantities are measured in $\!\kms$,  
\ie, the speed of light $c$ and the Hubble constant $H_0$ are set to unity. 
[We measure distances equivalently by $\!\kms\,$ or by $\!\hmpc$, where 
$h\equiv H_0/(100\kms{\rm Mpc}^{-1})$.]

 The slope of the TF relation, $b$, can be derived from cluster 
galaxies that are assumed to be at a common distance within each cluster.
The TF zero point, $a$, which fixes the absolute distance scale 
(in $\kms$, or $\hmpc$, still independent of the actual value of the 
Hubble constant), is free to be determined by minimizing residuals 
about a Hubble flow in a volume as large as possible.
The scatter about the mean TF relation is between $0.35$ and $0.45$ magnitudes, 
which translates to a random uncertainty of $17$-$21\%$ in the inferred 
distance.

The large-scale backbone of the Mark II data was the whole-sky set of 
$\sim\!500$ ellipticals and S0's dominated by Lynden-Bell \etal (1988, 
known as the ``7-Samurai" or ``7S" survey), supplemented by data from 
Dressler \& Faber (1991) and Lucey \& Carter (1988).
The more local neighborhood, out to $\sim 30\hmpc$, was dominated
by a set of spirals, mostly from Aaronson \etal (1982; A82).
In the Mark III catalog we have added the large southern sample of 
spirals by Mathewson, Ford \& Buchhorn (1992; MAT), the northern 
sample by Courteau \& Faber (Courteau 1996, 1997; CF),
the narrow-angle sample towards Perseus-Pisces by Willick (1991; W91PP),
and the whole-sky cluster sample by Han, Mould and collaborators (Han \&
Mould 1990, 1992; Mould \etal 1991; HMCL). We also revised
the A82 data set based on the uniform diameters and related revisions
by Tormen \& Burstein (1995).
The full catalog consists of $\sim\!2800$ spirals. 
Table 1 
summarizes the main properties and selection criteria of the
datasets that make up the Mark III catalog.
The number of galaxies actually used from each dataset
in the POTENT application of the present paper
is typically smaller than the number in the final published version
of the Mark III catalog because of removal of 
duplicate galaxies that are common to different datasets (see below).
On the other hand, a few galaxies that were removed from the final published
version based on large TF residuals are still present in the version
of the data used here. The effects of these slight differences in the 
data are not noticeable in the outcome.

Assuming that all galaxies trace the same underlying velocity field, 
the analysis of large-scale motions greatly benefits from merging the 
different samples into one, self-consistent catalog. In the Mark III 
catalog, the TF relations for each dataset were re-calibrated and 
merged into a homogeneous catalog for velocity analysis.
The treatment of the cluster datasets is described in Willick \etal (1995),
the field galaxies are calibrated and grouped in Willick \etal (1996),
and the final catalog is tabulated in Willick \etal (1997a).

\vbox{\vskip 9.7truecm}
\includegraphics{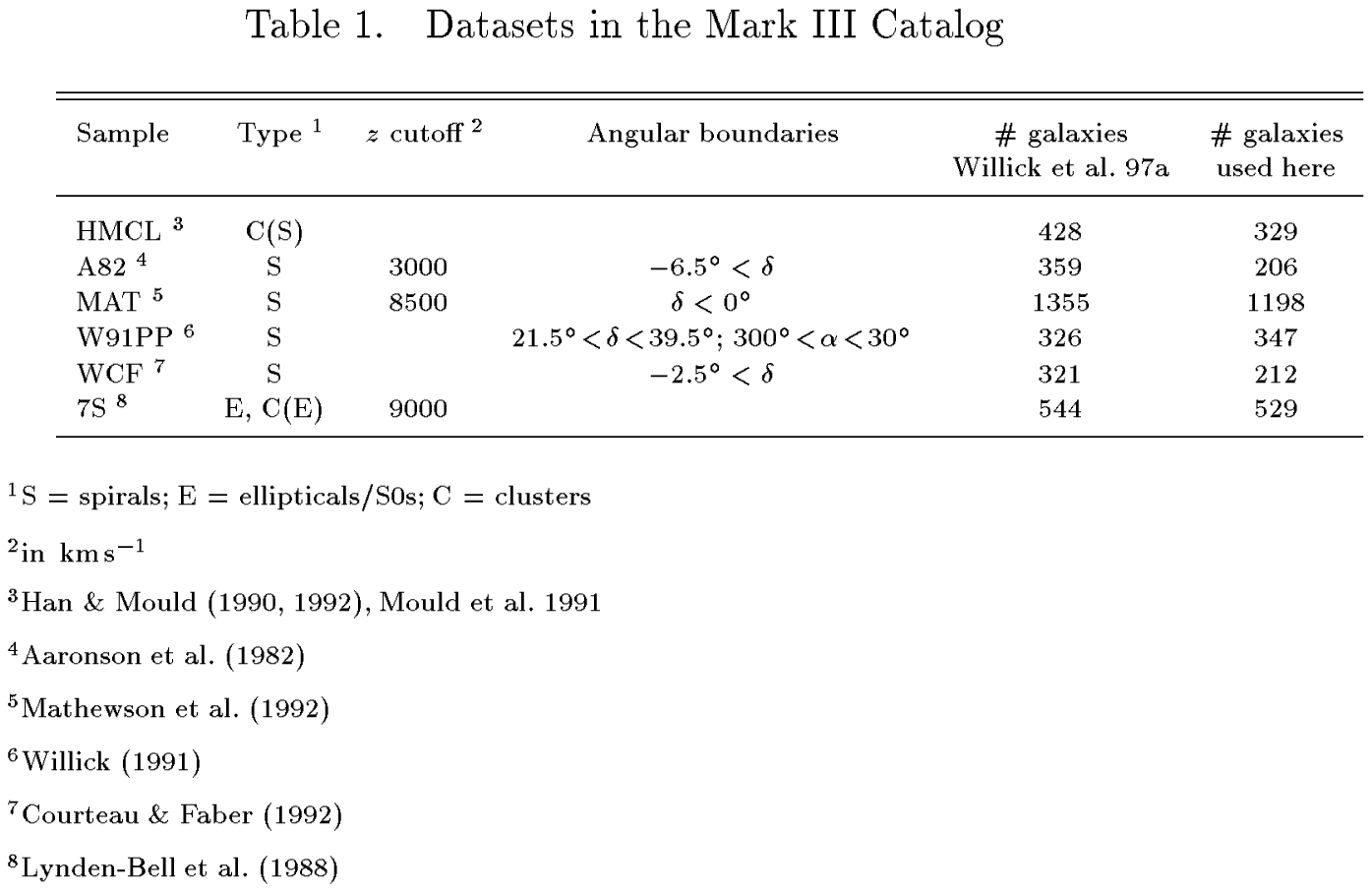}

Generating the Mark III catalog involved the following steps: 
\begin{list}{(\roman{mynum})}
            {\usecounter{mynum} \parsep 0in \itemsep .1cm \topsep -0pt} 
\item
Standardizing the selection criteria, \eg, rejecting galaxies of extreme 
inclination or low velocity parameter $\eta$, which  
exhibit large TF errors, and sharpening any redshift cutoff for the purpose of 
bias corrections.  
\item
Re-deriving a provisional TF calibration for each dataset using Willick's 
algorithm (1994), which simultaneously groups, fits and corrects for 
selection bias (see \se{mb_fmb}),
then verifying that inverse-TF distances to all clusters
are consistent with the forward-TF distances.
\item
Starting with one dataset (HMCL), 
adding each new set in succession using  
galaxies in common to adjust the TF  
zero points of the new set if necessary. 
\item
Retaining only one measurement per galaxy rather than averaging multiple
measurements to ensure well defined errors (but at the cost of losing
some 10\% of the spiral data) and using multiple observations for a
``cluster'' only if the membership duplication within that cluster
is small (\eg, $<\!50\%$).
\item
Including the ellipticals from Mark II, allowing for a slight zero-point 
shift of 3\% (based on a revised analysis of Kolatt \& Dekel 1994).  
\end{list}
Such an extensive calibration and merger procedure is {\it crucial} for 
reliable results --- in several cases it produced TF distances substantially 
different from those quoted by the original observers.

\begin{figure}[t!]
\vskip -0.5truecm
\centerline{\epsfxsize=6.0 in \epb{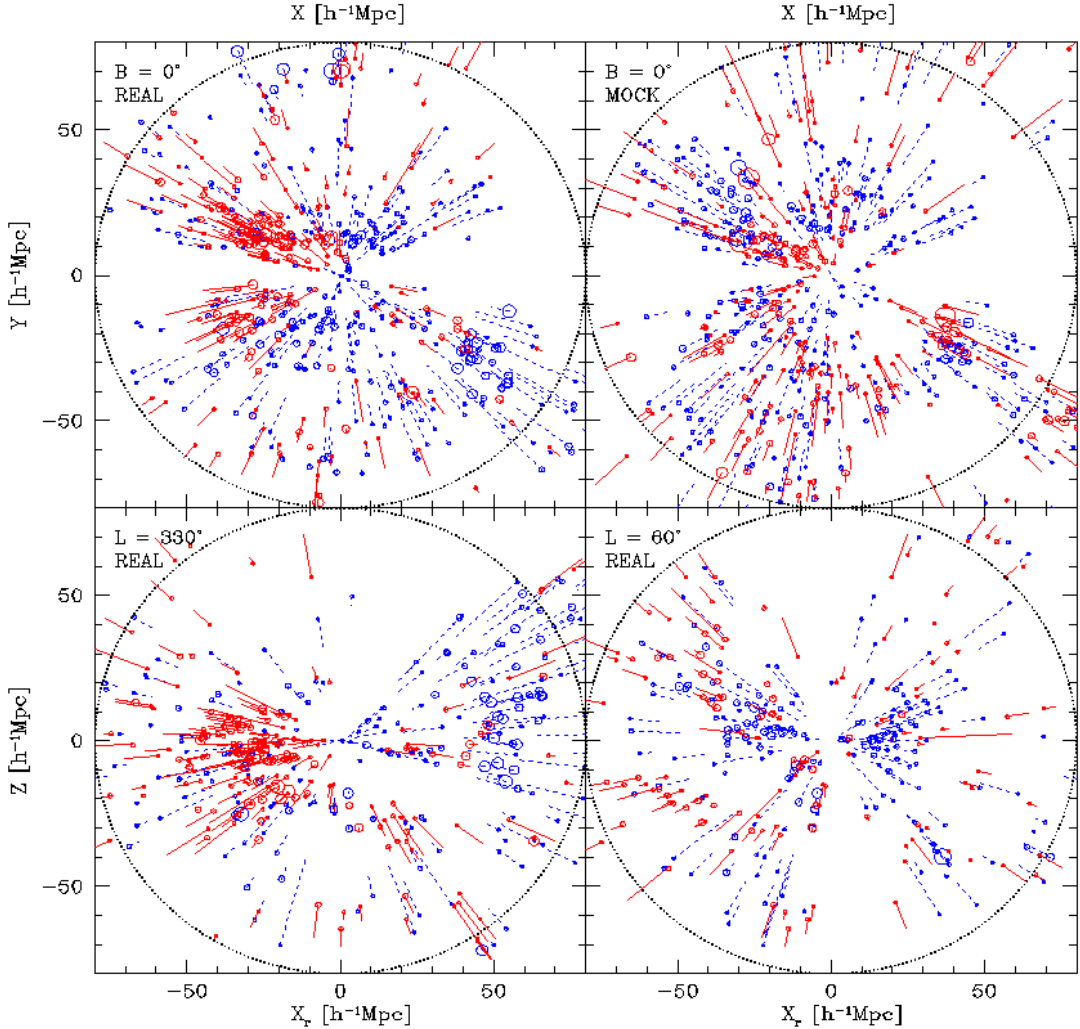}}
\vskip -0.5truecm
\caption{\protect\capt
Radial peculiar velocities of ``objects" in the Mark III
catalog. The real data are shown in slices of $\pm 20^\circ$
about three planes: the Supergalactic plane ($B=0^\circ$)
and two perpendicular planes, at $L= 330^\circ$ and $60^\circ$.
The coordinate $X_{\rm r}$ is either $X$, or $X$ rotated by $-30^\circ$
or $60^\circ$.
Mock data are shown in comparison, but only about the Supergalactic
plane (top-right panel).
Distances and velocities are in $\hmpc$.
The area of each symbol marking the object position is proportional to
the object richness in galaxies.
Solid and dashed lines distinguish outgoing and incoming radial
velocities.
The positions and velocities are corrected for Malmquist bias.
}
\label{fig:m3}
\end{figure}

The main purpose of grouping in the Mark III catalog is to reduce
the random error per object, and thus automatically reduce the resulting 
Malmquist bias (\S~\ref{sec:mb} below). The grouping procedure is described 
in detail in Willick \etal (1996). 
Galaxies were assigned to groups according to their proximity in angular
position and in redshift.  Proximity in inferred distance was used as 
a secondary criterion.
The final grouped Mark III catalog consists of $\sim\! 1200$ ``objects" ---
single galaxies, groups and clusters.

Figure~\ref{fig:m3}
illustrates the spatial distribution of objects in the Mark III catalog
and their radial peculiar velocities in three orthogonal slices.
One notices the poor coverage of the Galactic Zone of Avoidance
(ZoA, about $Y=0$) and at large distances.
The Supergalactic plane ($B=0^\circ$) and the plane $L=330^\circ$
both cut through the Great Attractor (GA, $X<0$)
and Perseus Pisces (PP, $X>0$ and $Y<0$) regions where the streaming
motions are large compared to the perpendicular plane, $L=60^\circ$.

Despite the careful effort made in the construction of the Mark III data,
some galaxies are left that could be regarded as outliers,
either because of observational errors or because they are indeed
eccentric galaxies not obeying the TF relation.
We make an additional effort to exclude such outliers using an
iterative procedure based on the deviation of the inferred object 
peculiar velocity from the smoothed underlying velocity field at the 
location of that object.
In each iteration, the data go through the smoothing procedure,
and an object at $\vx_i$ is rejected if its peculiar velocity $u_i$ 
deviates from the smoothed velocity $u(\vx_i)$ by more than 
$4(\Delta d + 200\kms)$, where the first term refers to the
random distance error of the object and the second is an estimate of 
the dispersion velocity of field galaxies.
In the end, only 3 galaxies are rejected from the Mark III data
by applying this criterion.

In a recent analysis (Willick \& Strauss 1998, VELMOD2, hereafter VM2), 
a revised calibration
has been proposed for the TF zero points in the Mark III datasets that 
cover the Perseus-Pisces region, based on maximizing the agreement
with the peculiar velocities predicted by the IRAS 1.2 Jy redshift survey.
We test below the effects of this revised calibration on the reconstructed
fields (\se{maps_vm2}).

\section{MOCK CATALOGS}
\label{sec:mock}

The POTENT method is tested using artificial catalogs based on 
an $N$-body simulation, which are described in detail in Kolatt \etal (1996).
We present here only a brief summary.

A special effort has been made to generate a simulation that mimics the actual
large-scale structure in the real universe about the Local Group, 
in order to take into account
possible dependencies of the errors on the underlying signal.
The present-day density field, smoothed with a Gaussian of radius 
$5\hmpc$ (G5), is taken to be the G5
density of IRAS 1.2 Jy galaxies as reconstructed by the method described in
Sigad \etal (1998), assuming $\Omega=1$ and no biasing ($\bi=1$).
The field is evolved back in time to remove non-linear effects
by integrating the Zel'dovich-Bernoulli equation of Nusser \& Dekel (1992).
Remaining non-Gaussian features are removed by rank-preserving 
``Gaussianization" (Weinberg 1991), and artificial structure on scales 
smaller than the smoothing length is added using 
the method of constrained realizations (Hoffman \& Ribak 1991), with the 
power spectrum of the IRAS 1.2 Jy survey (Fisher \etal 1993) as a 
prior model.  The resulting density field is fed as initial conditions to 
a PM $N$-body code (Bertschinger \& Gelb 1991), which then follows the 
forward non-linear evolution under gravity with $\Omega=1$.
The present epoch is defined by an rms density fluctuation
of $\sigma_8=0.7$ at a top-hat smoothing of radius $8\hmpc$,
consistent with the value observed for IRAS galaxies 
and $\bi=1$ (Fisher \etal 1994).
The periodic box of side $256\hmpc$ is simulated with a $128^3$ force grid
and $128^3$ particles.

\begin{figure}[t!]
\centerline{\epsfxsize=6.0 in \epb{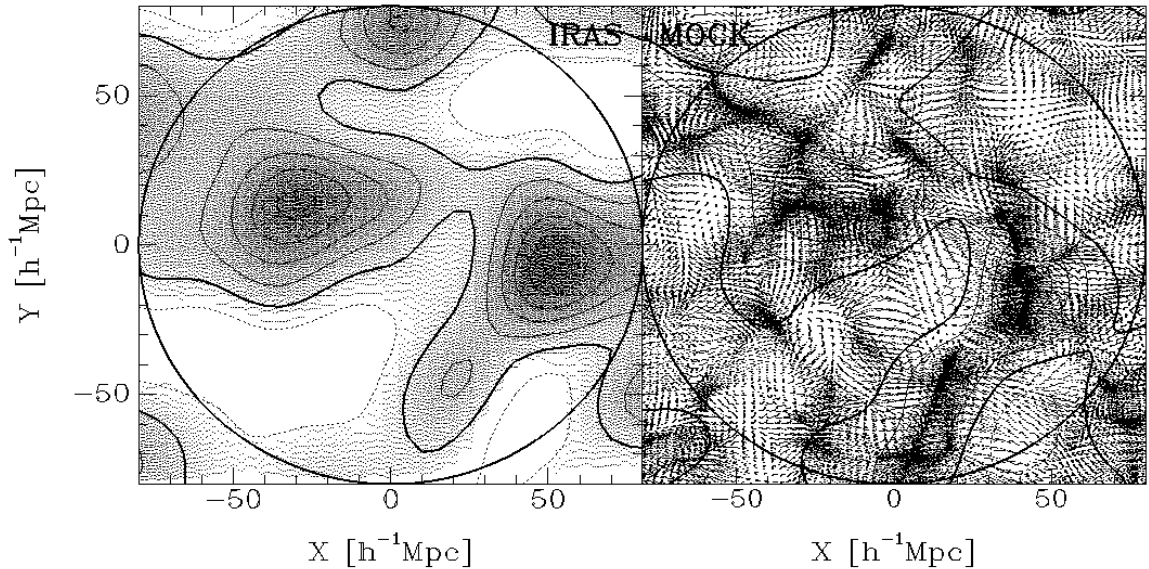}}
\vskip -0.2truecm
\caption{\protect\capt
The simulation in the Supergalactic plane.
The G12 density fluctuation field of IRAS 1.2 Jy galaxies (left)
is compared to the corresponding density field in the simulation
(right),
which is overlaid on top of the particle distribution in a slice of
thickness $\pm 10\hmpc$ about the plane.
The contour spacing is $0.2$, with positive contours solid,
negative contours dashed, and the zero contour heavy.
Distances are in $\hmpc$.
The LG is at the center. The GA is on the left and PP on the right.
}
\label{fig:mock_map}
\end{figure}

\Fig{mock_map} displays the particles in a slice of the simulation, of
thickness $\pm 10\hmpc$ about the Supergalactic plane.
It shows the familiar main features of the Great Attractor,
Perseus-Pisce, and the extended low-density region in between.
The fine sub-structure mimics the true rich clusters but it also
contains a certain random element. The G12-smoothed density field
is also marked, and is compared with the original, G12 fluctuation 
field of IRAS 1.2 Jy galaxies.

In a second step
that is repeated 10 times with different sets of random-numbers,  
``galaxies" are identified in the simulation, 
assigned their relevant physical properties, and then
``observed" to make mock catalogs that include the same errors
and selection effects as in the Mark III data.
Each of the $N$-body particles is considered a galaxy candidate and is
identified as an elliptical or a spiral depending on the local neighborhood
density of particles (following Dressler 1980).
Rich clusters are identified, mimicking the cluster samples in the real 
Mark III data, and the remaining particles are left as candidates for 
field galaxies.
The galaxies are assigned internal velocities $\eta$ drawn at {\it random\,}
from the observed $\eta$ distribution function (corresponding to the observed
``Schechter" galaxy luminosity function).  A Tully-Fisher (or $D_n$-$\sigma$)
relation is assumed, and absolute magnitudes $M$ are {\it randomly\,}
scattered about the TF value, $M_{\sss \rm TF}(\eta)$, following a Gaussian 
distribution 
of width appropriate to the corresponding subset of the Mark III catalog.
Field galaxies are selected in the angular regions corresponding
to each of the sub-samples, with the appropriate magnitude limits
and redshift cutoffs.
This procedure amounts to a random selection of galaxies (as well as 
their physical properties) that mimics the statistical properties of 
the observed sample. 
The only feature of the observational procedure that is not simulated
in this process is the error in the calibration of the TF relations,
including the uncertainties in matching the zero points of the individual 
data samples in the unified Mark III catalog.

These data are used to infer TF distances to all the galaxies in each
of the random mock realizations.
The ``observed" redshifts are taken to be the true velocities of the
particles in the simulation.
Finally, the galaxies selected are grouped using the same code used
for the real data, and then corrected for Malmquist bias as in
\se{mb} below, using the galaxy number density profile $n(r)$
as derived from randomly selected mock IRAS catalogs
(from Sigad \etal 1998).

The top-right panel of \Fig{m3} shows the distribution of objects and 
their radial peculiar velocities in one random mock catalog, projected from a 
slice of $\pm 20^\circ$ about the Supergalactic plane into the plane. 
It resembles statistically the corresponding map of the real data.

\section{STATISTICS FOR EVALUATING A RECONSTRUCTION}
\label{sec:eval}

Before we embark on a detailed description of the POTENT algorithm,
we first define the statistics needed for the testing procedure. 

We use the $N$-body simulation and mock catalogs for evaluating
the success of a reconstruction.  The target for reconstruction is 
a true field $\tr(\vx)$ (\eg, the density field, potential
field, or any component of 
the velocity field), smoothed with a given window directly from the 
underlying particles of the simulation.
The reconstruction from each 
mock catalog provides a corresponding POTENT field $\po(\vx)$. The quality of 
the reconstruction is evaluated by comparing these fields both 
locally and globally inside a given reference volume.
 
In the current paper, the comparison is first performed visually via maps in 
the Supergalactic plane out to $80\hmpc$, 
and then quantitatively at points of a uniform 
grid within a smaller reference volume.  
We focus here on the reconstruction 
of the mass-density field, smoothed with a Gaussian window of 
radius $10$-$12\hmpc$, within a reference volume of an effective 
radius $30$-$50\hmpc$ 
about the Local Group (see \se{err_vol} for a more specific definition
of the reference volume). 

For testing cases with noisy input,
POTENT is applied to each of $M$ ($\sim 10$) random mock catalogs to 
yield a series of $M$ noisy POTENT fields $\{ \po_m(\vx) \}_{m=1}^M$.  
The rms scatter of these fields 
about the true field $\tr(\vx)$ provides the error field $\sigtot(\vx)$.
This error contains both systematic and random components.
The systematic error field is given by  $\bar\po(\vx) - \tr(\vx)$,
where $\bar \po(\vx)$ is the average over the noisy fields $\{ \po_m(\vx) \}$
(we hereafter denote averaging over the noisy realizations by an over-line).
In test cases with no random selection of objects and no random errors, there 
is no need to average over realizations and the bias field is simply
$\po(\vx) - \tr(\vx)$.

The interpretation of the various statistics that are defined in this 
section will become clearer when associated with one 
of the scatter diagrams of $\po$ (or $\bar\po$) versus $\tr$ 
in the following sections.

The typical systematic error within the reference volume is estimated
by the rms of the residuals over the volume (\ie, over uniform grid points):
\be
\sigsys^2 \equiv \la [\bar \po(\vx) - \tr(\vx)] ^2 \ra \ .
\ee
(We hereafter denote spatial averaging by angular parentheses.)
A meaningful measure that allows a comparative evaluation of
reconstructions with different smoothings is
$\epssys \equiv \sigsys / \sigt$,
where $\sigt$ is the standard deviation of the true field.

The systematic error can be decomposed into global and local components
by performing a linear regression of $\bar\po$ on $\tr$: 
\be
\bar \po = s \tr + c \ .
\ee
The constant $c$ is typically small.
(If the volume is big enough such that cosmic scatter is negligible,
one can assume $\la \tr \ra =0$ and set $c$ to zero by enforcing
$\la \bar\po \ra =0$.)
The slope of the regression line, 
$s = \av{ (\bar\po- \av{\bar\po}) (\tr - \av{\tr}) } /\sigt^2$,
characterizes the {\it global\,} bias; a deviation of $s$ from unity can
be interpreted as an error in the effective smoothing of the reconstruction.
Note that such a global systematic error could be harmful 
when the POTENT output is used for measuring cosmological parameters 
or is compared to other data, and it should therefore be kept small. 
One can crudely correct for such a bias in retrospect, provided
it is sufficiently small (\eg, Dekel \etal 1993).

The scatter about the best-fit line characterizes the {\it local\,} bias:
\be
\sigl^2 \equiv \la [\bar \po(\vx) - s \tr(\vx) -c]^2 \ra \ .
\ee
It is a {\it bias\,} because the random noise has been removed
from $\bar \po$ by the averaging over the many mock catalogs.
Again, this quantity becomes of more general interest when scaled into 
$\epsl \equiv {\sigl /(s \sigt)}$.
Equivalently, the same local bias can be characterized by the
linear correlation coefficient, 
$r \equiv \av{ (\bar\po-\av{\bar\po}) (\tr - \av{\tr}) } /(\sigt \sigp)$,
where $\sigp$ is the standard deviation of the average POTENT field. 
The simple relation between these two equivalent measures of local bias is 
\be
\epsl^2 = r^{-2} -1 \ .
\ee
The whole systematic error is then characterized by the sum in quadrature
of the local and global components:
\be
\epssys^2  
= s^2(r^{-2}-1) + (s-1)^2 + {c^2 \over \sigt^2} 
\ee
(assuming $\av{T}=0$).

In each case, we list the parameters $s$, $r$ and $c$, plus the corresponding
$\sigt$, as measures of systematic errors inside a reference volume.
The other parameters can be derived from these using the above relations.
A perfect reconstruction is characterized by $s=r=1$ and $c=0$, while
a failure is diagnosed by a large deviations from these values.

The field of {\it random\,} errors, when relevant,
is derived from the series of noisy POTENT
fields $\{ \po_m(\vx) \}$ by the rms residual relative to the 
{\it average} field:
\be
\sigran^2(\vx) \equiv \overline{[\po_m(\vx) - \bar\po(\vx)]^2} \ .
\ee
The corresponding field of {\it total} error
is defined by the rms residual relative to the {\it true\,} field:
\be
\sigtot^2(\vx) \equiv \overline{ [\po_m(\vx) - \tr(\vx)]^2} \ .
\ee
Either $\sigran(\vx)$ or $\sigtot(\vx)$ can be used in further 
analysis using POTENT output.
The typical values of these errors within a reference volume
are estimated by spatial averaging:
\be
\sigran^2 \equiv \la \sigran^2(\vx) \ra \ , 
\ee
\be
\sigtot^2 \equiv \la \sigtot^2(\vx) \ra
= \sigsys^2 + \sigran^2 \ .
\label{eq:sigtot}
\ee
The parameters for evaluating the random and total errors are 
finally quoted as $\epsran\equiv\sigran/\sigt$ and 
$\epstot\equiv\sigtot/\sigt$.

In the above statistics, the volume averages are all unweighted; 
the grid points within the reference volume are treated as equals. 
However, further analysis using POTENT output may prefer to give 
more weight to points in which the reconstruction is of higher quality.
For example, in the POTIRAS comparison (Sigad \etal 1998),
the POTENT densities were weighted by $\sigtot^{-2}$,
including the random {\it and} the systematic errors.
In the tests presented here, however, we prefer to use $\sigran^{-2}$ as
weights and as flags of quality reconstruction.
This is because $\sigran$ does not explicitly depend on the true field,
and because the weighting by $\sigtot^{-2}$ would artificially
reduce the signal of systematic errors which we wish to identify.
We thus define weighted statistics, $\sw$, $\rw$, etc., in a way analogous 
to the above statistics $s$, $r$, etc., except that the volume averaging 
is replaced by {\it weighted\,} volume averaging, with weights $\sigran^{-2}$.
This weighting, like all the above statistics, is specific
to the given dataset, which is currently the Mark III catalog.

\section{POTENTIAL ANALYSIS AND DENSITY RECONSTRUCTION}
\label{sec:pot}


The most straightforward stage of POTENT is the potential
analysis starting from a smoothed radial peculiar velocity field, $u(\vx)$.
This stage
consists of applying the ansatz of potential flow, \eq{pot}, to recover
the velocity potential and the three-dimensional velocity field, $\vv(\vx)$,
and then deriving the underlying density fluctuation field $\delta(\vx)$ 
by an adequate approximation to GI in the mildly non-linear regime.
We test this part of the method assuming the true smoothed
velocity field as input. 

\vskip -0.5truecm \subsection{Potential Analysis}
\label{sec:pot_pot}

In practice, the smoothed radial velocity field is computed 
(\se{twf} below) at the points of a spherical grid of 24 equal 
radial bins out to $\Rm=80\hmpc$, 48 equal latitude ($B$) bins in 
the range $\pm \pi/2$, and $96 \times \cos B$ roughly-equal longitude 
bins between 0 and $2\pi$.  The potential at the origin is arbitrarily 
fixed to zero, and the potential at every other grid point is computed 
by cubic spline integration of $u(\vx)$ along the radial rays of the 
spherical grid, \eq{pot}.

The potential is then interpolated by linear cloud-in-cell (CIC) onto the
points of a cubic grid of $2.5\hmpc$ spacing (positioned such that there is
a grid point at the origin). 
The partial derivatives are computed by finite differencing
to yield the three-dimensional velocity components via 
$\vv\!=\!-\vnabla\Phi$.  The second partial derivatives of the potential
are computed in a similar way for the purpose of approximating
the density fluctuation field.

\vskip -0.5truecm \subsection{From Velocity to Density}
\label{sec:pot_vd}

In the linear approximation to GI, the underlying mass-density fluctuation 
field is derived from the velocity field via one more differentiation,
$\delta_0 = -f(\Omega)^{-1} \divv$, where $f(\Omega) \simeq \Omega^{0.6}$
(e.g., Peebles 1980).  However, the linear approximation is limited to 
the small dynamic range between a few tens of megaparsecs and the 
$\sim\!100\hmpc$ extent of the current samples.  The current sampling 
of galaxies enables reliable dynamical analysis with a smoothing 
radius as small as $\sim\!10\hmpc$ (or even smaller in more limited
regions), where $\vert\divv\vert$ reaches values
larger than unity such that non-linear effects play a role.
Therefore, the derivation of the density field requires an approximate
solution to the equations of GI in the mildly non-linear regime.
 
We appeal to the Zel'dovich (1970) approximation, which is known to be
a successful tool in the mildly non-linear regime. Substituting an Eulerian
version of the Zel'dovich approximation in the continuity equation yields
(Nusser \etal 1991)
\begin{equation}
\delta_c(\vx) = \Vert \pmb{I} - f^{-1} {\pa \vv / \pa \vx} \Vert -1 \ ,
\label{eq:delc}
\end{equation}
where the bars denote the Jacobian determinant and $\pmb I$ is the unit
matrix.  

\begin{figure}[t!]
\vskip 0.2truecm
\centerline{\epsfxsize=6.5 in \epb{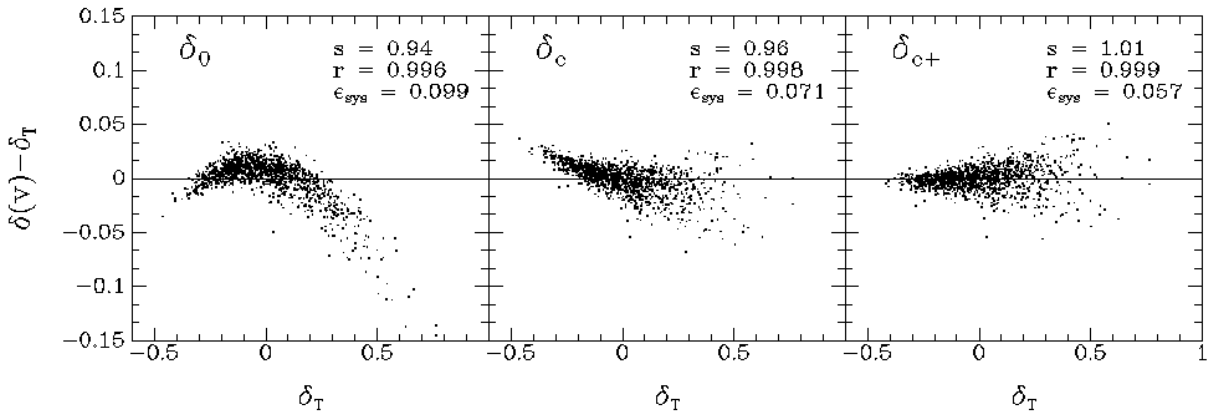}}
\vskip -0.2truecm
\caption{\protect\capt
Approximations to GI in the mildly non-linear regime.
The density fluctuation fields as derived from the partial derivatives
of the peculiar velocity field are compared point by point to the
true density field in the simulation ($\delt$). The smoothing is G12.
The mean spacing between the sampled points is $24\hmpc$.
The vertical scale is enlarged by a factor of $\sim 6$ compared to
the horizontal scale.
Error statistics are quoted (defined in \se{eval}).
Left: the linear approximation $\delta_0$.
Middle: the Zel'dovich-continuity approximation $\delta_c$.
Right: the improved approximation $\delta_{c+}$.
}
\label{fig:vtod}
\end{figure}

\Fig{vtod} (middle panel), based on $N$-body simulations (Ganon \etal 1998), 
demonstrates the performance of this approximation
in comparison with the linear approximation (left panel). 
The densities derived from the G12-smoothed peculiar velocity field are
compared at the points of a uniform grid to the true G12 density field in
the simulation. Note that the residuals are $\sim 0.05$ or less.
The error $\epssys$ improves from 0.099 to 0.071.
The approximation does a good job for $\delta \geq 0$, but it tends to be a
slight overestimate in the negative tail of $\delta < 0$,
and it becomes worse as one approaches $\delta \sim -1$
(see also Mancinelli \etal 1994; Mancinelli \& Yahil 1995).

This approximation can be improved as follows.  The Zel'dovich 
displacement is first order in $f^{-1}$ and $\vv$, and therefore 
the determinant in $\delta_c$ includes second- and third-order terms as well,
involving sums of double and triple products of partial derivatives:
\begin{equation}
\delta_{c} = -f^{-1} \divv
	     +f^{-2} \Delta_2
	     +f^{-3} \Delta_3 \ ,
\label{eq:delc-expand}
\end{equation}
where
\begin{equation}
\Delta_2(\vx)= \sum_{i < j} \left[\left({\pa v_i \over \pa x_j}\right)^2
- {\pa v_i \over \pa x_i}
{\pa v_j \over \pa x_j} \right] \ ,
\label{eq:d2}
\end{equation}
and
\begin{equation}
\Delta_3(\vx) = \sum_{i,j,k} \left[ 
{\pa v_i \over \pa x_i} {\pa v_j \over \pa x_k}{\pa v_k \over \pa x_j} -
{\pa v_1 \over \pa x_i} {\pa v_2 \over \pa x_j}{\pa v_3 \over \pa x_k} 
\right] \ ,
\label{eq:d3}
\end{equation}
in which $i,j,k$ run over the three cyclic permutations of $1,2,3$.
The $\delta_c$ approximation can be improved by slight adjustments 
to the coefficients of the three terms in \equ{delc-expand},
\begin{equation}
\delta_{c+} = -(1+\epsilon_1) f^{-1} \divv
	     +(1+\epsilon_2) f^{-2} \Delta_2
	     +(1+\epsilon_3) f^{-3} \Delta_3 \ .
\label{eq:delc+}
\end{equation}
These coefficients were empirically tuned to provide best fits for
a family of CDM
simulations of G12 smoothing over the whole range of $\delta$ values,
with $\epsilon_1=0.06$, $\epsilon_2=-0.13$ and $\epsilon_3=-0.3$.
\Fig{vtod} (right panel) demonstrates the improvement in $\delta_{c+}$ 
over $\delta_c$ for G12 smoothing.
The global systematic errors in the tails, that were apparent for $\delta_0$ 
and $\delta_c$, are now practically gone; $s=1.01$ and $r=0.99$.
The typical local error is down to $\epssys=0.057$.

The $\delta_{c+}$ approximation is found to be robust to quantities 
that are unknown
a priori such as (a) the value of $\Omega$ in the range $0.3-1.0$, 
(b) the shape of the power spectrum within the general CDM family,
allowing for a nonzero cosmological constant as well as a slight tilt in the
power spectrum on large scales 
(tested for the power index $n$ in the range $0.6-1.0$), 
and (c) the degree of non-linearity as determined by the fluctuation amplitude
and the smoothing scale (Ganon \etal 1998).
We adopt $\delta_{c+}$ as our standard approximation in POTENT.

\vskip -0.5truecm \subsection{Testing POTENT with Ideal Data}
\label{sec:pot_test}

We first use the simulation to evaluate the POTENT reconstruction from
ideal data with dense and uniform sampling and no distance errors.

The true, G-smoothed radial velocity field of the simulation, which serves
as input for this test, is computed as follows.
The three-dimensional field is first computed at the points of a fine 
cubic grid of $2\hmpc$ spacing by applying to the particle velocities
a Gaussian smoothing of equally small radius, $r_s=2\hmpc$.
The velocity field smoothed with a Gaussian of much larger radius $\Rs$,
say, G12, is then computed by FFT.
This G-smoothed velocity field is finally interpolated by CIC onto
the desired spherical grid of the potential integration, 
and the radial components serve as input for POTENT in the current
test.

\begin{figure}[t!]
\centerline{\epsfxsize=6.5 in \epb{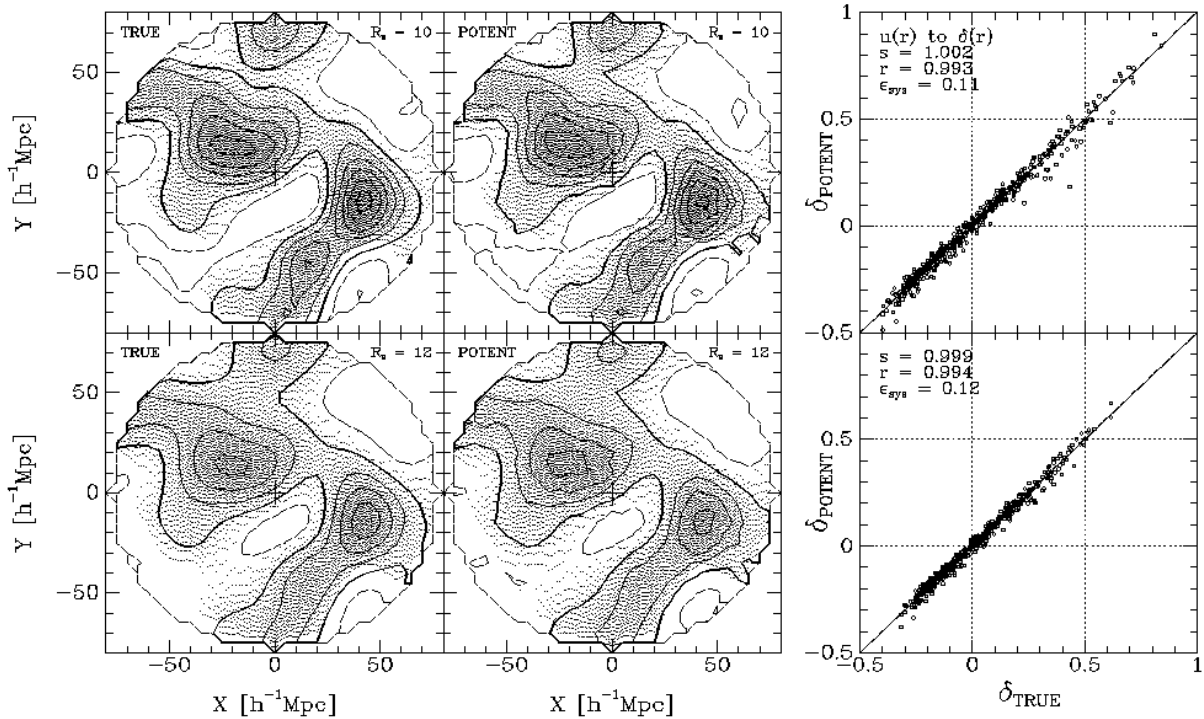}}
\vskip -0.2truecm
\caption{\protect\capt
Reconstruction from ideal mock data.
The mass density fluctuation field as reconstructed by POTENT
from the exact radial velocity field of the simulation, compared to
the true density field. Smoothing is G10 (top) or G12 (bottom).
Contours are as in \Fig{mock_map}.
A comparison to the true field is shown (right)
inside a sphere of radius $40\hmpc$,
at points sampled with a mean separation of $8.5\hmpc$.
Statistics measuring the typical systematic errors are quoted.
}
\label{fig:pot}
\end{figure}

Figure~\ref{fig:pot} compares the density fluctuation
field recovered by POTENT in this
ideal case, $\delp$, with the target true field of the simulation, $\delt$,
for both G10 and G12 smoothings.
The comparison is done both via maps in the Supergalactic plane and 
point by point on a cubic grid of spacing $5\hmpc$ inside a comparison
sphere of radius $40\hmpc$.
In order to make this scatter plot less crowded (and similar plots below), 
only a random subsample containing 20\% of the grid points is shown, 
with a mean separation of $8.5\hmpc$.

As explained in \se{eval}, the residuals in this scatter plot of 
$\delp$ versus $\delt$ are purely systematic errors (as the input is 
free of random noise in this case).
Their global and local components as measured by
$s$ and $r$ are shown in the figure and in 
Table 2. 

The success of the reconstruction in this case of no noise is excellent.
With $s$ deviating from unity by only a small fraction of a percent,
essentially no global bias is introduced by the potential analysis.
The small scatter of only $11-12\%$ of $\sigt$, which is reflected 
in the $<1\%$ deviation of $r$ from unity, is a result of the 
accumulating effects of
(a) numerical errors due to the finite grids used,
(b) scatter about the non-linear approximation $\delta_{c+}$,
and
(c) small deviations from potential flow at these smoothing scales.
The latter can be estimated by the rms of $\rotv$ in the box,
which is 4.4\% of the density $\sigt$ for G12, and 5.4\% for G10. 
All these errors are negligible compared to those associated with
computing the smoothed radial field $u(\vx)$ from the sparse,
non-uniform and noisy data. These are discussed next.

\section{THE SMOOTHING PROCEDURE}
\label{sec:twf}

The most difficult step in the POTENT procedure is the interpolation
and smoothing of the observed radial peculiar velocities $u_i$ at the inferred
object positions $\vx_i$, with errors $\sigma_i$, 
onto a radial velocity field at grid points,
$u(\vx)$, which serves as input for the potential analysis described
above.
The aim of this procedure is to mimic volume-weighted smoothing
of the three-dimensional velocity field $\vv(\vx)$
with a spherical Gaussian window of radius $\Rs$. 
In the unrealistic case of dense and uniform sampling of the three 
components of
$\vv(\vx)$ with random noise only, the desired smoothed velocity would have
simply been the best-fit bulk velocity of the data weighted by a spherical
Gaussian about the window center.  However, the limitations of the 
actual data introduce biases that present non-trivial complications.

The general idea of our smoothing procedure of the actual data about 
a grid point $\vx_c$ is that the smoothed radial velocity $u(\vx_c)$ 
is taken to be the value at $\vx\!=\!\vx_c$ of the radial component of
an appropriate {\it local\,} velocity model $\vv(\alpha_k;\vx\!-\!\vx_c)$ 
with free parameters $\{ \alpha_k \}$.  
These parameters are obtained by minimizing the weighted sum of residuals
\begin{equation}
-2 \ln {\cal L} \propto 
\sum_i W_i\, [u_i-\hat{\vx}_i\cdot\vv(\alpha_k;\vx _i)]^2 
\label{eq:sumv}
\end{equation}
within an appropriate local window function $W_i\!=\!W(\vx_i,\vx_c)$.  
The basic component of the window function
is a spherical Gaussian of radius $\Rs$,
$W_i \propto \exp [-(\vx_i-\vx_c)^2/2\Rs^2]$.
It is modified, together with the local velocity model,
in an effort to minimize the biases described below.
Note that if the errors were Gaussian and $W_i\!\propto\!\sigma_i^{-2}$,
then ${\cal L}$ would have been the formal likelihood;
it remains only an approximation to the likelihood in our case where 
the errors are closer to a log-normal distribution, and it may be 
carried further away from the likelihood by the additional weighting.

\vskip -0.5truecm \subsection{Tensor Window Bias} 
\label{sec:twf_wb}

In general, the radial directions from the origin (the Local Group) 
to the objects ($\hat{\vx}_i$) do not coincide with the radial 
direction to the window center ($\hat{\vx}_c$) at which we try to obtain
the smoothed radial velocity.
Therefore, unless the window radius is negligible compared to its 
distance from the origin ($\Rs\ll r_c$), 
the radial velocities cannot be averaged as scalars,
and one has to appeal to a fit of a local 3D velocity model, as 
described above.  The original \pot\ procedure of DBF used the simplest 
model within each window, \ie, a local bulk velocity with three parameters,
$\vv(\vx)\!=\!\vB$. 
For such a model the solution can be expressed explicitly in terms of 
a tensor window function, and the minimization procedure
can be replaced by a simple matrix inversion (BD; DBF; BFDFB).

Unfortunately, an attempt to use a model that does not have enough degrees 
of freedom for a proper fit of the variations of the velocity field 
within a window is likely to lead to a biased result. 
As an example, consider an infall towards a point attractor at 
the center of the smoothing window, and in particular the converging velocity
component in the plane transverse to the line of sight to the window center. 
For all tracers in that plane (except the window center),
the radial (line-of-sight) components observed  
are all negative, leading to a best-fit bulk velocity within the window that 
would be erroneously interpreted as a radial velocity towards the origin (LG). 
Similarly, a transverse outflow pattern
would be interpreted as a radial outflow away from the origin.  
Another bias arises from the fact that the tensorial correction to 
the spherical window has a conical symmetry, weighting more heavily 
objects that lie along radial rays that are closer to the radial ray 
through the window center,
and thus distorting the effective window shape.
We term these systematic effects ``window bias" (WB).

\begin{figure}[t!]
\vskip -1.2truecm
\centerline{\epsfxsize=4.0 in \epsfbox{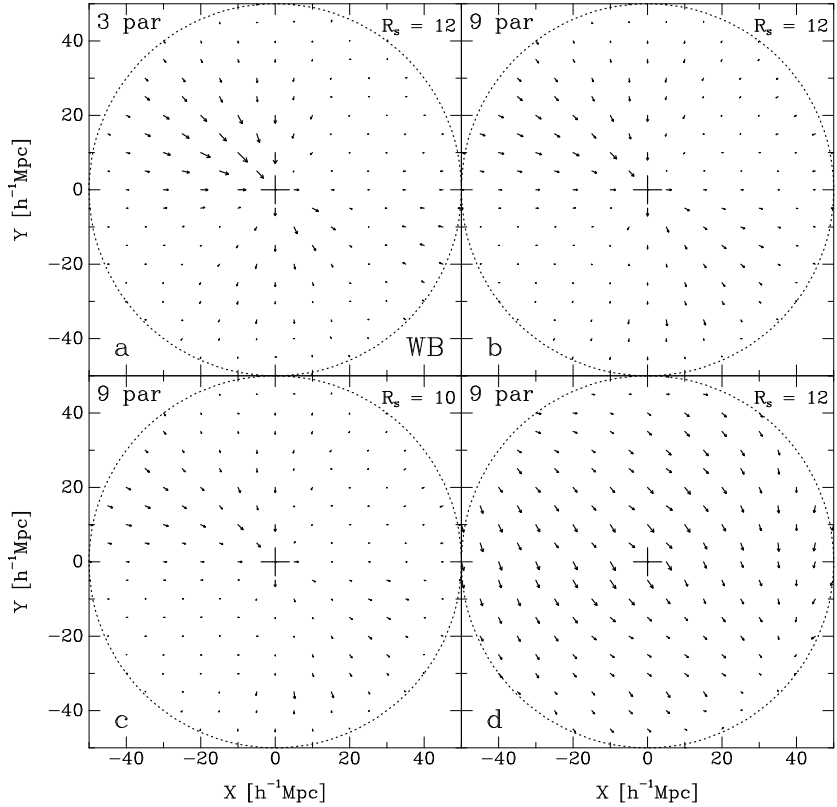}}
\vskip -1.8truecm
\caption{\protect\capt
Window bias in the velocity field in the Supergalactic Plane
as a function of the local velocity model and the smoothing scale.
The input data are the true high-resolution radial velocities of the
simulation, sampled uniformly at grid points of spacing $6\hmpc$.
Distances and velocities are in $\hmpc=100\kms$.
(a) Radial bias in a 3-parameter fit with G12 smoothing.
(b) Remaining radial bias in a 9-parameter fit with G12 smoothing.
(c) Remaining radial bias in a 9-parameter fit with a smaller window,
G10.
(d) Same as (b), but showing the remaining bias in the projected 3D
velocity field.
}
\label{fig:wb_u}
\end{figure}

Figure~\ref{fig:wb_u} (panel a) shows the WB in $u(\vx)$
in the Supergalactic plane, resulting from G12 bulk-velocity 
(3-parameter) smoothing of unperturbed and uniformly-sampled 
radial velocities from the simulation.  In the converging GA region 
(upper-left quadrant), the WB is as high as $-350\kms$. In the outflowing
void in the foreground of PP (bottom right quadrant), it is about $+200\kms$.

\begin{figure}[t!]
\vskip -0.2truecm
\centerline{\epsfxsize=6.5 in \epb{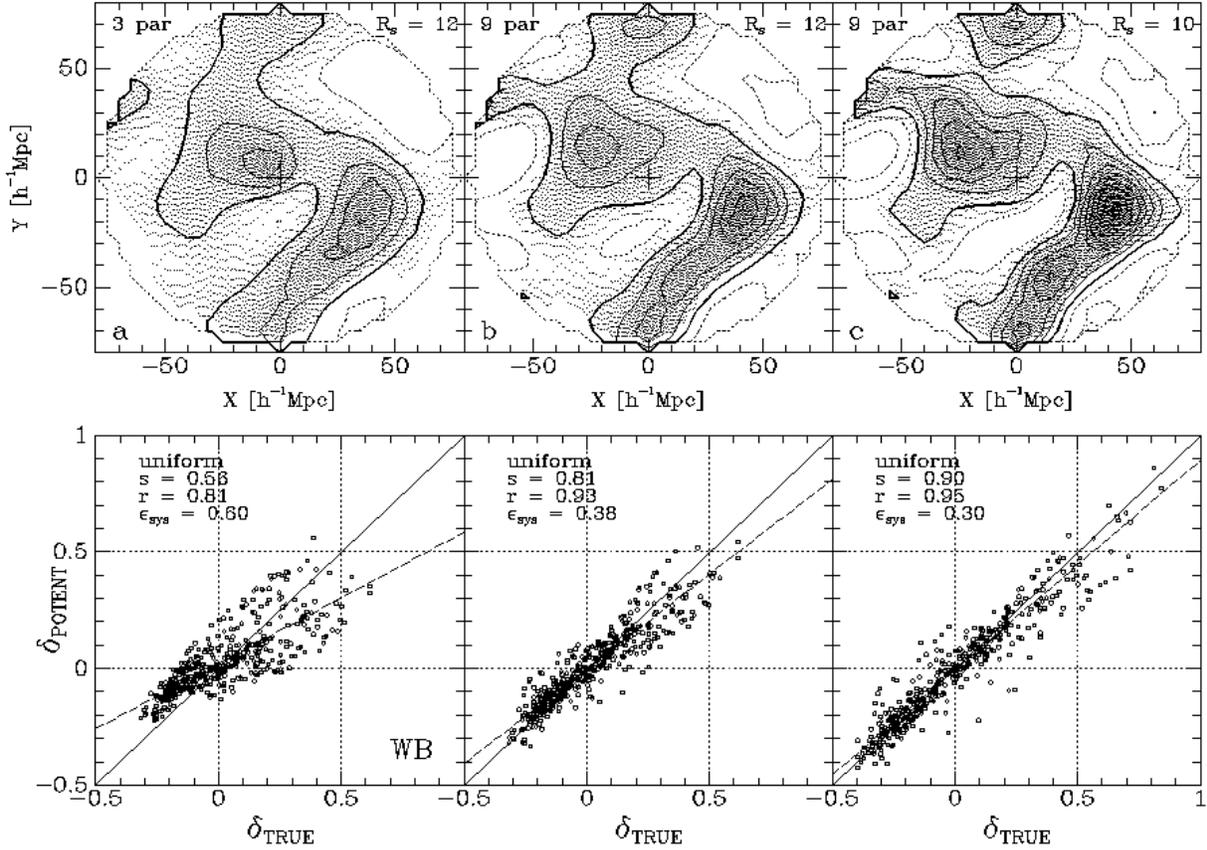}}
\vskip -0.2truecm
\caption{\protect\capt
Window bias in the recovered density field. Mock unperturbed and
uniformly sampled data as in \Fig{wb_u}.
Top: maps in the Supergalactic plane, to be compared to the true fields
with the corresponding smoothings (\Fig{pot}, left panels).
Contours are as in \Fig{mock_map}.
Bottom: point-by-point comparison with the true smoothed density field
inside a sphere of radius $40\hmpc$.
Error statistics are marked.
(a) bulk-velocity fit with 3 parameters, G12.
(b) 9-parameter fit, G12.
(c) 9-parameter fit, G10.
}
\label{fig:wb_d}
\end{figure}

Figure~\ref{fig:wb_d} (left panels) shows the resulting WB in the 
density field, both via a contour map of the biased field in the 
Supergalactic plane and a point-by-point comparison with the true 
density field inside a sphere of radius $40\hmpc$.
A comparison of the map of the recovered density field with the map of 
the true density field (same smoothing; \Fig{pot}, left panels) shows that 
the structures become severely distorted, with the density differences 
between the GA peak and the LG, and between the PP peak and the LG, 
erroneously reduced by factors larger than two.
The global and local biases are characterized by $s=0.56$ and $r=0.81$,
adding up to a large total systematic error of $\epssys=0.59$ relative
to $\sigt$.
The WB is thus a severe systematic error that must somehow be 
reduced.\footnote{This WB has been ignored in certain other applications
of POTENT-like procedures to other data, \eg, da Costa \etal (1996), 
leading to inaccuracies in the reconstruction.} 

The WB can be reduced by introducing shear into the local model velocity field,
\begin{equation}
\vv(\vx)=\vB + \bar{\bar{\vL}} \cdot (\vx-\vx_c) \ ,
\label{eq:high}
\end{equation}
with $\bar{\bar{\vL}}$ a symmetric tensor, which ensures local irrotationality. 
The zeroth-order, bulk-velocity model of 3 parameters is thus extended
into a first-order velocity model of 9 parameters.
The additional free components tend to ``absorb" most of the bias, 
leaving the value of the model velocity at the window center,
$\vv(\vx_c)=\vB$, less biased.

\Fig{wb_u} (right panels) and \Fig{wb_d} (middle panels)
demonstrate the improvement in the WB when the 9-parameter 
model is used. The bias in the velocity field is reduced by about a 
factor of two, the global bias in the density field is improved to $s=0.81$,
and the local bias is much improved to $r=0.93$, such that the total
systematic error in density is down to $\epssys=0.38$.
Adding an additional quadratic term to the model turns out not to 
lead to a significant further improvement. This means that, for the
current data, one has to live with the level of WB remaining in the 
9-parameter fit, which, for G12 smoothing, is still at the level of
$-150\kms$ at the worst point near the GA.
 
Figures \ref{fig:wb_u} and \ref{fig:wb_d} demonstrate that 
the WB can also be reduced by reducing the window size. 
With G10 smoothing, the global bias in the density field is significantly
improved to $s=0.90$, while the local bias is slightly improved 
to $r=0.95$, yielding together $\epssys=0.30$
(\Fig{wb_d}, right).
A window significantly smaller than G10 is impractical with the current
data; sparse sampling and large distance errors would limit the
reconstruction to the very local neighborhood of Virgo and the LG.

Even for G12 smoothing, sampling and distance errors restrict use of 
the 9-parameter model.  The danger of a high-order model is that it 
tends to pick up undesired small-scale noise --- a problem that becomes 
severe at large distances, where distance errors are large. 
Furthermore, the poor sampling at large distances may make it difficult 
to find enough data points ($>9$) within the effective volume of
the window to constrain a 9 parameter model.  Fortunately, at these 
large distances ($r \gg \Rs$) the WB decreases independently of the 
sampling and errors, so one can return there to the simple 
local bulk-velocity fit.  Based on experimenting with the mock catalogs, 
we find the optimal procedure to be a 9-parameter fit out to 
$r=40\hmpc$, a 3-parameter fit beyond $60\hmpc$, and a smooth 
transition region between these radii where we adopt a linear 
weighted mean of the two kinds of fit, smoothly varying from one to the
other.

\vskip -0.5truecm \subsection{Sampling-Gradient Bias}
\label{sec:twf_sb}

\begin{figure}[t!]
\centerline{\epsfxsize=5.0 in \epb{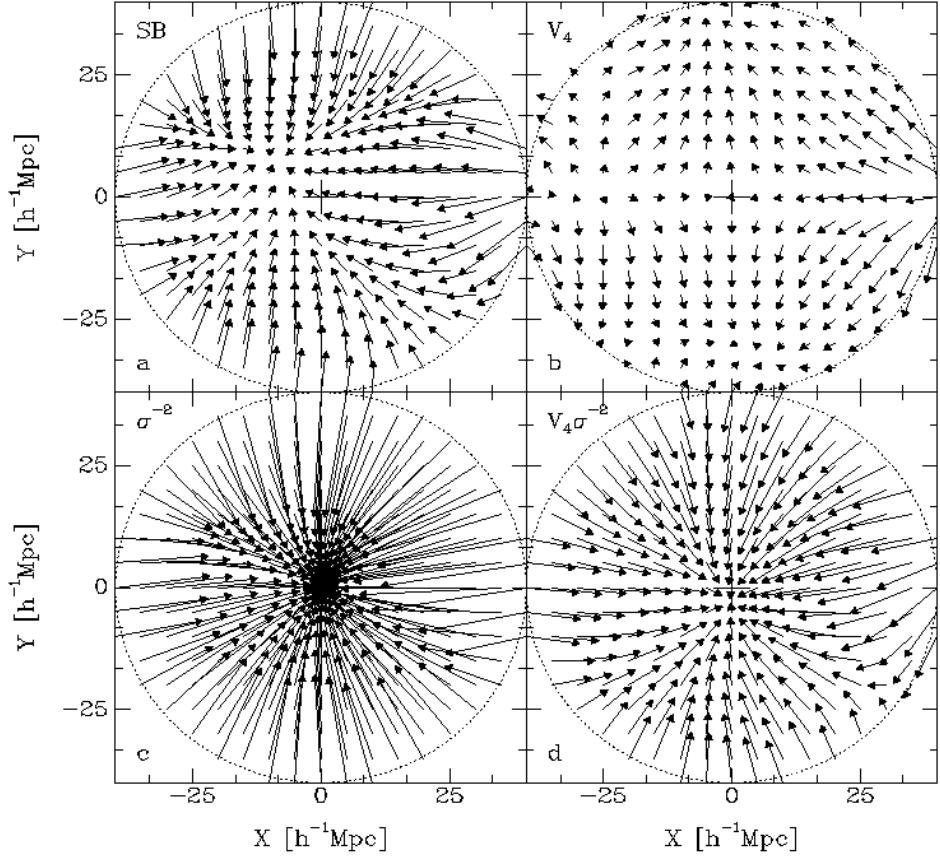}}
\caption{\protect\capt
The origin of sampling-gradient bias.
The displacements of the centers of mass of galaxies
in a $12\hmpc$ Gaussian window from the window centers.
The object positions and weights are from the Mark III catalog.
Shown is the inner Supergalactic plane out to a distance of $40\hmpc$.
(a) Unweighted, refering to the original SB.
(b) Volume-weighted by $V_4$, to minimize SB.
(c) Error-weighted by $\sigma^{-2}$, for optimal treatment of random
errors.
(d) A compromise: weighted by both volume and errors.
}
\label{fig:sb1}
\end{figure}

\begin{figure}[t!]
\centerline{\epsfxsize=5.5 in \epb{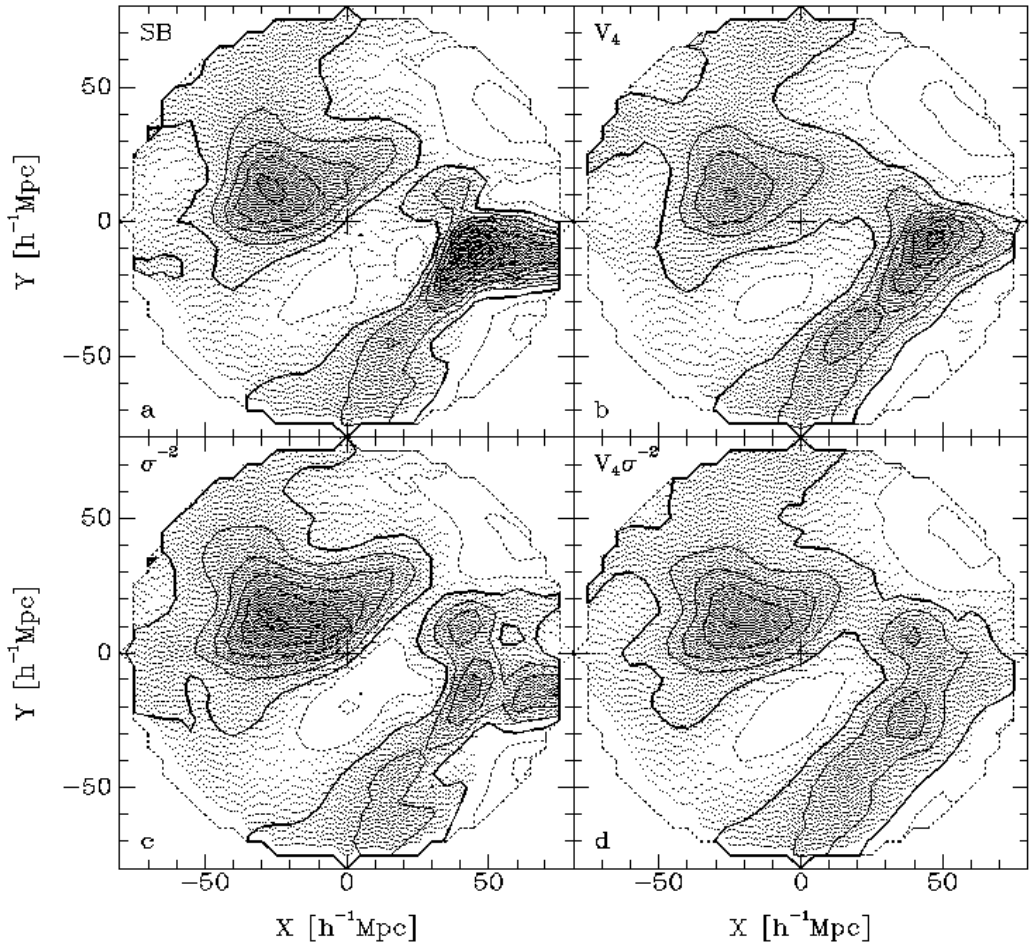}}
\caption{\protect\capt
Sampling-gradient bias in the recovered density field
in the Supergalactic plane.
The input data are the true high-resolution radial velocities of the
simulation, sampled non-uniformly at positions that mimic
the spatial distribution of objects in the Mark III catalog.
Contours are as in \Fig{mock_map}.
The four panels are as in \Fig{sb1}.
}
\label{fig:sb2}
\end{figure}

\begin{figure}[t!]
\vskip -0.2truecm
\centerline{\epsfxsize=5.5 in \epb{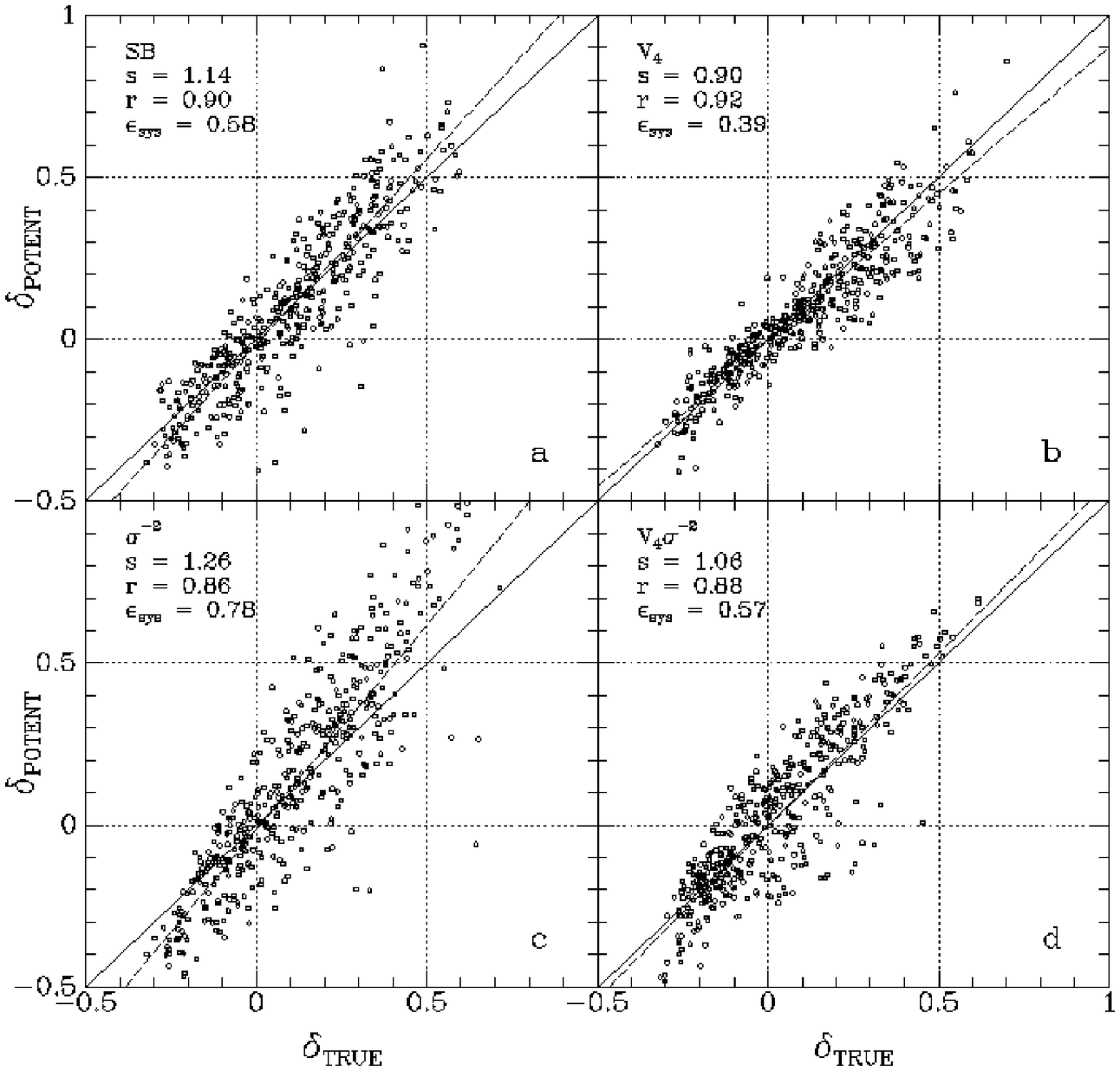}}
\vskip -0.2truecm
\caption{\protect\capt
Sampling-gradient bias inside a sphere of radius $40\hmpc$.
Mock unperturbed and non-uniformly sampled data as in \Fig{sb2}.
The recovered density field is compared to the true G12 field at the
points of a cubic grid, and the statistics measuring the typical
systematic errors are quoted.
The four panels are as in \Fig{sb1}.
}
\label{fig:sb3}
\end{figure}

If the true velocity field is varying within the effective window, 
non-uniform sampling introduces a sampling-gradient bias (SB), 
which has been evaluated analytically in DBF. The smoothing is 
galaxy-weighted whereas the aim is equal-volume weighting. 
In other words, the smoothed velocity as computed so far really
refers to the Gaussian-weighted center of mass of the galaxies 
instead of the geographical window center.
 
Figures~\ref{fig:sb1}, \ref{fig:sb2} and \ref{fig:sb3} address the SB. 
\Fig{sb1} illustrates the origin of the effect in the Supergalactic plane 
by showing the projected displacements of the window-weighted centers of 
mass of the Mark III galaxies from the window centers 
(positioned at the points of a cubic grid of spacing $5\hmpc$).
These displacements are responsible for SB once the velocity field 
is varying within the window.  
In order to isolate the effects of SB (and remaining WB) from the effects 
of random errors, the POTENT procedure with G12 smoothing
is applied to each of 10 special mock catalogs that mimic 
the nonuniform sampling of the Mark III objects, but with the 
{\it true}, unperturbed distances and velocities of the simulation.
In these mock catalogs, the random properties assigned to the galaxies
affect only the selection, not the distances or velocities.
\Fig{sb2} shows maps of the average recovered density field in the 
Supergalactic plane, while \Fig{sb3} compares this field to the true G12 field 
at grid points inside a sphere of radius $40\hmpc$.

Panels (a) in these three figures refer to the uncorrected bias in the 
case of a spherical G12 window with no additional weights.
The displacements in \Fig{sb1}a are typically comparable to the radius of 
the Gaussian window, and they in general follow the large-scale sampling 
gradient towards the origin. The resulting density map in \Fig{sb2}a is 
distorted accordingly, with the peaks of GA and PP overestimated, especially 
near the ZoA.  The local systematic error within $40\hmpc$, \Fig{sb3}a, 
is typically $\epssys=0.58$.

To correct for SB, we wish to weight each object in 
\eq{sumv} by the local volume that it ``occupies", $V_i$.  A simple 
way to estimate $V_i$ (see DBF) is via the inverse of the local density 
at the object as estimated crudely by the cube of the distance $R_n$ 
to its n-th neighboring object, $V_i\!\propto\!R_n^3$. 
For the current sampling density we find best results with $n \sim 4$.
Panels (b) in the three SB figures demonstrate the significant reduction 
in SB when the Gaussian window is weighted by $V_4$.
The displacements of the centers of mass in \Fig{sb1}b are of order of only
a few $\!\hmpc$, much smaller than in panel a.  The density map in 
\Fig{sb2}b becomes much more similar to the true field, \Fig{pot} (left), 
than panel a. The typical systematic error in \Fig{sb3}b is
only $\epssys=0.39$, and the global systematic error, $s=0.90$,  
is reminiscent of the remaining WB, \Fig{wb_d}b, bottom panel.

The volume weighting can be improved further at the expense of making it 
more elaborate.  For example, one can assign cells from a fine grid to 
neighboring objects and weight each cell by the value of the window 
function at the cell position. In this fancier procedure, a given 
object is weighted differently for windows centered at different positions.  
In practice, in view of the much larger distance errors discussed below, 
the simple $R_4$ procedure is presently adequate, 
and there is not much gain in applying the more elaborate procedure
(which may however prove useful for future, better sampled data).

The tentative conclusion for the current Mark III sample is that the $V_4$
volume-weighting procedure could successfully reduce SB to negligible 
levels within 
the region where at least a few objects reside in the effective central 
region of the window. This region typically extends out to $60\hmpc$
from the Local Group outside the Galactic Zone of Avoidance.
However, we shall see next that the need to deal with random distance errors
prevents an optimal SB correction.
 
The same $R_n(\vx)$ field also serves later (\se{err_vol})
as a useful diagnostic 
flag for poorly sampled regions, and as a criterion for excluding such
regions from quantitative analyses.  The displacement of the 
weighted center of mass from the window center likewise serves as a 
complementary flag for regions of high SB.

\vskip -0.5truecm \subsection{Weighting by Random Distance Errors}
\label{sec:twf_err}

Each of the systematic errors in the POTENT analysis can in principle 
be corrected in a satisfactory way. However, the dominant errors in our 
reconstruction are the random errors due to scatter in the 
distance indicators and measurement errors. The random errors 
are particularly large at large distances, where both the error per 
object is big and the sampling is sparse such that shot noise becomes 
a major factor.  

\begin{figure}[t!]
\centerline{\epsfxsize=5.5 in \epb{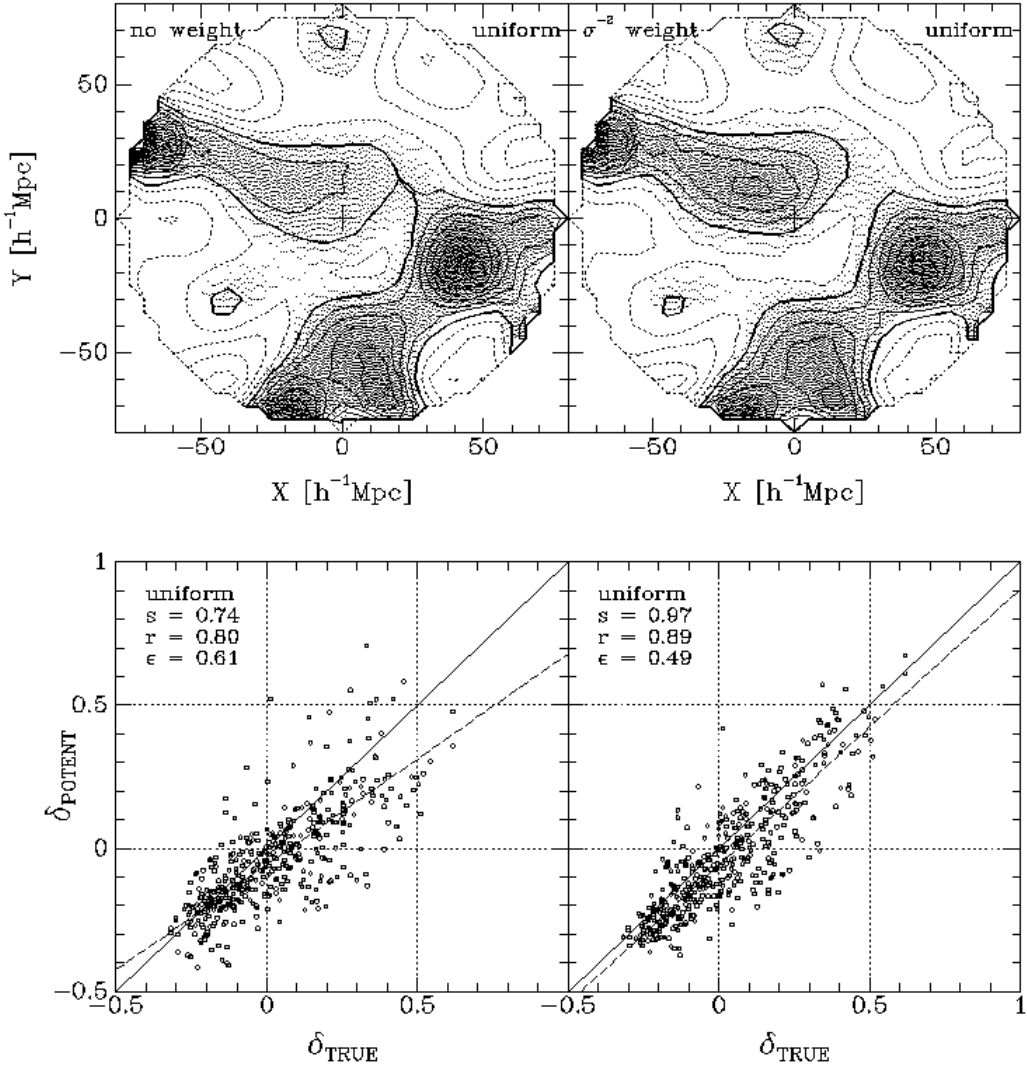}}
\vskip -0.1truecm
\caption{\protect\capt
The effect of error weighting on one mock realization of perturbed
velocities.  The input sampling is uniform at the points of a grid of
spacing
$6\hmpc$, thus eliminating SB.
The input velocities are at true distances,
with the redshifts randomly perturbed instead of the distances,
thus eliminating Malmquist bias.
Top: maps in the Supergalactic plane
of the recovered density field with and without error weighting,
to be compared to the map of the true field (\Fig{pot}, bottom left).
Contours are as in \Fig{mock_map}.
Bottom: corresponding point-by-point comparisons of the recovered fields
with the true density field of the simulation inside a sphere of
radius $40\hmpc$.
Error statistics are quoted.
}
\label{fig:sig_wei}
\end{figure}

The standard way to reduce the effect of this roughly Gaussian noise 
and obtain the most probable smoothed field would be to 
weight the contribution of each object inversely by the variance 
of its distance (velocity) measurement error, \ie,
$W_i\!\propto\!\sigma_i^{-2}$.  
In Figure~\ref{fig:sig_wei}, we isolate the effect of random velocity
errors on the G12-recovered density field from one special mock catalog, and 
demonstrate the improvement resulting from the straightforward error 
weighting by $\sigma^{-2}$. In this case, the sampled galaxies are 
distributed {\it uniformly\,} at the points of a cubic grid 
of spacing $6\hmpc$ inside a sphere of radius $100\hmpc$, thus eliminating
SB if no additional weighting is applied.
Random Gaussian perturbations of rms $20\%$ of the distance 
are assigned to the 
redshifts rather than the distances, thus avoiding the Malmquist bias (and 
the additional SB) that would be introduced by perturbing the distances
(see \se{mb} below). 
The sampling spacing is chosen to be similar to the mean spacing 
in the well-sampled regions of the Mark III catalog (\eg,
within a sphere of radius $22.5\hmpc$ about the LG and in the GA region),
such that the resulting effect of the random errors on the smoothed
fields in the mock realization is characteristic of that in the well-sampled
regions of the real data.

The density maps of \Fig{sig_wei} are to be compared to the true G12 
field (\Fig{pot}, bottom left or \Fig{bias_maps}a, below). 
With no error weighting, the GA density peak is
underestimated and so is the depth of the local void, while the PP density 
peak is overestimated. The error weighting improves all three
effects. The mean field recovered with error weighting still
deviates from the true field because the input is after all severely
perturbed, the sampling is not dense enough to eliminate
shot noise, and the error weighting itself introduces sampling-gradient bias
(that was absent in the unweighted recovery).
A point-by-point comparison of the recovered and true fields
is shown inside a sphere of radius $40\hmpc$.  The statistics
quoted refer in this case to the random errors only (when no
weighting is applied, left panels) plus the artificial SB that is introduced
by the $\sigma^{-2}$ weighting (right panels).
With weighting,
the regression slope drastically improves from $s=0.74$ to $0.97$,
the local scatter improves from $r=0.80$ to $0.89$, and the
total relative error is down from $\epstot=0.61$ to $0.49$. 
We conclude, in this case of uniform sampling and random velocity errors 
only, that the error weighting leads to a significant improvement 
in the recovery.

Unfortunately, in the real case of non-uniform sampling, the error 
weighting somewhat spoils the volume weighting that has been carefully designed 
to minimize the sampling-gradient bias.  
The SB introduced by the $\sigma^{-2}$ weighting alone (without $V_4$ 
weighting) is shown in panels (c) of Figures~\ref{fig:sb1}-\ref{fig:sb3} 
(which address the SB without random errors).
In particular, the $\sigma^{-2}$ weighting tends to strongly bias the 
smoothed velocity towards the velocities in those parts of each window 
that are closer to the LG, where the errors are smaller and the sampling 
is typically denser. It may also bias the results towards the velocities 
of nearby clusters that may have small errors even if they lie relatively 
far from the central region of the window.
The peaks in the density map of \Fig{sb2}c are too high and distorted, and the
statistical measures show large global and local biases
($s=1.26$, $r=0.86$) with a total systematic error of $\epssys=0.78$. 

The standard compromise adopted here (as in BDFDB) is to weight 
simultaneously by the volume weights {\it and\,} the inverse 
square of the random distance errors.
Panels (d) of Figures~\ref{fig:sb1}-\ref{fig:sb3} show the SB 
resulting in this case. With $\epssys=0.57$,
the bias is larger than in the case of weighting
by $V_4$ alone, and is only slightly better than the raw SB bias with 
no weighting at all. However, given the requirement imposed by the random 
errors to weight by $\sigma^{-2}$, the combined weighting is a 
reasonable compromise.
Thus, the value of the adopted window function at the position of object $i$
about the window center $\vx_c$ is of the form
\be 
W(\vx_i,\vx_c) \propto
V_i\, \sigma_i^{-2}\, \exp [-(\vx_i-\vx_c)^2/ 2\Rs^2] \ ,
\label{eq:window}
\ee
up to a normalization factor.
By not allowing the $V_i$ values to differ from their mean value
by more than a factor of five (say), 
the deviation of the smoothed velocity from the most likely signal given 
the noisy data is kept limited to a reasonable range, while the 
SB is still significantly reduced compared to the case of weighting
by $\sigma_i^{-2}$ only.  

Two additional comments regarding the error weighting are appropriate. 
First, as mentioned above, we try to keep the smoothing radius $\Rs$ constant
throughout the volume for the purpose of statistical spatial uniformity that 
allows straightforward direct comparison with theoretical models or
other data of uniform smoothing.  However, if desired, one could easily 
vary $\Rs$ such that the random errors of the recovered fields are kept 
roughly at a constant level throughout the volume (DBF).

Second, the error weighting typically distorts the spherical window 
shape and reduces its effective volume, thus mimicking
smoothing with higher resolution. This effect turns out to roughly
compensate for other effects that cause over-smoothing, such as SB
in empty regions (\se{err_anal}).

After introducing Malmquist bias in the next section, we will address
in \se{errors} the errors of all sources combined.

\section{MALMQUIST BIAS CORRECTION}
\label{sec:mb}

The random scatter in the distance estimator is a source of distinct
systematic biases in the inferred distances and peculiar velocities,
which are generally termed ``Malmquist bias" but should be
carefully distinguished from each other as described below
(\eg, Lynden-Bell \etal 1988; DBF; Willick 1994; 1998).
We describe three different ways of correcting for Malmquist bias
and compare the results.
The Malmquist correction is actually applied to the peculiar-velocity data
in a preliminary stage, before they are fed into the main POTENT analysis,
but we discuss it only here, after discussing the rest of the method,
because it is a complication that
is relatively independent of the other steps of the POTENT procedure,
and we prefer to test it with the full POTENT machinery in hand.

\vskip -0.5truecm \subsection{Forward TF Correction}
\label{sec:mb_fmb}

The calibration of the ``forward" TF relation 
$M_{\sss \rm TF} (\eta)$ is affected
by a {\it calibration\,} bias (or {\it selection\,} bias).
An apparent magnitude limit in the selection of the sample used for
calibration at a fixed {\it true} distance (\eg, in a cluster)
tilts the forward TF regression line of $M$ on $\eta$ towards bright
magnitudes $M$ at small $\eta$ values.  This bias may extend to all values of
$\eta$ when objects over a large range of distances are used for the
calibration.  This bias is inevitable when the dependent quantity (here $M$)
is explicitly involved in the selection process, and it occurs to a
certain extent even in the ``inverse" TF relation 
$\eta_{\sss \rm TF} (M)$ due to
weak dependencies of selection on $\eta$.  The calibration bias
can be properly corrected once the selection function is known, as explained,
\eg, in Willick \etal (1995), and we assume hereafter that the given Mark III
TF parameters are unbiased (but see \se{maps_vm2}).

Even after the selection bias in the TF calibration is properly corrected,
the TF-inferred distance, $d$, and therefore the mean peculiar velocity
at a given $d$, suffers from an {\it inferred-distance} bias, which we term
hereafter ``Malmquist" bias or ``MB".

This bias can be reduced by grouping, and corrected in a statistical way.
We devote a separate paper (Eldar, Dekel \& Willick 1999)
to the issue of Malmquist-bias in inferred-distance space, both in
forward and inverse TF analyses, where we describe in more detail
our correction procedures and test them with realistic mock data
based on N-body simulations.  Only a brief summary is provided here.

Malmquist bias can be quantified as follows.  If the magnitude $M$ is
distributed normally for a given log-velocity parameter $\eta$, with standard
deviation $\sigma_m$, then the TF-inferred distance $d$ of a galaxy at
a true distance $r$ is distributed log-normally about $r$,
with relative error $\Delta=(\ln 10 /5) \sigma_m$. 
Given $d$, the expectation value of $r$ is (\eg, Willick 1998):
\be
E(r \vert d)=
{ \int_0^\infty r P(r\vert d) {\rm d}r
\over
\int_0^\infty P(r\vert d) {\rm d}r } =
{ \int_0^\infty r^3 n(r)\ {\rm exp}
\left( -{[{\rm ln}(r/d)]^2 \over 2\Delta^2} \right) {\rm d}r
\over
\int_0^\infty r^2 n(r)\ {\rm exp}
\left( -{[{\rm ln}(r/d)]^2 \over 2\Delta^2} \right) {\rm d}r } \ ,
\label{eq:fmb}
\ee
where $n(r)$ is the number density of galaxies in the underlying distribution
from which the galaxies were selected for the catalog 
(using $m$ and $\eta$ alone).
Strictly, $n(r)$ depends on the particular
line of sight to each galaxy. The deviation of $E(r\vert d)$ from $d$
is the Malmquist bias.

The ``homogeneous" part (HM) refers to a constant $n(r)$ and
arises from the geometry of space ---
the inferred distance $d$ underestimates $r$ because it is more likely
to have been scattered by errors from a larger true distance $r>d$
than from a smaller true distance $r<d$, because the corresponding
volumes are proportional to $r^2$.  Another contribution to the HM bias
arises from the fact that the distribution of $d$ about
$r$ is not symmetric (it is $\log d$ that is symmetric about $\log r$).
Quantitatively, if $n\!=\!const$, then \eq{fmb} reduces to
\be
E(r\vert d) = d\, e^{3.5 \Delta^2} \ .
\label{eq:HMB}
\ee
If all objects have the same error,
the inferred distances are simply multiplied by a constant factor,
increasing the distance by, \eg, about 10\% for $\Delta=0.17$.
This is equivalent to a global change of the zero point $a$
of the forward TF relation, \eq{FTF}.
Grouped objects have smaller errors, and their corrections are smaller.
The HM bias has been corrected this way on a regular basis (\eg, in
Lynden-Bell \etal 1988).

Fluctuations in $n(r)$ are responsible for the ``inhomogeneous" Malmquist bias
(IM), which is worse because it systematically enhances the amplitude
of the inferred density perturbations (and, consequently, the inferred
value of $\Omega$).
If $n(r)$ is varying slowly in the range $\pm \Delta r$ about $r$,
and if $\Delta\ll 1$, then \eq{fmb} reduces to
\be
E(r\vert d) = d\, \left[ 1 + 3.5\Delta^2
+ \Delta^2 \left( {d \ln n \over d \ln r}\right)_{r=d} \right] \ ,
\label{eq:imbapprox}
\ee
explicitly showing the dependence on $\Delta$ and on the gradient
of $n(r)$.  To illustrate the origin of IM bias, consider a lump of
galaxies at one point $r$ with zero true peculiar velocity, $u\!=\!0$.
Their inferred distances are randomly scattered to the foreground and
the background of $r$.  With all galaxies having the same redshift $z=r$,
the inferred values of $u$ on either side of $r$ mimic a spurious infall
towards $r$, which is then interpreted by the dynamical analysis
as an attractor with a spurious overdensity at $r$.

In the forward-TF Mark III data for \pot\ analysis,
the standard correction for IM bias consists of two steps.
First, the galaxies are properly grouped to ``objects" as discussed above
(\se{m3}; Willick \etal 1996),
reducing the distance error of each group of $N$
members to $\Delta/\sqrt N$ and thus significantly weakening the bias.
The noisy inferred distance of each object, $d$, is then replaced by
$E(r\vert d)$ of \eq{fmb}, based on an assumed galaxy density profile
$n(r)$ and with additional corrections for redshift limits in the data
and for the grouping, as follows.

The practical uncertainty in the IM correction procedure is in the
assumed function $n(r)$ along each line of sight.
In principle, this could be obtained from the POTENT-recovered
mass density itself through an iterative procedure under certain assumptions
about how galaxies trace mass. However, the resolution provided by the Mark
III data is not sufficient at large distances, where the IM bias is large and
an accurate correction is needed.
Instead, we approximate $n(r)$ from the high-resolution galaxy distribution
in redshift surveys.  For spirals, we use the galaxy density field in
real space as recovered from the IRAS 1.2 Jy redshift survey, which
is dominated by spirals. This recovery was done by
Sigad \etal (1998) with G5 smoothing via a power-preserving filter
assuming $\Omega=1$, no biasing $b=1$, and mildly non-linear corrections based
on Nusser \etal (1991).
For ellipticals and S0's, we use a similar density field as recovered
by Hudson (1995) from a survey of optical, mostly early-type galaxies.
(This IM correction is similar but not identical to that published
in Willick \etal 1997a.)

The redshift limits $z_c$, which are present in some of the Mark III datasets,
are approximated as cutoffs in $n(r)$ at a given distance $r_c$.
For most datasets, where the cutoff is beyond $8500\kms$, we adopt $r_c=z_c$.
For the Aaronson \etal (1982) data, which are truncated at a heliocentric
redshift $z_c=3000\kms$, we compute $r_c$ under the assumption of a constant
bulk flow $\vB$ across the sampled volume,
\ie, $r_c = z_c - \vB\cdot\hat{\vx}$ in the direction $\hat{\vx}$.
Based on our results in \se{bulk},
we adopt for this correction a bulk flow of amplitude $350\kms$
in the direction $l=310^\circ$, $b=10^\circ$.
(The actual bulk flow used in this MB correction has only a small effect
on the measured bulk flow from the corrected data. Anyway, by using in the MB
correction the same bulk flow as finally obtained from the corrected data,
the correction is made self-consistent.)

When the data are grouped, the density profile $n(r)$ which enters the
integrals in \eq{fmb} has to be multiplied by another correction factor.
In principle, the distances of all groups of a given richness, which have
the same relative error, should be corrected using specifically
just the number density
of such groups. Because of the limited number of objects, we
simplify the procedure and distinguish only between grouped and
ungrouped galaxies.  The density profile that enters \eq{fmb}
for ungrouped galaxies is multiplied by $f_{\rm ug}(r)$,
the fraction of ungrouped galaxies in the vicinity of $r$ along the
given line of sight. For groups, the number density is multiplied by
$1-f_{\rm ug}(r)$.  Because of the sparse sampling, we use for
$f_{\rm ug}(r)$ the spherical average within each dataset.
This is one source of inaccuracy in the correction.

\vskip -0.5truecm \subsection{Comparison to Inverse TF Correction in Inferred-Distance Space}
\label{sec:mb_imb}

\begin{figure}[t!]
\centerline{\epsfxsize=5.5 in \epb{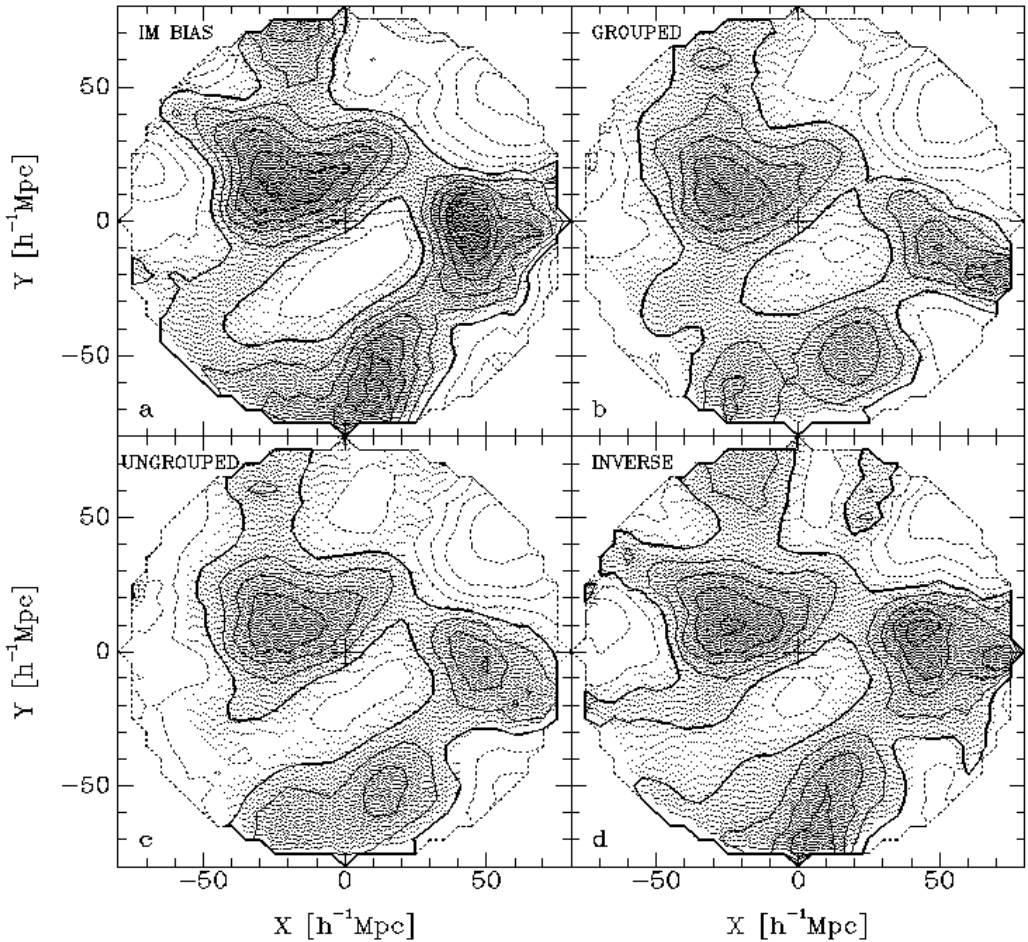}}
\caption{\protect\capt
Malmquist bias corrections.
Maps of POTENT G12 $\delta$ fields averaged over 10 random mock
catalogs, corrected for homogeneous and inhomogeneous Malmquist bias in
three
different ways.
The maps are to be compared to the true G12 field (\Fig{pot}, bottom
left).
Contours are as in \Fig{mock_map}.
(a) Forward correction for HM bias only,
(b) Forward IM-bias correction applied after grouping,
(c) Forward IM-bias correction applied with no grouping,
(d) Inverse IM-bias correction in inferred-distance space.
}
\label{fig:mb_maps}
\end{figure}

\begin{figure}[t!]
\centerline{\epsfxsize=5.5 in \epb{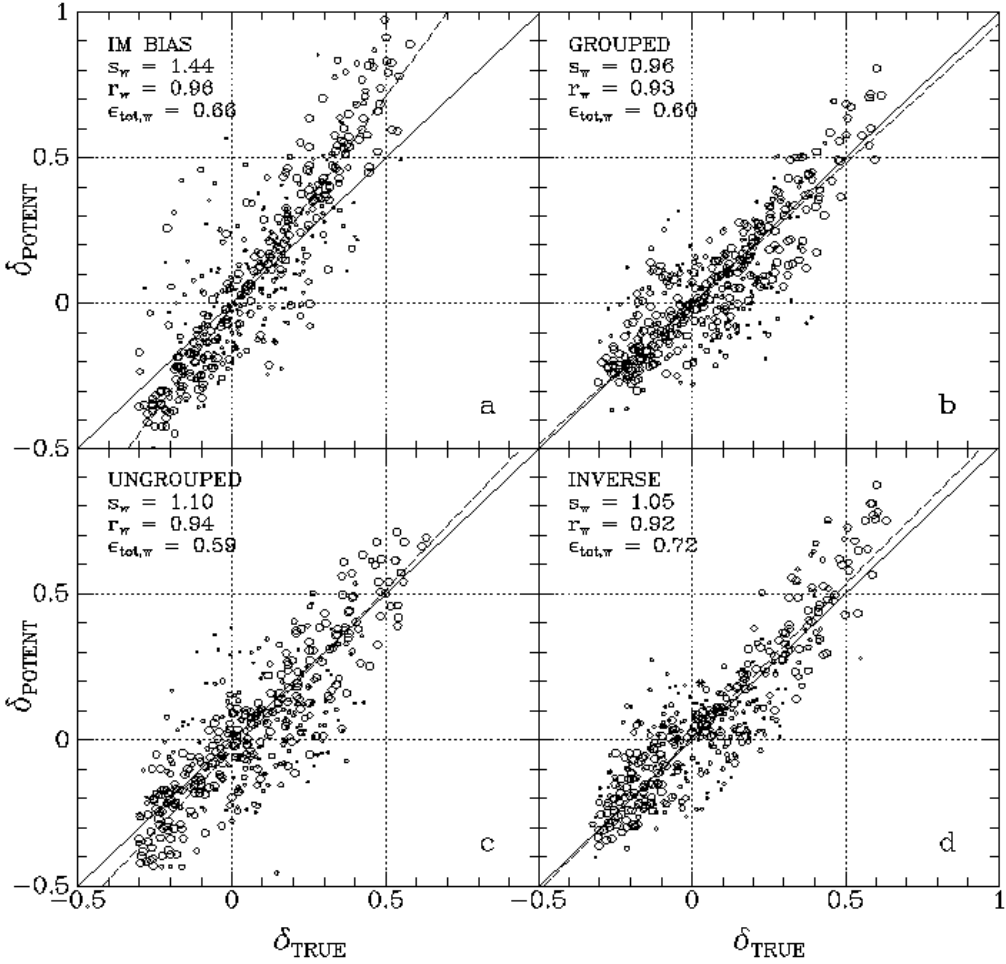}}
\vskip -0.2truecm
\caption{\protect\capt
Malmquist bias corrections.
Point-by-point comparisons of the {\it average\,} G12 $\delta$ fields as
recovered from 10 mock catalogs with the
different correction procedures of \Fig{mb_maps} and the true field.
The area of each symbol
is inversely proportional to the error $\sigd^{2}$ as
determined by the dispersion over the random realizations (\se{errors}).
The weighted regression lines are shown, and the corresponding
weighted statistics are quoted (defined in \se{eval}).
}
\label{fig:mb_scat}
\end{figure}

\def\dinv{d_{\rm inv}}
\def\Dinv{\Delta_{\rm inv}}
Distances, $\dinv$, can alternatively be inferred via the {\it inverse\,}
TF relation between the log-velocity parameter $\eta$ and the
magnitude $m$, namely,
\be
\eta_{\sss \rm TF} (M) = a_{\rm inv} - b_{\rm inv} M, \quad {\rm where} \quad
M=m-5 \log \dinv \ ,
\label{eq:ITF}
\ee
by identifying $\eta_{\sss \rm TF} (M)$ with the observed $\eta$.
The Malmquist bias is corrected in this case by Eldar, Dekel \& Willick (1999)
following the method proposed by Landy \& Szalay (1992). Instead of the
externally supplied density run $n(r)$, the sufficient input in this case 
is the number density of galaxies $\tilde n(\dinv)$, which is in 
principle derivable from the sample itself.  Under the assumption 
that the selection is independent of $\eta$,
the expectation value of the true distance $r$, given $\dinv$, is
\begin{equation}
E(r|\dinv ) = \dinv\ e^{3.5 \Dinv^2}\
{\tilde n(\dinv\, e^{\Dinv^2}) /\tilde n(\dinv ) } \ ,
\label{eq:imb}
\end{equation}
where $\Dinv \equiv (\ln 10/5)\, \sigma_\eta / b_{\rm inv}$
is the relative error in $\dinv$ (in which $\sigma_\eta$ is the scatter in the
inverse TF relation), and is assumed to be small. 
In practice, the sparseness of the sample makes the derivation of
$\tilde n(\dinv)$ non-trivial, and this is the main source of error in this
MB correction procedure.

Figures~\ref{fig:mb_maps} and \ref{fig:mb_scat} demonstrate, in maps
and point-by-point comparisons with the true $\delta$ field,
the success of the full POTENT reconstruction including the 
full Malmquist-bias correction done in three different ways:
a forward correction via \eq{fmb} with a preliminary
grouping procedure, a forward correction without grouping,
and an inverse correction via \eq{imb}.
Shown as a reference is the case where no special correction was 
applied beyond the simple forward correction for HM bias.       
The input data are the fully perturbed mock catalogs, and shown in the
figure are the average fields over 10 realizations.
All the IM-bias corrections turn out to be reasonably good, 
with a global bias of only a few percent ranging from $\sw=0.95$ to $1.10$,
and a local bias measured by $\rw=0.92$ to $0.94$.  
(Note that once the fully perturbed mock catalogs are used we started
quoting the ``weighted" statistics, as defined at the end of \se{eval} 
and using the error weights $\sigd=\sigran$ specified in \se{err_vol} below;
this is motivated by the assertion that subsequent analyses using POTENT 
output may use similar weights.) 
The total error is at the level of $\sim 60\%$ and $70\%$ of 
$\sigt$ for the forward corrections and the inverse correction
respectively.
The ``grouped" correction seems to be overall the best in terms of
these statistics, but the differences in the quality of the
reconstructions are small. In fact, the ranking of the methods
should be different in different regions, depending on the sampling
within each data set and on the true field itself.
In \se{maps_mb} below, \Fig{gsi}, we show a similar comparison of the
three MB correction methods as applied to the real data.
It is important to realize that the effect of the IM correction of the
grouped Mark III data on the resultant density fluctuation with G12 smoothing
is less than 20\% even at the highest peaks.
The bottom line is that the correction procedure reduces any remaining
effects of Malmquist bias to the level of a few percent in the density field.
The similarity between the results based on the different correction methods
confirms the validity of our M-bias correction procedures.

It is worth mentioning that, in principle,
Malmquist bias can be avoided altogether by performing the inverse TF
analysis in redshift space using a parametric model for the velocity
field (the ``Schechter" method,
\eg, Davis, Nusser \& Willick 1996; Blumenthal, Yahil \& Dekel 1999).
However, this is done at the expense of a more
complicated procedure, it involves other subtle biases, and it is based on
a variable effective smoothing which does not
straightforwardly provide a uniform reconstruction in real space.

\section{REMAINING ERRORS IN POTENT}
\label{sec:errors}

Having addressed the main sources of errors in POTENT one by one
and having tried to minimize their effects on the recovered fields,
we conclude the discussion of the method by evaluating its performance
under realistic conditions, in which all the different sources of error
are present simultaneously. In this section we use the fully perturbed
mock catalogs described in \se{mock}. Error flags are used to
define standard ``reference volumes" within which the quantitative evaluation 
is pursued. In the following text and figures we refer mostly to the 
``weighted" version of the evaluation statistics (\se{eval} and below),
but we list both the unweighted and weighted
statistics in 
Tables 2 and 3. 

\vskip -0.5truecm \subsection{Reference Volumes}
\label{sec:err_vol}

\begin{figure}[t!]
\centerline{\epsfxsize=5.5 in \epb{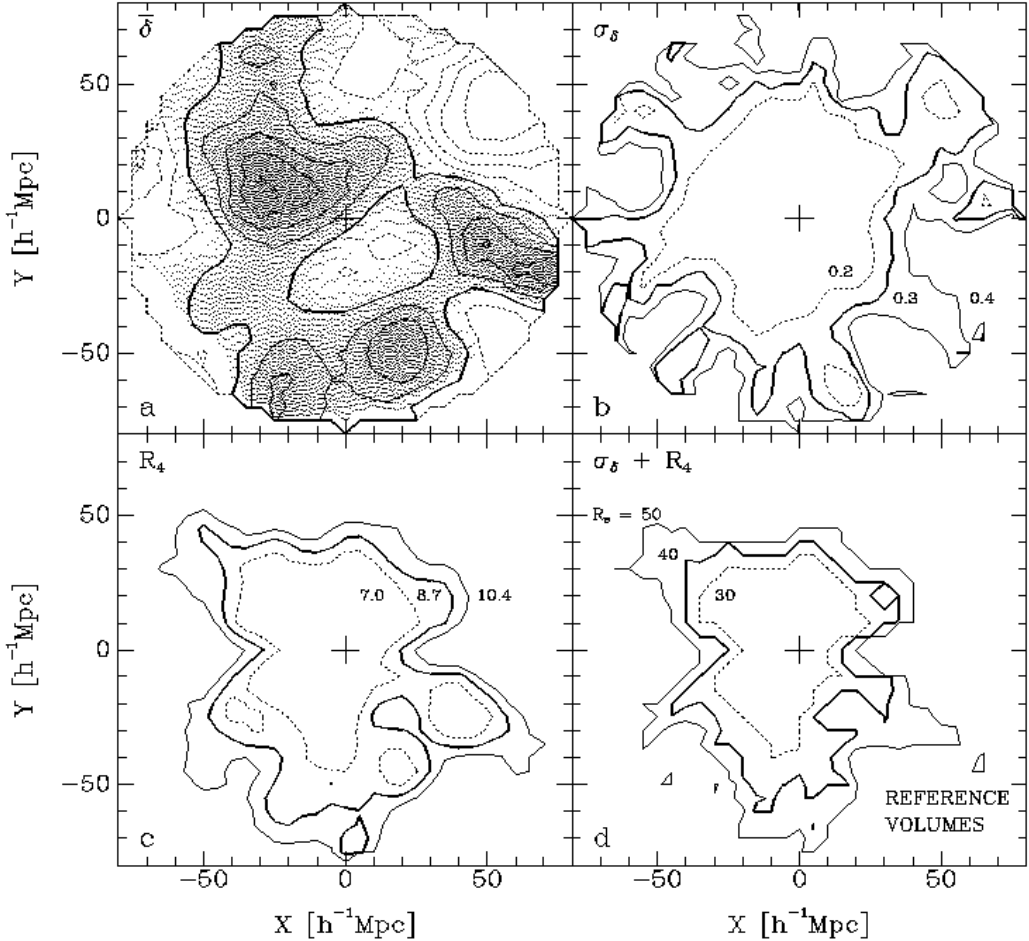}}
\vskip -0.2truecm
\caption{\protect\capt
Reference volumes defined by errors in the Supergalactic plane.
The maps are derived from G12 POTENT reconstructions of 10 mock
catalogs.
(a) The average $\delta$ field in the mock catalogs.
Contours are as in \Fig{mock_map}.
(b) The standard deviation of the 10 $\delta$ fields.
Contours from inside out are for $\sigd = 0.2$, 0.3 and 0.4.
(c) The average $R_4$ field. Contours from inside out refer to
$R_4= 7.0$, $8.7$ and $10.4\hmpc$.
(d) Combined contours, of the pairs
$(\sigd,\, R_4)= (0.2,\, 7.0),\ (0.3,\, 8.7),\ (0.4,\,10.4)$,
corresponding to our reference volumes with effective radii of
$\Re= 30$, 40, and $50\hmpc$ respectively.
}
\label{fig:vol_maps}
\end{figure}

The errors in the recovered fields are assessed empirically
from the 10 random-realization mock catalogs, as explained in \se{eval}.
These are the mock catalogs described in \se{mock},
which fully mimic the sampling and perturbed distances in the Mark III catalog.
POTENT is applied to each of the mock catalogs, and the error at each
grid point is the rms value over the realizations of $\delta$ there,
either about their average ($\sigran$) or about the true density
field ($\sigtot$).  In the present paper, we denote by $\sigd$
the $\sigran$ of the density field.
The velocity error $\sigv$ can be computed similarly, component by component.
This error estimate based on mock catalogs from simulations replaces
the error estimate of DBF, which was based on a series of perturbed
realizations of the real data. The new procedure avoids an artificial
Malmquist-like bias that was problematic in the old procedure
(see discussion in Dekel \etal 1993), and it includes the shot noise
due to discrete sampling.

As an additional criterion for the quality of the reconstruction at
each grid point, we use the same ``emptiness" parameter $R_4$ used
for volume weighting (\se{twf_sb}), which also serves as
a crude measure of sampling-gradient bias.
The error flags $R_4$ and $\sigd$ are not independent everywhere;
they can be either correlated or anti-correlated in different places.
We find the correlation coefficient of $R_4$ and $\sigd$
inside spheres of radii $40$ and $80\hmpc$ to be $r = 0.50$ and $0.46$
respectively (where $r=0$ refers to no correlation).
As an example of a local anti-correlation between these error flags, consider
empty regions such as the ZoA, in which the data weighted by the
window function may be dominated by the velocities of many objects
outside the central region of the window. The derived smoothed
velocity in the empty region can thus be quite stable to random variations
in these many inputs, despite the fact that hardly any local information
is provided. In this case, the sampling bias introduces additional smoothing,
which causes $\sigd$ to be artificially small. This provides additional
motivation for using $R_4$ as a second criterion for an a priori evaluation
of the quality of the reconstruction at a given point.

\begin{figure}[t!]
\centerline{\epsfxsize=6.5 in \epb{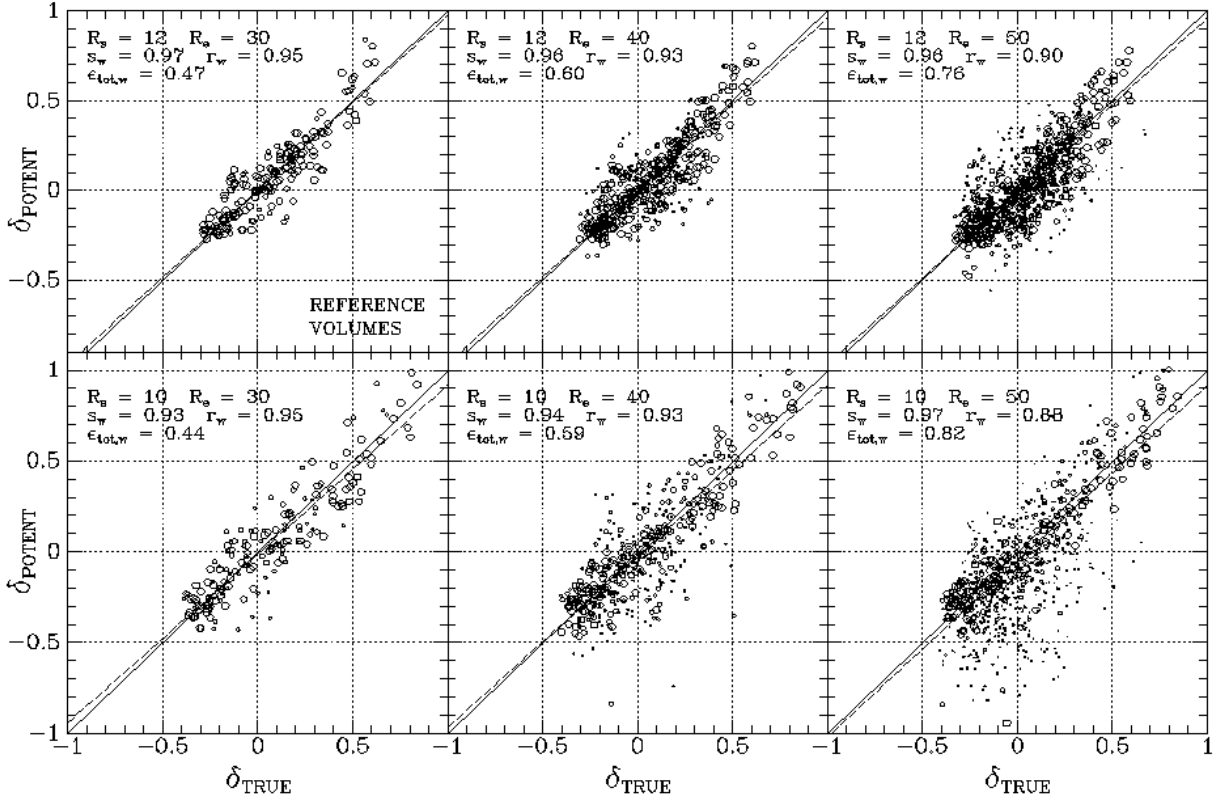}}
\vskip -0.2truecm
\caption{\protect\capt
Biases in different reference volumes.
Average reconstructed $\delta$ fields from 10 mock catalogs
are compared to the true field of the simulation point by point
inside reference volumes of effective radii $\Re=30$, 40 and $50\hmpc$.
The smoothing scale is $\Rs=12$ or $10\hmpc$.
Symbols are as in \Fig{mb_scat}.
The weighted regression line is shown, and the corresponding
weighted statistics are quoted.
The total error is the sum in quadrature of the systematic
and random errors (\eq{sigtot}).
}
\label{fig:vol_scat}
\end{figure}

\Fig{vol_maps} shows maps in the Supergalactic plane
of $\sigd$ (G12) and $R_4$, and a map of boundaries defined by pairs
of values:  $\sigd < \sigdm$ and $R_4 < \R4m$.
We define the ``effective radius" of the volume encompassed by a given
error surface as the radius of a sphere that has the same volume.
For example, with $\sigma_{\delta, {\rm max}}=0.3$ and
$\R4m = 8.7\hmpc$, the effective radius is $\Re = 40\hmpc$.
High-quality reconstruction is limited to effective radii
of order $40\hmpc$, while medium-quality reconstruction can extend out to
$\sim 60\hmpc$ or more in certain directions, such as the parts of the GA
that lie outside the ZoA.

\Fig{vol_scat} shows the systematic errors, via point-by-point comparisons 
of the average reconstructed $\delta$ fields over the mock realizations 
and the true density field of the simulation.  The comparison is made 
inside three different volumes of comparison, of effective radii 
$\Re=30$, 40, and $50\hmpc$.  The criteria are as in \Fig{vol_maps}. 
For the G10 smoothing, the error criteria are
$(\sigd,\, R_4)= (0.25,\,7.5),\, (0.35,\,10.7)$ and $(0.45,\,13.5)$
for the same effective radii respectively.
The global bias is consistently small in this range of smoothings and
reference volumes, ranging from $\sw=0.97$ to $0.93$.
The local bias naturally increases as the volume becomes larger
and the smoothing length becomes smaller.
For both smoothings, the total error becomes larger than 70\% of $\sigt$
only near $\Re = 45\hmpc$ or beyond.
Note in particular the large scatter of points with large errors
in the $\Rs=10$, $\Re=50$ case.

\vskip -0.5truecm \subsection{Error Analysis}
\label{sec:err_anal}

We are now in a position to conclude the error analysis of POTENT
based on the mock catalogs.
Tables 2 and 3 
summarize the performance of POTENT with mock data
of gradually increasing complexity. In each case, the
various measures of errors in the reconstructed mass-density
field are given in comparison with the true density field,
as defined in \se{eval}.
The statistics reported in 
Table 2 
treat all the grid points
within the reference volume equally,
while the statistics reported in 
Table 3 
weight the data at each
grid point by the local random POTENT error there, $\sigd^{-2}$.
These tables provide a full summary of the testing results.
They refer to corresponding sections and figures, where appropriate.

\begin{figure}[t!]
\centerline{\epsfxsize=5.5 in \epsfbox{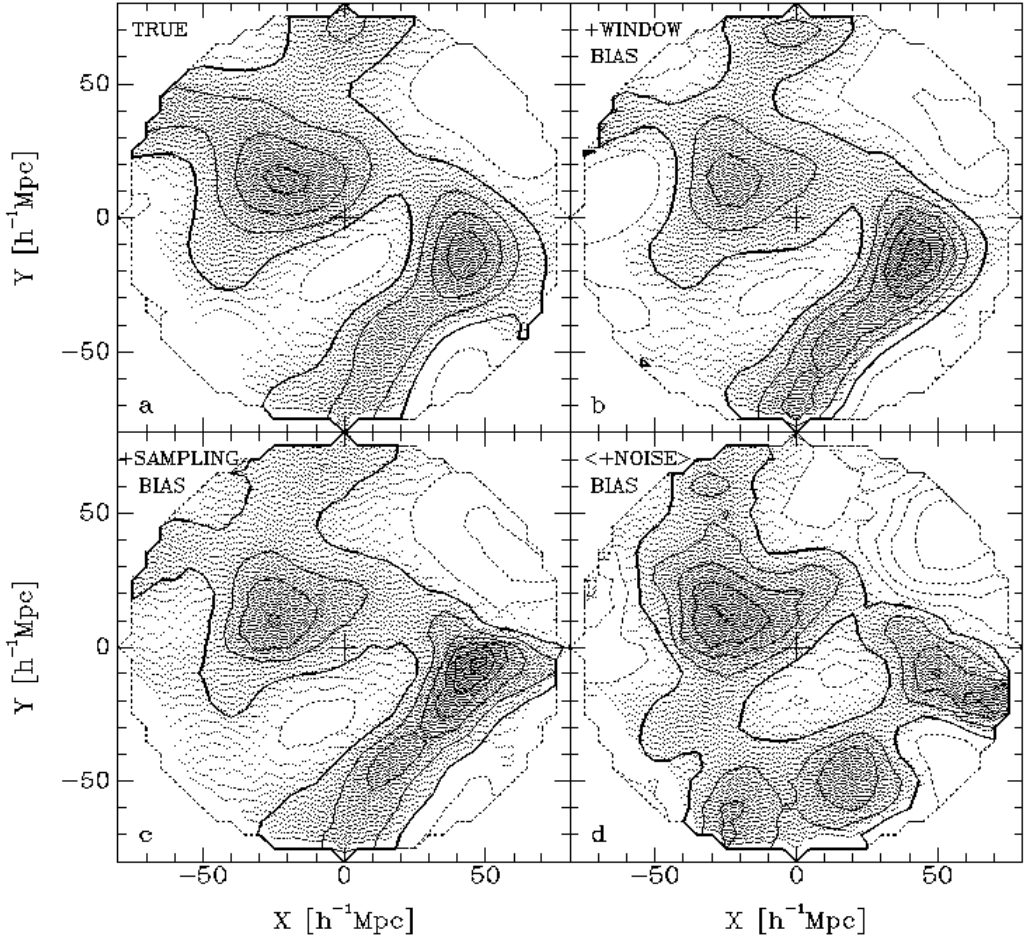}}
\caption{\protect\capt
A summary of systematic errors in POTENT. Shown are maps of the G12
density fluctuation field $\delta$ in the Supergalactic plane
of the mock catalogs.
Contours are as in \Fig{mock_map}.
(a) The true field of the simulation
(same as \Fig{pot}, bottom-left panel).
(b) The effect of window bias as shown by a POTENT reconstruction
from noiseless radial velocities sampled uniformly at dense grid points
of spacing $6\hmpc$, smoothed with an unmodified Gaussian window
and using a 9-parameter fit (as in \Fig{wb_d}b, \se{twf_wb}).
(c) The effect of sparse and non-uniform sampling as demonstrated
by the average field of 10 POTENT reconstructions from noiseless
mock catalogs, with a window corrected for equal-volume weighting,
$V_4$ (\se{twf_sb}).
(d) The total systematic error field, including the effect of distance
errors, as shown by the average field of 10 POTENT reconstructions from
noisy
mock catalogs, corrected for Malmquist bias
(\se{mb}), with additional $\sigma^{-2}$ weighting (\se{twf_err}).
}
\label{fig:bias_maps}
\end{figure}
\begin{figure}[t!]
\centerline{\epsfxsize=5.5 in \epsfbox{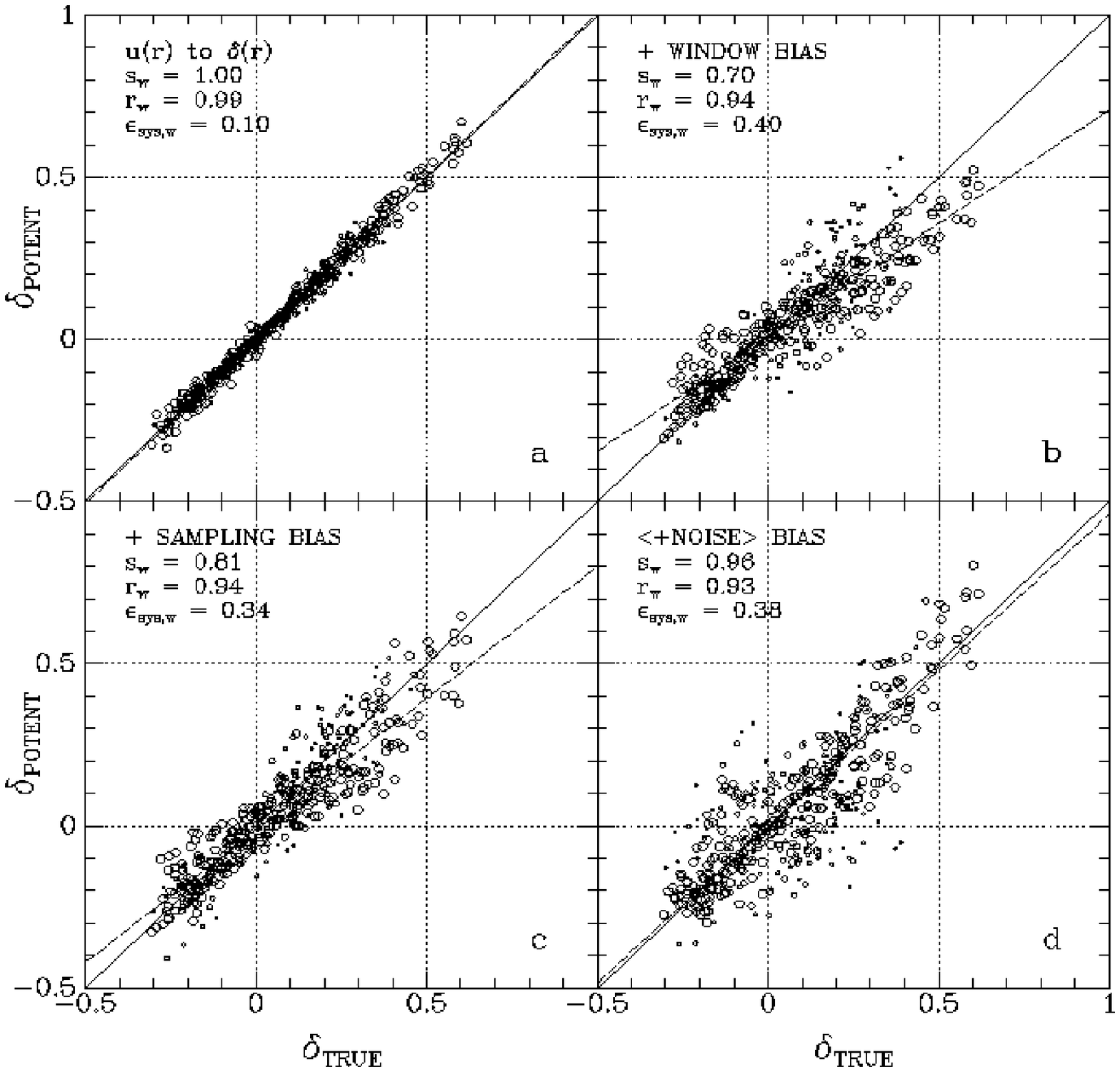}}
\vskip -0.2truecm
\caption{\protect\capt
A Summary of systematic errors in POTENT. Shown are point-by-point
comparisons of the G12 reconstructed density fields from the mock
catalogs
and the true density field of the simulation.
The comparison is made at points of a cubic grid of spacing $5\hmpc$
within a sphere of radius $40\hmpc$.
(a) POTENT reconstruction from perfect input of the exact radial
components of the true G12 velocity field
(same as \Fig{pot}, bottom-right panel).
Panels (b), (c) and (d) are the same as the corresponding panels of
\Fig{bias_maps};
panels (c) and (d) refer to the average reconstructed fields of 10
noiseless and noisy mock catalogs respectively.
Symbols, regression lines and statistics are as in \Fig{mb_scat}.
}
\label{fig:bias_scat}
\end{figure}

The remaining {\it systematic errors\,} after the optimization of POTENT 
are shown in Figures~\ref{fig:bias_maps} and \ref{fig:bias_scat}.
In accordance with the tables,
we show a sequence of POTENT output density fields in which the 
complexity of the mock input data drawn from the simulation grows gradually
as follows:
\begin{list}{}
            {\parsep 0in \itemsep .1cm \topsep -0pt} 
\item[(a)]
The input consists of the radial components of the {\it true\,} {\it G12\,}
velocities on a spherical grid, as in \se{pot_test} 
(relevant only to \Fig{bias_scat}).
\item[(b)]
The input is the true radial velocities sampled densely and {\it uniformly\,}
and then G12 smoothed by POTENT with 9 parameters nearby and 3 parameters at 
large radii. The result shows the remaining window bias (\se{twf_wb}).
\item[(c)]
The input of each of 10 realizations consists of the true radial velocities 
sampled at random, {\it sparsely\,} and {\it non-uniformly\,}, 
mimicking the statistical properties of the Mark III sample. 
The average result shows the additional effect of sampling-gradient 
bias remaining after volume weighting by $V_4$ (\se{twf_sb}).
\item[(d)]
The input of each of 10 realizations consists of fully {\it perturbed\,} 
positions and (therefore perturbed) radial velocities 
sampled at random to mimic Mark III, as described in \se{mock}.
The average result shows the additional systematic effect of random 
errors remaining after our correction for Malmquist bias (\se{mb}) 
and the additional weighting by $\sigma^{-2}$ (\se{twf_err}).  	
\end{list}

\vbox{\vskip 10truecm}
\includegraphics{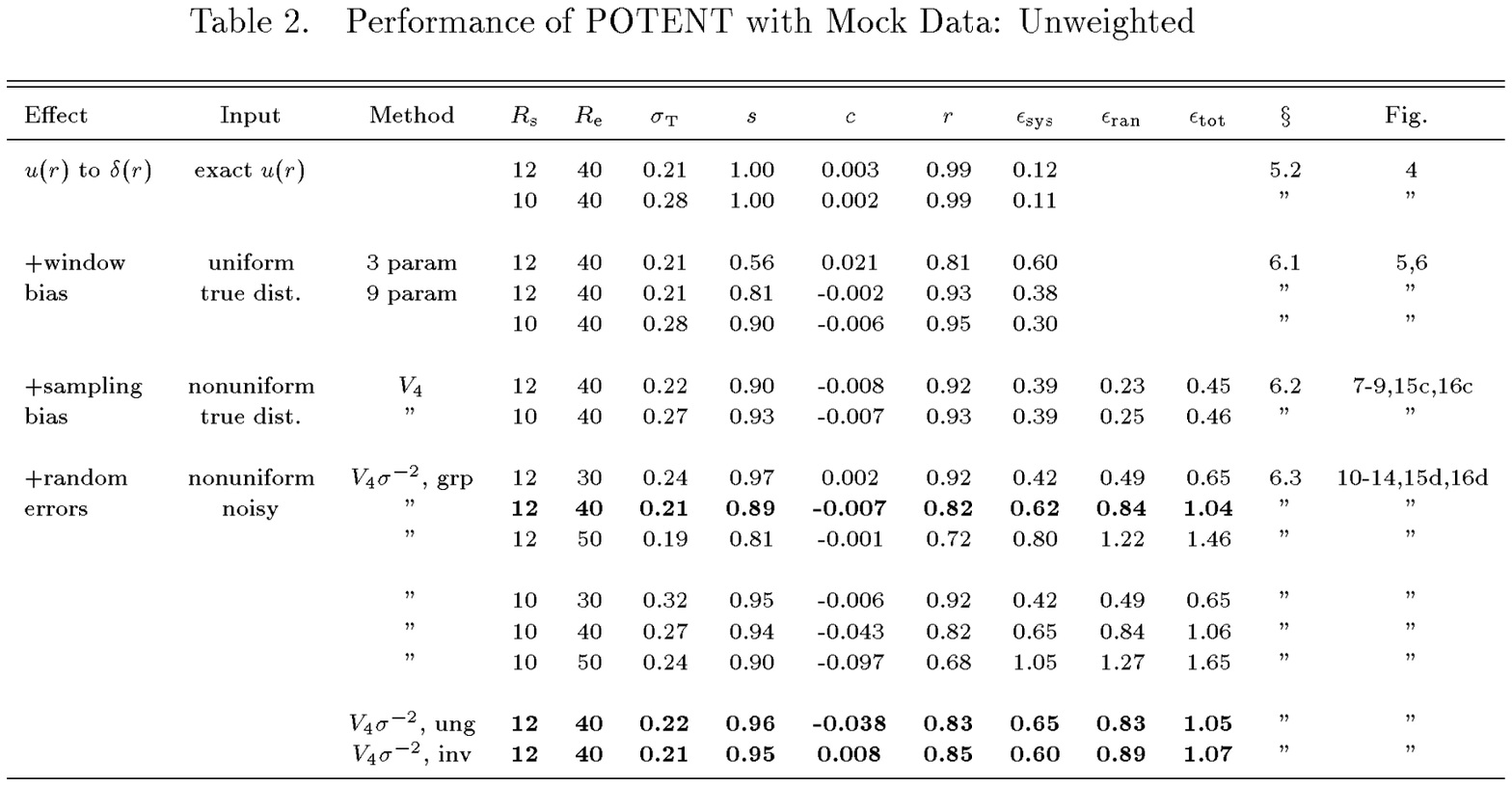}

\vbox{\vskip 6truecm}
\includegraphics{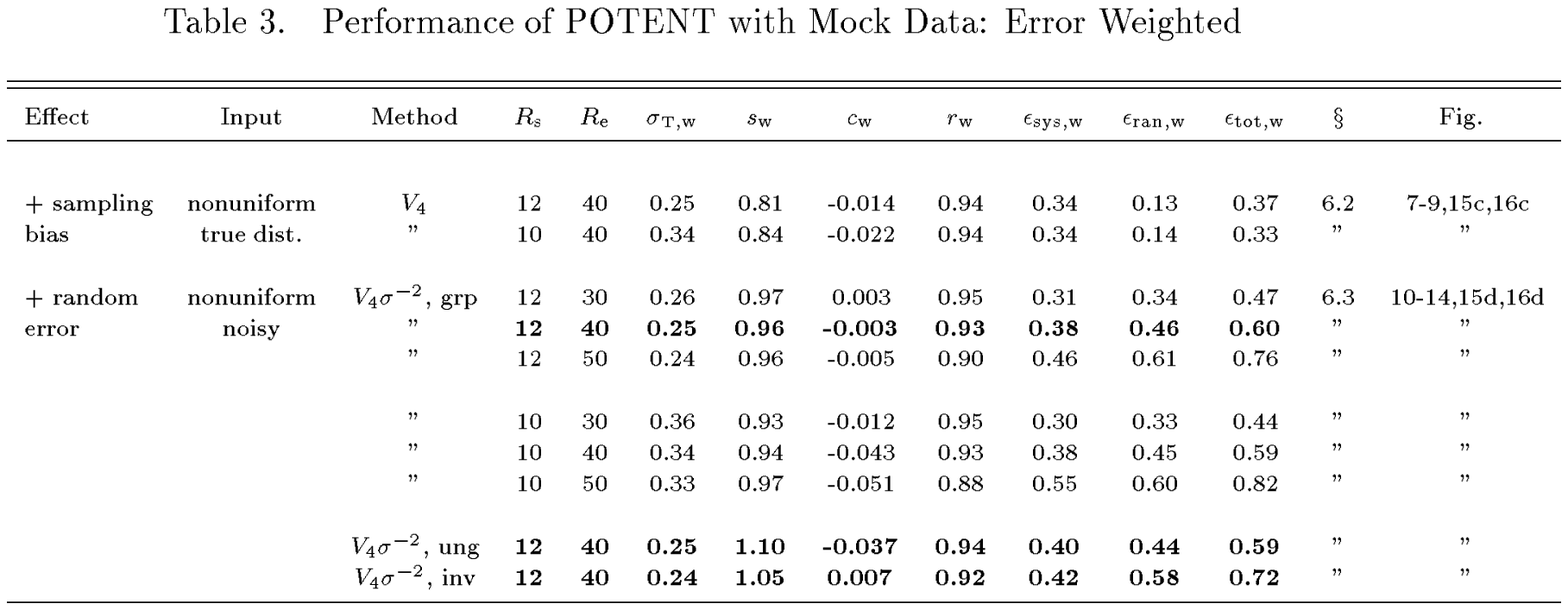}

As the mock data evolve from the ideal case to the realistic noisy case, 
the quality of the reconstruction gradually deteriorates.
The maps deviate steadily from the true map, and the local residuals
in the scatter diagrams grow accordingly. The corresponding weighted
correlation coefficients (defined in \se{eval}) 
are $\rw=0.99$, $0.94$, $0.94$, $0.93$.
The fact that the only significant drop in $\rw$ occurs between cases 
(a) and (b) indicates that 
the biases due to the non-uniform sampling and the 
random errors are well corrected while the remaining local bias is 
dominated by the window bias.
 
The global bias is characterized by $\sw=1.00$, $0.70$, $0.81$, $0.96$
respectively. There is in fact a significant improvement after the 
introduction of non-uniform sampling and noise in the data due to the 
appropriate corrections in the smoothing scheme.
This is due to the fact that the
under-smoothing induced by the $V_4 \sigma^{-2}$ weighting roughly 
compensates for the over-smoothing introduced by the window bias
(as mentioned at the end of \se{twf_err}).
The final global bias of only $4\%$ (for the weighted fit within the 
reference volume) is very promising for further analysis involving
linear regression (\eg, Sigad \etal 1998).

\begin{figure}[t!]
\centerline{\epsfxsize=5.5 in \epb{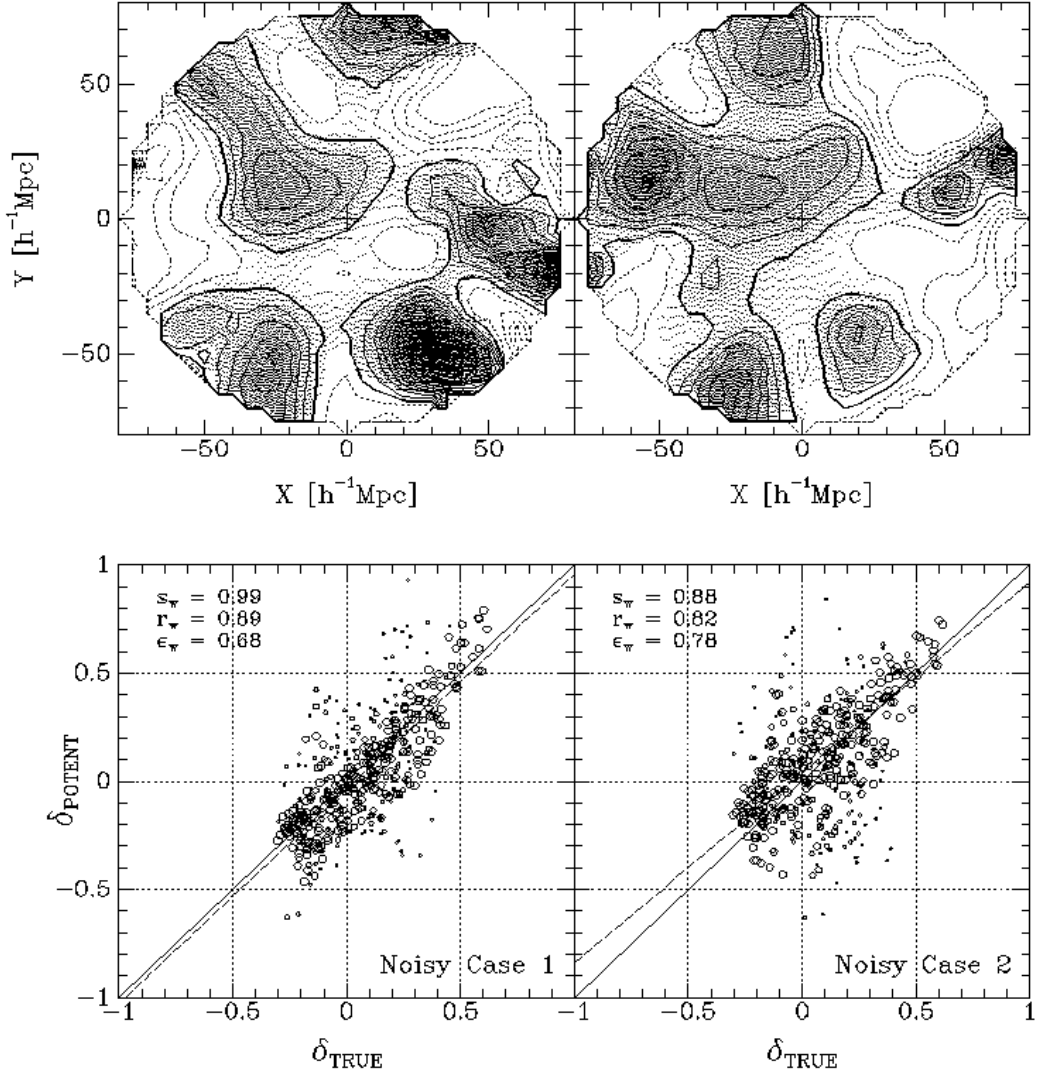}}
\caption{\protect\capt
Two individual noisy realizations.
Supergalactic maps of the reconstructed G12 $\delta$ fields from two
individual mock catalogs and the corresponding point-by-point
comparisons to the true field within $\Re=40\hmpc$.
Left: an example of a relatively good reconstruction.
Right: an example of a relatively bad reconstruction.
Contours and symbols are as in previous figures.
Here, the statistics measure random errors in the individual
realizations.
}
\label{fig:two_mocks}
\end{figure}

The random errors remain the dominant source of uncertainty.
To illustrate their possible effects, we show in 
Figure~\ref{fig:two_mocks} two noisy examples --- the ``best" and
the ``worst" G12 reconstructions among the 10 random noisy mock catalogs.
The maps are to be compared to the true field (\Fig{bias_maps}a).
Note that the reconstruction in the GA region is quite robust,
while in PP it is more sensitive to the noise.
The errors can be evaluated in these two cases by
$\sw=0.99$, $0.88$ and $\rw=0.89$, $0.82$.
These are the same statistics defined in \se{eval}, but applied
here to each of the two catalogs, not the average.
The scatter in these individual reconstructions reflects both
random and systematic errors and is therefore larger than the scatter in
the mean reconstructed fields of \Fig{vol_scat}, which was dominated by
the local systematic error at each grid point.

\begin{figure}[t!]
\centerline{\epsfxsize=5.5 in \epb{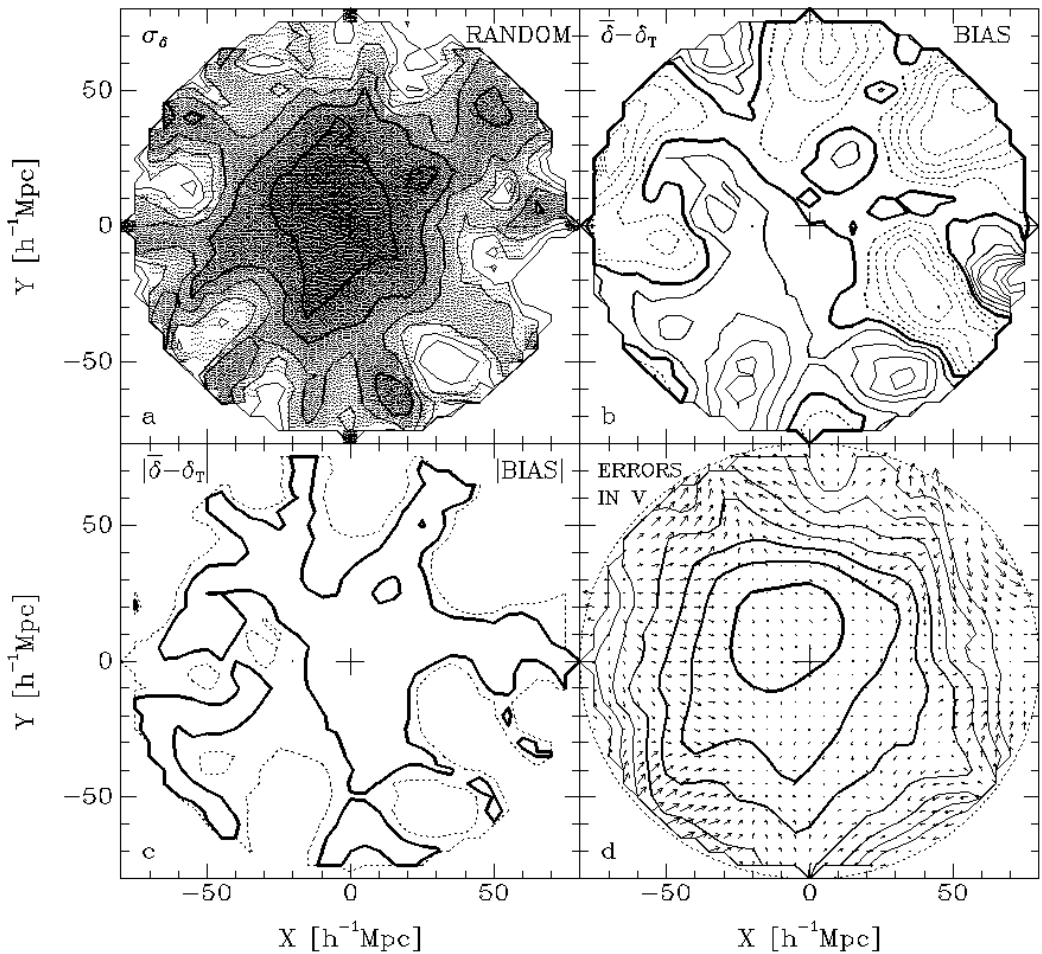}}
\vskip -0.1truecm
\caption{\protect\capt
Error fields in the Supergalactic plane based on mock catalogs.
(a) The random error in the density field, $\sigd$ (as in
\Fig{vol_maps}b).
    Contours are 0.1, 0.2, 0.3, etc.
(b) The systematic error in the density field, $\bar\delta - \delt$.
    Contours are 0, $\pm0.1$, $\pm0.2$, etc. (solid and dashed curves
    for positive and negative bias respectively).
(c) The absolute value of the systematic error in the density.
    Contours are 0.1, 0.2, etc.
(d) Errors in the velocity field. The systematic-error field is marked
    by arrows. The random-error field $\sigv$ is mapped by contours of
    $100\kms$, $200\kms$, etc.
}
\label{fig:error_maps}
\end{figure}

Supergalactic-plane maps of the error fields derived from the mock catalogs
are shown in \Fig{error_maps}.
These error fields should be used when interpreting the POTENT density 
and velocity fields of the real data, such as the ones shown below 
(\Fig{maps_sgp}). 
The error fields are crucial for any quantitative analysis using
POTENT output.

In order to evaluate one-point correlations between pairs of error fields 
of different types and other local quantities, we compute for each pair 
the linear correlation coefficient $r$ inside the reference volume of 
effective radius $40\hmpc$.
For densities,
the magnitudes of the random errors ($\sigd$) versus the systematic errors
give $r=0.38$ --- a positive but rather weak correlation.
The magnitudes of the total error field ($\sigtot$) versus the true density
fluctuation field $|\delt |$ show only a weak correlation, $r=0.08$ 
(with a slight correlation for the systematic component, $r=0.07$, 
and a slight anti-correlation for the random component, $r=-0.18$).
Finally, the distance dependence of the errors in the recovered density 
field can be characterized by their correlation coefficients with distance: 
$r=0.83$, 0.40 and 0.77 for the magnitudes of
the random, systematic and total errors respectively.

Our elaborate error analysis can be summarized by the following few numbers.
Within the reference volume of effective radius $40\hmpc$, the  rms 
systematic error in the local G12 density fluctuation field
$\delta$ is $\pm 0.13$, and the corresponding random error is $\pm 0.18$,
adding up in quadrature to a total (unweighted) error of $\pm 0.22$.  
The weighted total error in this volume is $\pm 0.15$.

\section{POTENT RECONSTRUCTION FROM REAL DATA}
\label{sec:maps}

We are now ready to apply POTENT to the real Mark III data.
While showing the reconstruction of our cosmological neighborhood, 
we will also have additional opportunities to address some of the 
systematic errors discussed above.

The final POTENT Mark III maps of the G12 smoothed fields of projected
three-dimensional peculiar velocity and of mass density fluctuations
are presented in Figures~\ref{fig:maps_sgp} to \ref{fig:3D}.
\Fig{maps_sgp} shows maps of the POTENT recovery in the Supergalactic
plane.
\Fig{surface} helps in visualizing the density field in
the Supergalactic plane via a surface landscape plot in which 
height is proportional to the density fluctuation.
\Fig{maps_slices} adds the fields in two planes parallel
to the Supergalactic plane, at $Z=-25$ and $+20\hmpc$.
Finally, \Fig{3D} provides a three-dimensional view of the
density fluctuation field via the surface of constant density 
fluctuation $\delta=0.4$.
The density fluctuation shown in all these plots is computed 
assuming $\Omega\!=\!1$ in \eq{delc+},
but it can be easily generalized to any reasonable value of $\Omega$.
Recall that the linear correction is simply a scaling by $f(\Omega)^{-1}$.

\vskip -0.5truecm \subsection{The Supergalactic Plane}
\label{sec:maps_sgp}

\begin{figure}[t!]
\centerline{\epsfxsize=6.5 in \epsfbox{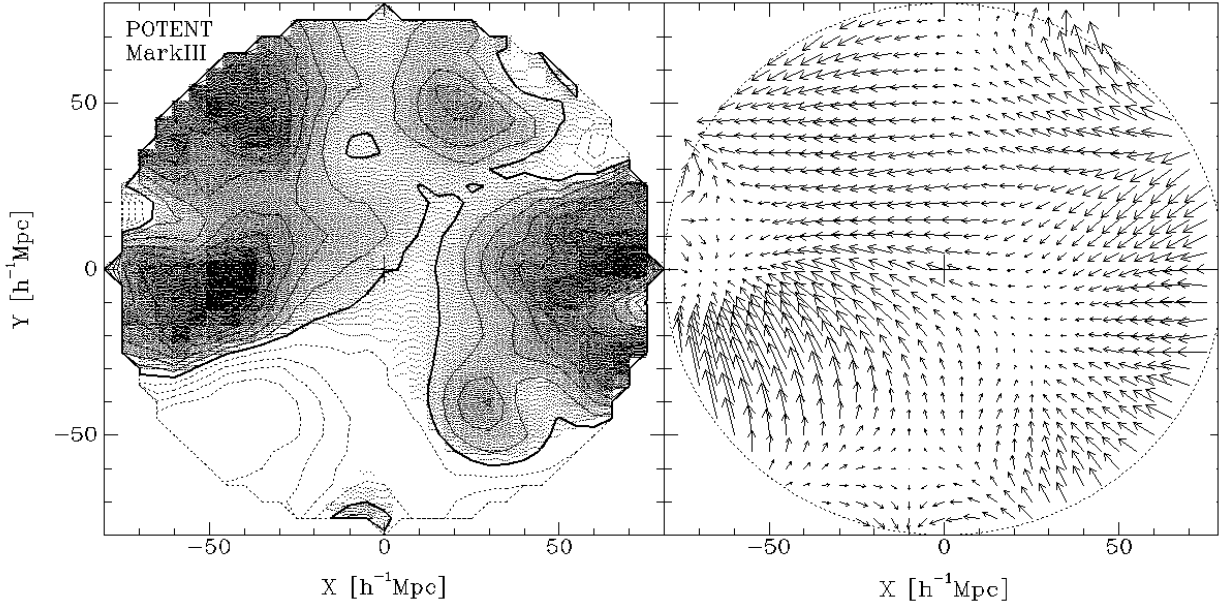}}
\vskip -0.1truecm
\caption{\protect\capt
POTENT recovery of real data.
The fluctuation fields of velocity and mass density in the
Supergalactic plane as recovered by POTENT from the Mark III peculiar
velocities with G12 smoothing.
Left: As in previous density maps, the contour spacing
is 0.2, the heavy contour marks $\delta=0$, and the dashed contours
correspond to $\delta<0$. The grey scale is proportional to $\delta$.
Right: The three-dimensional velocity field relative to the CMB frame
is projected onto the Supergalactic plane.  Distances and velocities
are in $\!\hmpc$ or $100\kms$.
The LG is at the center.  The GA is on the left, PP on the right,
and a large local void lies in between.
}
\label{fig:maps_sgp}
\end{figure}

\begin{figure}[t!]
\vspace{4.0truecm}
\includegraphics{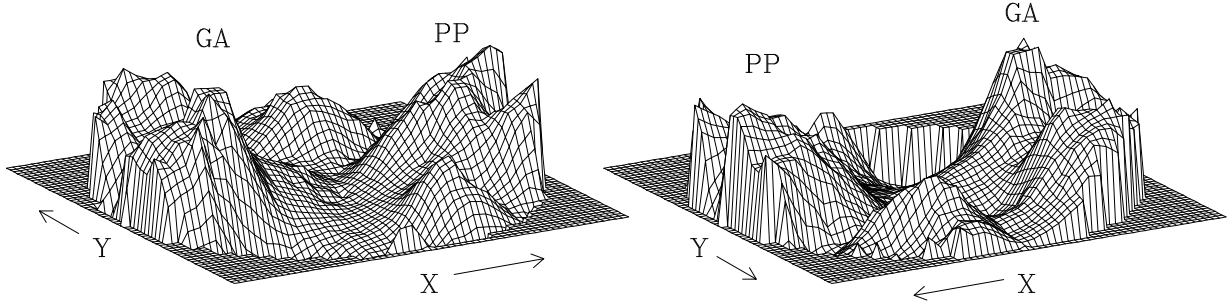}
\caption{\protect\capt
The POTENT density field in the Supergalactic plane out to $80\hmpc$,
as in Fig~\ref{fig:maps_sgp},
shown as landscape maps observed from two different
directions.  The height of the surface is proportional to $\delta$.
Note the attractors GA and PP, and the extended void in between.
}
\label{fig:surface}
\end{figure}

The velocity map in \Fig{maps_sgp} shows a clear tendency for motion
from right to left, in the general direction of the LG motion
in the CMB frame (which is $L,B\!=\!139^\circ,-31^\circ$ 
in Supergalactic coordinates).
The bulk velocity within $60\hmpc$ is roughly $300-350\kms$ towards
($L,B\!\approx\!170^\circ,-10^\circ$) (\se{bulk}),
but the flow is not coherent over the whole volume sampled, \eg,
there are regions in front of PP (bottom right) and behind the GA
(far left) where the $XY$ velocity components vanish, \ie, the streaming
relative to the LG is opposite to the bulk flow direction.
The velocity field shows local convergences and divergences which
indicate strong density variations on scales about twice as large
as the smoothing scale.
The G12-smoothed velocity at the LG is 
$547\kms$ towards $(L,B) = (161^\circ, -19^\circ)$,
compared to the unsmoothed LG motion relative to the CMB
of $627\kms$ towards $(L,B) = (139^\circ, -31^\circ)$
(Lineweaver \etal 1996; Yahil \etal 1977)

The Great Attractor (with G12 smoothing and $\Omega\!=\!1$) is a broad
density ramp of maximum height $\delta\!=\!1.4\pm 0.3$ located near
the Galactic plane $Y\!=\!0$ at $X\!\approx\!-40\hmpc$.  The GA extends
through the Hydra and Centaurus clusters
towards Virgo near $(X,Y)\!\approx\!(0,10)$ (the ``Local Supercluster"),
towards Pavo--Indus--Telescopium (PIT) across the Galactic plane to the
south ($X\!<\!0$ and $Y\!<\!0$), 
and towards the Shapley concentration behind the GA
($X\!<\!0, Y\!>\!0$). The structure at the top is related to the
``Great Wall'' of Coma, with $\delta\approx0.6$.  The Perseus--Pisces
structure, which dominates the bottom-right quadrant, peaks near Perseus with
$\delta\!=\!1.0\pm0.4$.  PP extends towards the Southern Galactic
Hemisphere (Aquarius, Cetus), coinciding with the ``Southern Wall"
as seen in redshift surveys.  Underdense regions separate the GA and PP,
extending from bottom-left to top-right.  The deepest region in the
Supergalactic Plane, with $\delta\!=\!-0.8\pm0.3$, roughly coincides
with the galaxy void of Sculptor (Kauffman \etal 1991).

\vskip -0.5truecm \subsection{Outside the Supergalactic Plane}
\label{sec:maps_3D}

\begin{figure}[t!]
\centerline{\epsfxsize=6.5 in \epb{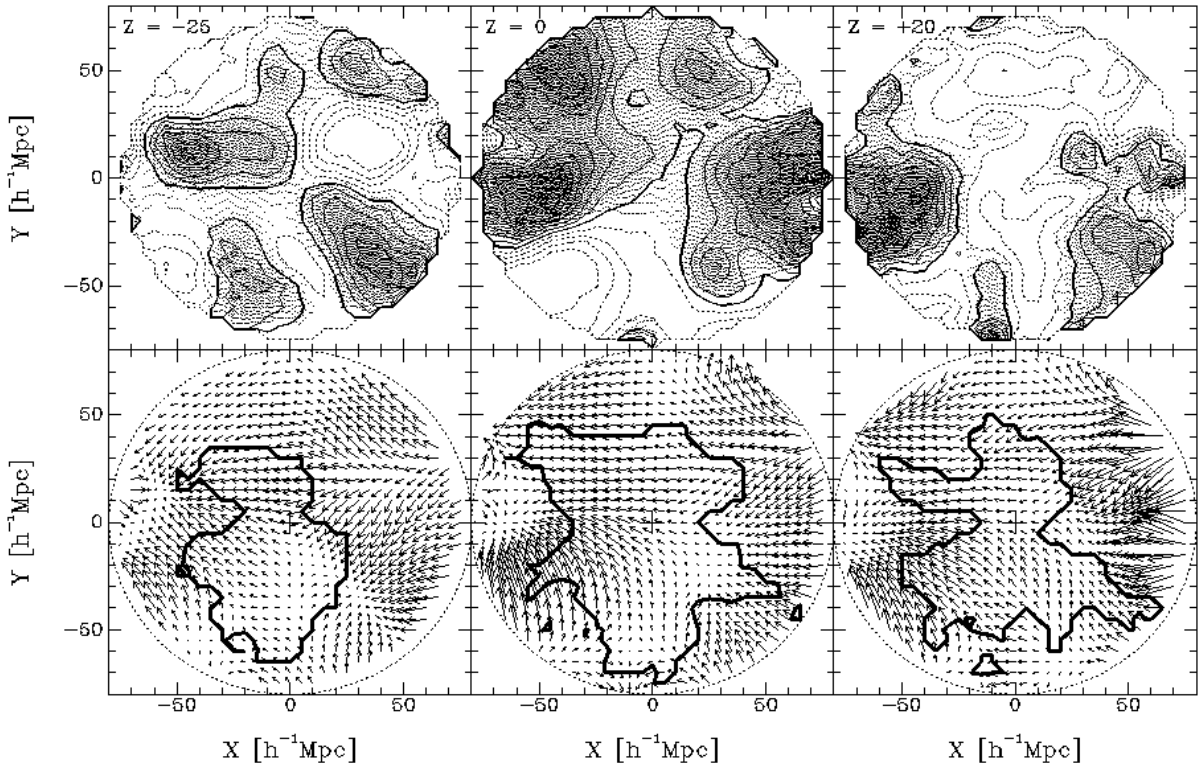}}
\vskip -0.2truecm
\caption{\protect\capt
Same as \Fig{maps_sgp} but for two additional planes parallel to
the Supergalactic plane, at $Z=-25$ and $+20\hmpc$.
The heavy contours on top of the velocity fields are cuts through
the surface $\sigd < 0.4$ and $R_4 < 10.4\hmpc$, which has an
effective radius of $50\hmpc$.
}
\label{fig:maps_slices}
\end{figure}

We have focused so far on the fields in the Supergalactic plane,
where the structure is rich, featuring large attractors and big voids.
However, the data and the analysis span the three-dimensional space
about the LG, and in particular regions away from the Supergalactic plane.
The two slices shown in \Fig{maps_slices} above and below the Supergalactic
plane demonstrate the continuity and extent of the two big structures,
the GA and PP, and how the large void stretches between them.
The bulk flow is seen in all three slices, which means that it 
is valid throughout most of the volume analyzed.
The combined error contour of $\Re=50\hmpc$ 
($\sigd < 0.4$ and $R_4 < 10.4\hmpc$) encompasses parts of the GA and  
PP, but it excludes the parts of these structures that lie near the \zoa.
The big streams on the right beyond PP are clearly noise features.

\begin{figure}[t!]
\centerline{\epsfxsize=6.0 in \epsfbox{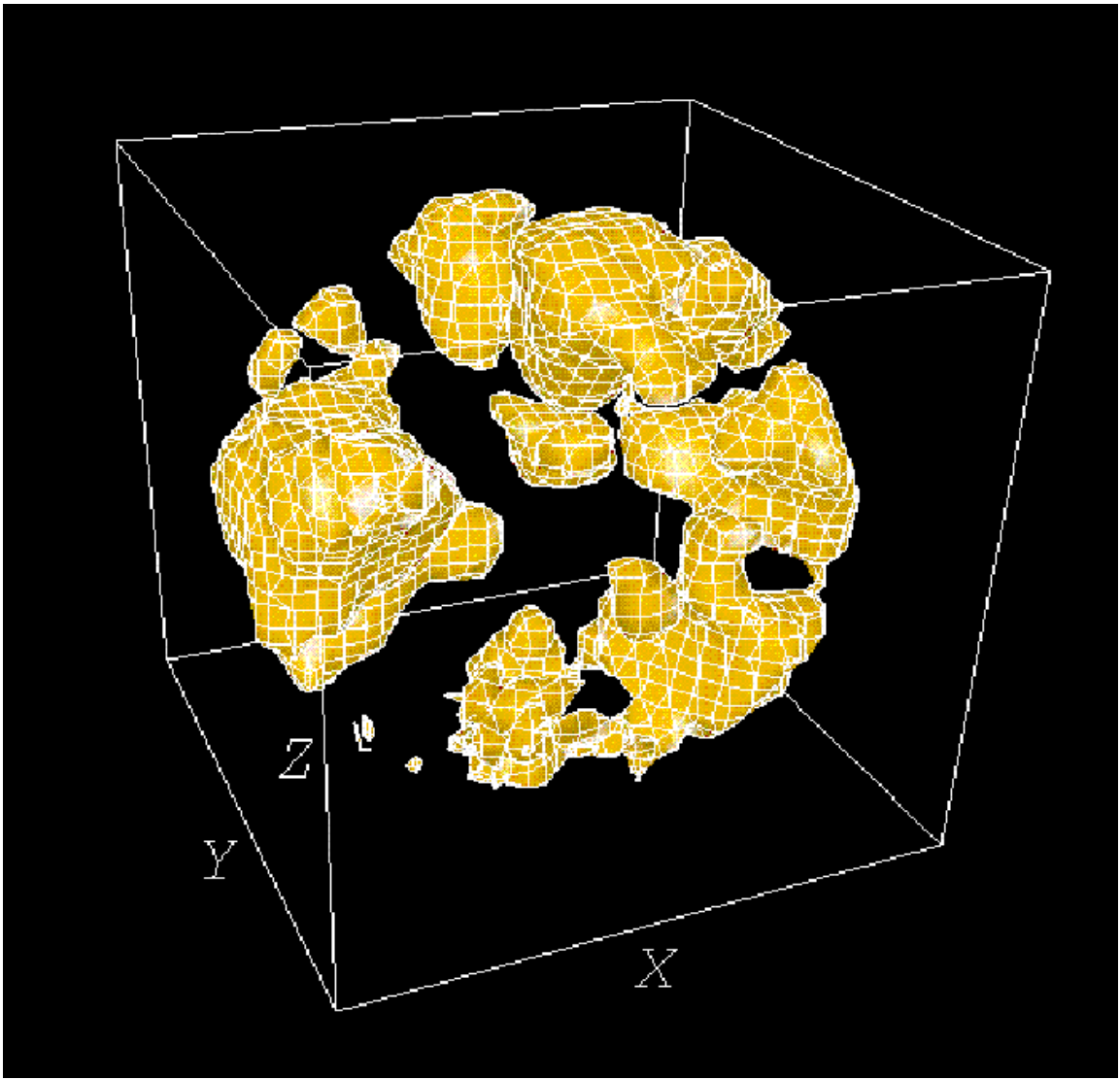}}
\caption{\protect\capt
The POTENT G12 $\delta$ field in three dimensions. Shown is the surface
$\delta=+0.4$. The box marks the Supergalactic coordinates $X,Y,Z$;
the Supergalactic plane is the equatorial plane $Z=0$
and the LG is at the center of the cube.  The GA is the big structure
on the left and PP is the big complex structure extending above
and below the Supergalactic plane on the right,
with the big void in between.
}
\label{fig:3D}
\end{figure}

The three-dimensional structure in the whole volume is
also illustrated in \Fig{3D}, which shows the surfaces of $\delta=+0.4$.
The GA and PP extend dramatically into the two sides
of the Supergalactic plane and so does the big void between them. 
While the GA appears to be a coherent big structure,
PP develops a more complex structure outside the Supergalactic plane
and connects to other super-structures.

\vskip -0.5truecm \subsection{Issues in Local Cosmography}
\label{sec:maps_cosmography}

Following are a few comments about issues of interest in the local 
cosmography.

 One can still find in the literature statements questioning the 
existence of the GA (\eg, Rowan-Robinson 1993), which simply reflect ambiguous
definitions for this entity.  A GA clearly exists in the sense that
a dominant feature in the local peculiar-velocity field is a coherent
convergence, centered near $X\!\approx\!-40$.  Whether or not the 
associated density peak has an exact counterpart in the galaxy
distribution is another question.  The GA is ambiguous only in the sense
that the correlation observed between the mass density inferred
from the velocities and the galaxy density in redshift surveys, though
generally good, is not perfect (\eg, Sigad \etal 1998).  
This might reflect underestimated 
errors in the analyses, a nontrivial biasing relation between galaxies 
and mass, and/or incomplete coverage of the ZoA region by existing
redshift surveys (see below).

\begin{figure}[t!]
\centerline{\epsfxsize=6.0 in \epb{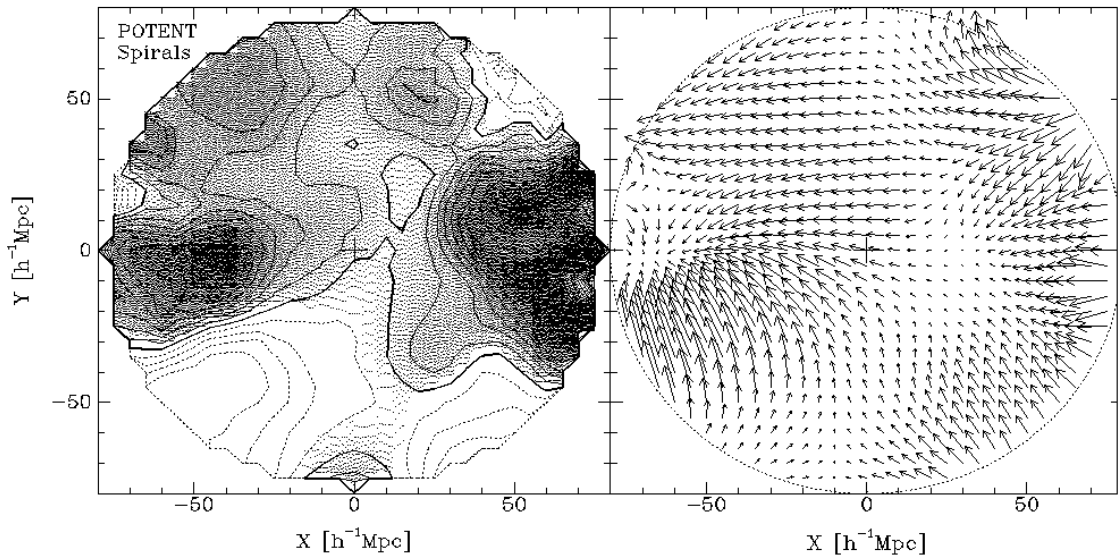}}
\vskip -0.2truecm
\caption{\protect\capt
POTENT recovery from spirals only.
The G12 fluctuation fields of velocity and mass density in the
Supergalactic plane as recovered by POTENT from the Mark III peculiar
velocities of spirals only. Compare to \Fig{maps_sgp}.
}
\label{fig:spirals}
\end{figure}

The GA was originally discovered by Lynden-Bell \etal (1988) from the
velocities of elliptical and S0 galaxies. Kolatt \& Dekel (1994)
found that the flows of spiral galaxies and early-type galaxies are
consistent with each other, but Mathewson, Ford \& Buchhorn (1992) 
expressed some doubts concerning the GA convergence in their spiral data.
However, these data now dominate the Mark III data in that region.           
\Fig{spirals} shows Supergalactic-plane maps of the POTENT velocity and 
density fields derived from {\it spirals only}. 
Comparison to the fields recovered from all the data, \Fig{maps_sgp},
shows only minor differences. The GA and the other main features in our
cosmological neighborhood are clearly present in a consistent way
in the velocities of the tracers of {\it all types}. 

It is also now possible to correct another misimpression as to the
amplitude of the local GA inflow. Lynden-Bell \etal (1988) originally modeled
the entire flow pattern within $40\hmpc$ of the LG as a spherical
inflow onto an at-rest GA. The amplitude of this flow at the LG was 
$\sim 500\kms$, indicating a rather high mass for the GA.
The analysis below (\se{bulk}, \Fig{tidal}) decomposes the velocity field 
into its locally produced, divergent part and an externally produced, 
tidal part. It is seen there that only about half of the GA velocity is
due to the GA and the local distribution of mass. Thus, very crudely,
the mass excess associated with the GA should be only about half of that 
envisioned by Lynden-Bell \etal

Other debated cosmographic issues in the Supergalactic plane,  
are the nature of the flow behind the GA and the velocity 
of PP relative to the LG.  The current POTENT analysis indicates that,
although there are signs of a relative local backflow towards the GA center, 
the flow behind the GA in the CMB frame is dominated by a continuing 
outflow, roughly towards the Shapley concentration (see \se{bulk} below).
The region between the LG and PP shows small CMB velocities,
namely, a velocity of recession away from the LG, but the central regions
of PP seem to be roughly at rest relative to the LG. The back side of PP
appears to be flowing towards PP, as expected, and is approaching the LG.
However, the flows beyond the GA and PP are only marginally detected by 
the current 
data and POTENT analysis, at the $\sim 1.5\sigma$ level in terms of the 
random uncertainty.  Furthermore, the remaining freedom in the zero point 
of the distance indicators (see \se{maps_vm2} and \se{disc_errors})
permits adding a Hubble-like peculiar expansion velocity to the whole 
ensemble, which could balance the 
flows at the back of the GA and in PP.  Thus, these issues remain 
partly debatable until settled by future data of better sampling in these
outer regions.

 To what extent should one believe the reconstruction in the Galactic Zone
of Avoidance (\zoa), in which the sampling is very poor?  
The velocities observed 
at medium Galactic latitudes on the two sides of the \zoa\ are used 
as probes of the mass in the \zoa.  The interpolation is based on 
the assumed irrotationality, where the recovered transverse components 
allow a reconstruction of the mass-density.  However, while the 
sampling-gradient bias can be corrected where the width of the \zoa\ 
is smaller than the smoothing length, the result could be severely 
biased in places where the unsampled region is larger.  With G12 smoothing 
in the Mark III data, the interpolation is suspect of being severely biased
in $\sim\!50\%$ of the \zoa\ at $r\!=\!40\hmpc$ (where $R_4\!>\Rs$), 
but the interpolation is fairly reliable, for example, 
in the closer and highly populated near side of the GA.  
Indeed, a deep survey of optical galaxies at low Galactic latitudes 
(Kraan-Korteweg \etal 1996) as well as data from the ROSAT X-ray satellite 
(Boehringer \etal 1996) revealed the very rich Abell cluster A3627  
centered at $(l,b,z)\!=\!(325^\circ,-7^\circ,45\hmpc)$,
which is near the center of the broad peak of the GA as predicted earlier by 
POTENT from Mark III at $(320^\circ,0^\circ,40\hmpc)$ 
(Kolatt, Dekel \& Lahav 1995).
This is in agreement with the theoretical expectation that rich clusters
(originating from small-scale, high-amplitude peaks) prefer to
form near the centers of larger superstructures (large-scale, low-amplitude
peaks).
In fact, it has become clear recently that the whole ZoA region between 
the Centaurus clusters and PIT contains 
enhanced galaxy density hidden behind heavy extinction, as high as
12 V magnitudes at $b = -1.5^\circ$ along the line connecting A3627 and 
Centaurus (Woudt 1998).

\vskip -0.5truecm \subsection{Sensitivity to Malmquist Bias: Forward vs. Inverse}
\label{sec:maps_mb}

\begin{figure}[t!]
\centerline{\epsfxsize=6.5 in \epb{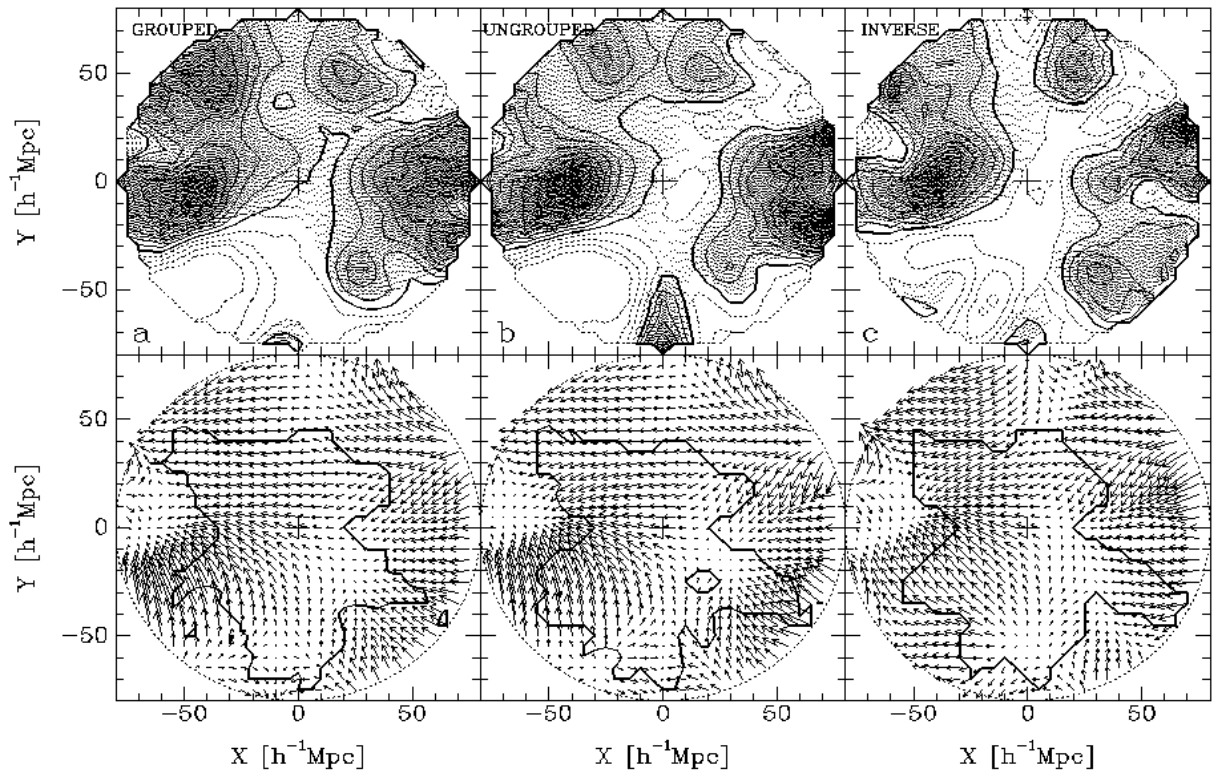}}
\vskip -0.2truecm
\caption{\protect\capt
Three Malmquist-bias corrections of real data.
Maps in the Supergalactic plane of the G12 mass-density and velocity
fields from the Mark III data, corrected for Malmquist bias in three
different ways as in \Fig{mb_maps}.
Contours are as before.
(a) forward correction after grouping,
(b) forward correction with no grouping,
(c) inverse correction in inferred-distance space.
The error contours refer to an effective radius of $\Re=50\hmpc$.
}
\label{fig:gsi}
\end{figure}

Sensitivity to Malmquist bias is explored again in
\Fig{gsi}, which shows maps in the Supergalactic plane of the
reconstructed fields corrected in three different ways,
as in \Fig{mb_maps} of \se{mb_imb} for the mock catalogs.
Shown are Malmquist corrections in a forward TF analysis
via \eq{fmb} with (standsrd) and without a preliminary stage of
grouping, and the result of an inverse TF analysis in inferred-distance 
space, where the correction is done via \eq{imb}.
(We do not have inverse-TF distances for the field ellipticals in 
Mark III, which constitute $\sim 16\%$ of the sample. In order to allow a
comparison with the forward corrections of the whole sample, 
we added in panel (c) the forward-corrected elliptical distances
to the inverse-corrected spiral distances.)

The maps are quite similar in their gross features.
The GA has slightly higher density in the ``ungrouped" correction
compared to the ``grouped" correction, while the ``inverse" correction
is intermediate between the two, but the difference at the highest peak
is only $\sim 20\%$. There are differences in the fine details of PP,
especially in the ``inverse" correction compared to the ``forward" cases,
but the overall amplitude and shape within the error contour shown
are not very sensitive to the grouping, and are not very different
in the ``inverse" case.
The void between GA and PP is somewhat wider and deeper in the
``inverse" and ``ungrouped" cases compared to the ``grouped" case.
These differences in the fine details of the three maps can be
interpreted as the typical residual error in the Malmquist-bias
correction.

\vskip -0.5truecm \subsection{Sensitivity to TF Calibration}      
\label{sec:maps_vm2}

\begin{figure}[t!]
\centerline{\epsfxsize=6.5 in \epb{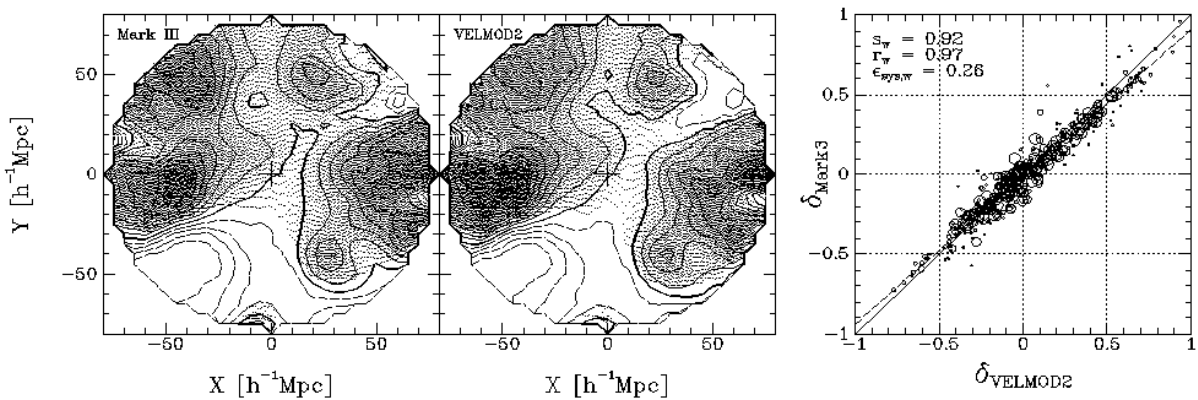}}
\vskip -0.2truecm
\caption{\protect\capt
Alternative Zero-point Calibrations of Mark III.
The G12 POTENT density field as reconstructed from the original Mark III
data is compared with the density field as reconstructed from
the alternative VM2 calibration of the same TF data.
The comparison is made via maps in the Supergalactic plane and via
a point-by-point comparison within a reference volume of effective
radius $40\hmpc$. Contours, symbols, and statistics quoted are as
before.
}
\label{fig:vm2}
\end{figure}

The {\it relative\,} TF calibration of the various datasets that constitute the 
Mark III catalog suffers from dome uncertainty. The original
relative calibration was done within the Mark III catalog itself based on 
galaxies that are common to more than one dataset.
An alternative way to match the datasets is by requiring maximum agreement
between the velocity field extracted from Mark III and the galaxy
density field extracted from a whole-sky redshift survey, such as
the IRAS 1.2 Jy catalog, under the assumption of a simple, linear biasing
relation between the densities of galaxies and the underlying mass. 
A comparison of the corresponding density fields within $\sim 40\hmpc$
has indicated an acceptable goodness of fit using the original calibration 
of Mark III (Sigad \etal 1998). A similar conclusion has been obtained in the
VELMOD high-resolution velocity-velocity comparison that was limited to an 
even smaller reference volume (Willick \etal 1997b). 
However, velocity-velocity comparisons 
of those two catalogs show a lower goodness of fit at larger radii 
(Davis, Nusser \& Willick 1996; Willick \& Strauss 1998). 
In particular, the VM2 analysis suggests a somewhat different relative TF 
calibration for the datasets of the Mark III catalog.  
The main proposed change boils down to the  
zero points of the datasets that cover the PP region. It works
in the sense of reducing the distances by a multiplicative factor 
of 7--8\%, thus reducing any inwards back-flow behind PP.
(As far as the global zero point is concerned, in the original calibration 
we eventually reduced it by 3\% in order to minimize the rms peculiar 
velocities of POTENT within $40-50\hmpc$, while
no such correction is needed with the VM2 calibration.)

However, the mass density fluctuation is a function of the partial 
derivatives of the velocity field, which are only weakly sensitive to the 
TF zero point. Hence, changes of the proposed magnitude are expected
to have only little effect on the density maps.
This is confirmed in \Fig{vm2}, which compares the POTENT density 
fields as reconstructed from the original calibration and from the VM2 
calibration of the same Mark III data. 
The differences are minor: in the revised calibration, the density of PP 
outside the ZoA is somewhat higher, and the void between the LG and PP 
is slightly wider. 
The small systematic difference within the reference volume of effective 
radius $40\hmpc$ is characterized by $\sw=0.92$ and $\rw=0.97$.
These differences should serve as a crude estimate of the error associated 
with the relative zero-point calibration in the Mark III catalog; 
it is small compared to the other errors.  The difference is larger
in the bulk velocity at large radii (\se{bulk}).
A new systematic whole-sky survey of TF data in a shell between 
redshifts $4500$ and $7000\kms$ (Courteau \etal 1999, SHELLFLOW, hereafter
SHF) should help reduce this uncertainty and thus allow a quantitative 
analysis in a larger reference volume.

\section{BULK VELOCITY}
\label{sec:bulk}

The simplest statistic that can be used to characterize 
the peculiar-velocity field is the bulk velocity (in the CMB frame)
within a given volume.
The POTENT output allows a reliable measurement of the bulk velocity
within top-hat concentric spheres about the Local Group out to radii 
of $50$-$60\hmpc$,
beyond which the errors are large and uncertain (both for the Mark III
data and for any other currently available TF whole-sky galaxy survey). 
One could alternatively compute the bulk velocity by a direct fit to the raw 
data of radial peculiar velocities of the individual galaxies,
but the advantage of using the output of POTENT for this purpose is that it 
automatically provides equal-volume averaging and it avoids many of the
biases associated with a direct fit. The disadvantage is that 
the $12\hmpc$-Gaussian smoothing significantly affects 
the resultant bulk flow out to radii of $20-30\hmpc$.

\begin{figure}[t!]
\vskip -2.3truecm
\centerline{\epsfxsize=5.0 in \epsfbox{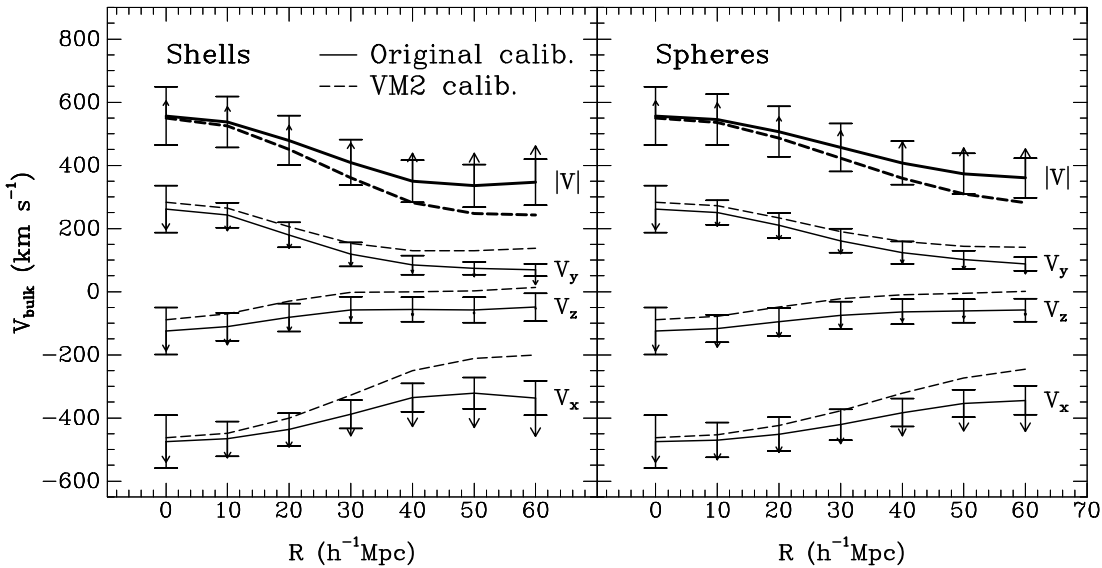}}
\vskip -7.7truecm
\caption{\protect\capt
The bulk velocity in the CMB frame
in concentric shells or spheres of radius $R$ about the
Local Group.  It is derived by equal-volume vector
weighting of the G12 uniform POTENT output. Shown is the magnitude
of the bulk velocity and its three components in Supergalactic
coordinates.
Also shown are the random (error bars) and systematic (arrows) errors,
derived from a series of mock realizations.
Shown in comparison are the results from the original calibration of the
Mark III TF data (solid) and from the proposed VM2 calibration (dashed).
}
\label{fig:bulk}
\end{figure}

Random errors in the bulk velocities are determined by the rms 
scatter in the results obtained from the series of random noisy mock 
catalogs (\se{mock}). 
Systematic errors in the bulk velocity are determined by comparing
the average over the mock catalogs with the true G12-smoothed bulk velocities
in the simulations.  In order to feed POTENT
with mock data that have a bulk flow as large as the one observed
(the large-scale bulk velocity in the simulation is artificially small because
of the periodic boundary conditions), we added to the 
mock data an artificial component of bulk flow such that the total 
matches the flow obtained from the real data in a sphere of radius 
$60\hmpc$, namely $356\kms$ in the direction $(L, B) = (171^\circ, -7^\circ)$.
\Fig{bulk} shows the resultant bulk velocity as a function of radius $R$,
both in shells and in spheres, for the original calibration of
Mark III and for the proposed VM2 re-calibration.

Perhaps the most solid useful number from this analysis is the 
bulk flow relative to the CMB within the sphere of radius $50\hmpc$:
$V_{\rm 50}= 370 \pm 110\kms$ towards $(L,B)=(165^\circ,\, -10^\circ)$ 
for the original calibration, and
$V_{\rm 50}= 305 \pm 110\kms$ towards $(L,B)=(150^\circ,\, 0^\circ)$ 
for the VM2 re-calibration of the same data.
[The corresponding directions in Galactic coordinates are
$(l,b)=(305^\circ,\, 14^\circ)$ and $(314^\circ,\, 30^\circ)$.]
This bulk velocity is dominated by the negative $X$ component, 
which represents a coherent flow of
much of the local region, including parts of PP, towards the GA and the 
Shapley Concentration behind the GA at roughly twice the distance.
The error quoted includes both the random and systematic effects 
as determined from the mock catalogs. The magnitudes of these two
components are comparable, 
as can be seen in \Fig{bulk} where they are shown separately.  
The cosmic scatter, which is not included,
could be of comparable magnitude to the quoted error, 
depending on the cosmological model assumed (including for example   
the power spectrum of fluctuations and the cosmological density parameter 
$\Omega$).

These results are quite similar to the results obtained
by BDFDB from POTENT applied to the Mark II data,
$V_{\rm 40}=388 \pm 67\kms$ towards $(L,B)=(177^\circ,\, -15^\circ)$.
Note that the errors are estimated more accurately in the current 
analysis, and the results are more reliable now out to $50$ or even $60\hmpc$.
The current results from POTENT Mark III are also consistent with the results
of Courteau \etal (1993) based on a direct likelihood analysis from 
preliminary Mark III radial velocities, in which a certain effort was made
to approximate equal-volume weighting:
$V_{\rm 60}=360 \pm 40\kms$ towards $(L,B)=(177^\circ,\, -23^\circ)$.
Note, again, that the errors are computed in different ways and they have
different meanings.

For comparison, the bulk velocity within the sphere of radius $50\hmpc$
for the SFI catalog (Wegner \etal 1999, Haynes \etal 1999) 
analyzed via POTENT in a similar way is 
$V_{\rm 50}= 310 \pm 80\kms$ towards $(L,B)=(148^\circ,\, -13^\circ)$
(Eldar \etal 1999).  
A direct fit to the SFI radial velocities yields
$V_{\rm 65}=200 \pm 65\kms$ towards $(L,B)=(151^\circ,\, -17^\circ)$
(Giovanelli \etal 1998). The errors are estimated in yet a different way.
These results all agree within the $1\sigma$ error.
The VM2 calibration of Mark III also shows a bulk velocity similar to 
SFI shell by shell near $50\hmpc$ and beyond, and they are both
smaller there than the bulk velocity from the original Mark III calibration.
However, this is where the errors begin to become
large and the bias corrections are uncertain in both catalogs.

\begin{figure}[t!]
\vskip -2.2truecm
\centerline{\epsfxsize=5.0 in \epsfbox{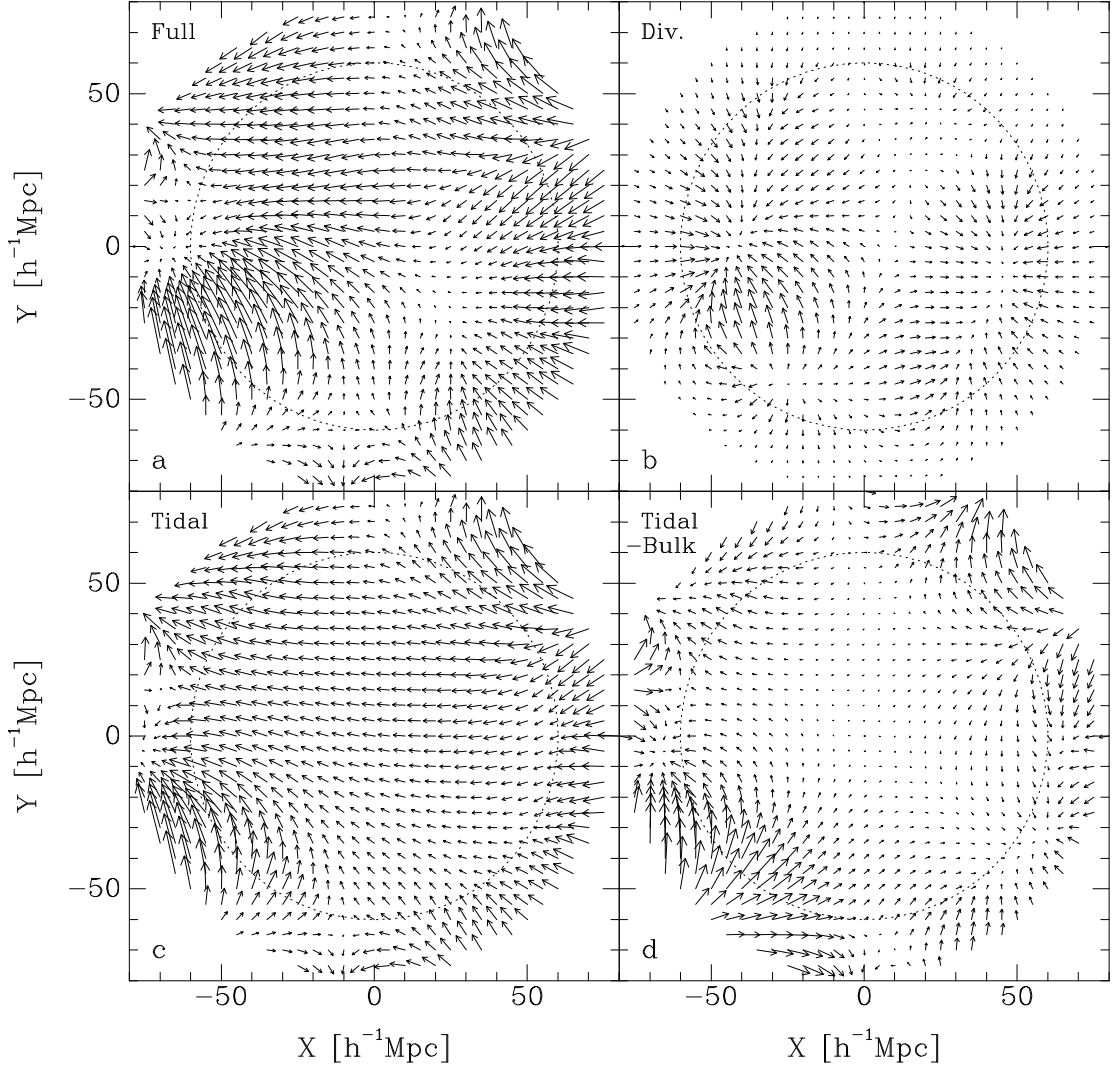}}
\vskip -0.8truecm
\caption{\protect\capt
A decomposition of the velocity field in the Supergalactic plane (a)
into its ``divergent" component, due to the mass fluctuations within
$60\hmpc$ (b),
and its ``tidal" component, due to the mass outside this sphere (c).
The residual left after subtracting the bulk from the tidal component
includes quadrupole and higher moments in the tidal field (d).
Velocities and distances are measured in $\hmpc$.
}
\label{fig:tidal}
\end{figure}

Where are the main sources of this bulk velocity? 
How much of it is contributed from the local distribution of mass, 
\eg, the GA, and how much is due to the mass distribution at larger 
distances? What are the main features of the velocity field on top of the
bulk velocity?  To begin to address these questions, we 
decompose the velocity field into two components: the ``divergent" 
component arising from the mass fluctuation field within the sphere of 
radius $60\hmpc$ about the LG, and the ``tidal" component arising from 
the mass distribution outside this sphere. 
The decomposition is performed as follows. First, the ``linear"
density fluctuation field is 
recovered from the total velocity field using the linear relation,
$\delta_0= -f(\Omega)^{-1}\, \divv$. Then, assuming $\delta_0 = 0$
outside the sphere of radius $60\hmpc$, the divergent component
of the velocity is computed by the inverse relation,
\be
\vv(\vx)={f(\Omega)\over 4\pi}   \int \delta_0(\vx')\, 
{\vx'-\vx \over | \vx'-\vx |^3 }\, d^3x' \ .
\label{eq:linear}
\ee
The tidal component is then obtained by subtracting the
divergent component from the total velocity.
This decomposition is illustrated in \Fig{tidal}.

The divergent component clearly shows the dominant infall patterns
into the two attractors,
GA and PP, or, equivalently, the outflow from the elongated void between them.
There is little bulk flow in this component,
only $12 \pm 35\kms$ within the sphere of radius $60\hmpc$. 
The tidal component is dominated by a bulk flow, 
of $370 \pm 125\kms$ in the direction $(L,B)=(165^\circ,\, -10^\circ)$.
As expected, most of the bulk velocity within $60\hmpc$ is due to the 
mass distribution outside this volume.  When this bulk velocity is
subtracted off, the residual tidal field shows hints of a quadrupole moment
aligned with the direction to Shapley.
A detailed investigation along these lines, using both Wiener and POTENT
reconstructions and considering the errors, 
is described by Hoffman, Eldar, Zaroubi \& Dekel (1999). 


\section{DISCUSSION}
\label{sec:disc}

Before concluding this paper, we provide a discussion of issues 
concerning the remaining errors (\se{disc_errors}),
the use of POTENT results for cosmology (\se{disc_implications}), 
alternative methods of reconstruction from peculiar velocities 
(\se{disc_methods}),
and related results from other data (\se{disc_sfi}). 

\vskip -0.5truecm \subsection{Errors}
\label{sec:disc_errors}

As described in \se{maps_vm2}, a possible source of systematic error 
in the Mark III data is the simultaneous
calibration of the TF distance indicators of the different datasets,
and in particular the relative zero-pointing, which may be responsible 
for artificial differential Hubble-like peculiar flows in different 
datasets at different directions, 
and thus, \eg, for a spurious bulk flow. We find that the uncertainty, 
as expressed by the difference between the original calibration of
Mark III and the IRAS-motivated VM2 calibration of the same data, 
or by the apparent differences with the independent calibration of 
related data in SFI, is limited in practice to the 
flow at large distances, especially in the PP direction, and it hardly 
propagates into the inner regions where the 
reconstruction of the mass density is reliable and useful for further 
quantitative analysis.  The ongoing SHF survey of TF galaxies 
in a shell between $4500$ and $7000\kms$ will help to determine the
relative zero points and hopefully put the remaining uncertainty to rest.
 
The global zero point remains subject to cosmic scatter.
Based on the current wisdom regarding the cosmological model and 
the fluctuation power spectrum, the zero-point estimate is probably 
good to an accuracy of about $5\%$ (in Hubble-like peculiar velocity).
Using the recovered POTENT fields in a certain volume which is
our best bet for a ``fair sample" given the Mark III data, 
we can re-calibrate the zero point in retrospect such that either the mean 
density fluctuation vanishes or the rms of the peculiar velocities relative to 
the Hubble flow is minimized.  The advantage of using the POTENT fields
for this purpose rather than the raw galaxy radial velocities 
lies in the statistical spatial uniformity of the POTENT fields
that are sampled at the points of a uniform grid.
We find that, in order to satisfy $\la \delta \ra =0$ inside $40\hmpc$,
we need to decrease the TF distances by only 3\% ($\pm 0.7\%$) compared to
the zero point originally determined for Mark III (Willick \etal 1995; 1997a).
The determination of the zero point is based on assuming $\la \delta\ra=0$
locally; the true universal value can only be determined by comparing to the
Hubble flow in a much larger volume.
Preliminary hints for a local ``Hubble Bubble" of radius $\sim 70\hmpc$
encompassed by the Great Walls (defined by redshift surveys) 
come from a sample of accurate (5-8\%) peculiar 
velocities of supernovae type Ia extending to much larger distances
(Zehavi \etal 1998). These data
may in fact indicate that the Hubble expansion within the volume 
sampled by Mark III is $\sim 6\%$ faster than the global expansion, 
such that the local mean density in that volume is $10$ to $20\%$ 
lower than the universal mean (Zehavi \etal 1998).

The Malmquist bias, which used to be a systematic error of major concern
in previous analyses, has been dealt with in three different ways,
yielding satisfactory similarity between the results. The window 
bias is now well understood and an effort has been made, for the first time,
to minimize it properly. The sampling-gradient bias has been
understood since DBF and an effort has been made to minimize it.  
The current POTENT procedure provides a simple compromise recipe 
to handle these biases simultaneously.  
Reliable estimates of the remaining biases
are provided, and they should be used in any following quantitative 
analysis using the POTENT fields.  Our understanding of these biases 
should guide future survey selection, with a special emphasis on 
uniform spatial coverage.

The mock catalogs used for testing the method and evaluating the errors
were based on a simulation of $\Omega=1$ and $\sigma_8=0.7$
with a resolution of $\sim 2\hmpc$. If $\sigma_8$ is higher or $\Omega$ 
is lower, then the errors are expected to be larger and the
reconstruction more difficult. Such cases should be tested with 
appropriate simulations. New constrained-realization simulations
of much higher resolution and a variety of values of $\Omega$ are
in progress (Lemson \etal$\!$, in progress).  They will allow us to better 
test the POTENT procedure under more difficult conditions.

With the systematic errors reasonably under control, 
the main source of remaining uncertainty
is due to the random errors, which mostly arise
from the intrinsic scatter in the distance indicator,
accompanied by sparse sampling and measurement errors.
This uncertainty can be reduced by denser sampling and especially by
using more accurate distance indicators (see Willick 1998 for a review). 

Several new TF/\dns\ surveys are in progress. 
The SHF survey will obtain uniform TF data for an added full-sky
sample of 300 field spirals with $4500\leq z \leq 7000\ \kms,$
and will thus make possible a definitive calibration of existing TF data.
Several deeper, full-sky surveys are focusing  
on cluster galaxies at redshifts $\gsim 8000\ \kms$
(Giovanelli \etal 1997a,b, SCI; Hudson \etal 1998, SMAC; Willick 1998,
LP10K) and hold the promise of constraining
bulk flows on very large ($\gsim 100\hmpc$) scales.
A survey of ellipticals at similar 
distances in two large patches 
of the sky (Wegner \etal 1996, EFAR), 
will soon be complemented by a whole-sky survey
that will replace the original ``7-Samurai" survey with denser sampling
(da Costa \etal 1999, ENEAR).

Most promising in the long run are distances to Supernovae type Ia, 
with estimated errors of only $5-8\%$ and potential coverage of a very 
large volume
(Hamuy \etal\ 1995; Riess, Press \& Kirshner 1995a,b; Perlmutter \etal\ 1997).
The density of sampling, down to the mean separation of galaxies,
is limited only by the effort put into long-term supernovae searches
and the time needed to accumulate a large sample. Preliminary attempts
for a low-resolution POTENT analysis out to $\sim 100\hmpc$ are
already in progress.
 
On small scales, the sampling of Mark III galaxies nearby is dense
enough for a POTENT reconstruction with a resolution higher than G12, 
which can resolve better the Local Supercluster (in preparation). 
The distances to early type galaxies based on surface-brightness fluctuations 
(Tonry \etal 1997), with their higher accuracy, can be of significant 
contribution here.

\vskip -0.5truecm \subsection{The Study of Cosmological Implications}
\label{sec:disc_implications}


Extracting cosmological information from the peculiar velocity data is possible
in several different ways, some of which do not involve POTENT reconstruction
at all. However, the statistical spatial uniformity of the 
POTENT fields offers certain unique opportunities 
to constrain the cosmological parameters and the initial fluctuations.
The main applications involve 
(a) measuring $\Omega$ directly from POTENT data alone,
(b) obtaining cosmological constraints from a combination of  
POTENT data and CMB fluctuations, 
and (c) comparing POTENT fields to redshift surveys
for measuring $\Omega$ via the parameter $\beta$, which is contaminated by 
galaxy biasing. These methods were reviewed, for example, in Dekel (1994; 1998),
Dekel, Burstein \& White (1997), and in numerous conference proceedings. 
Following is a brief summary.

The POTENT fields enable direct constraints on $\Omega$ subject only to
the assumptions of Gaussian initial fluctuations and GI growth of 
fluctuations; they are independent of the complex issue of galaxy 
density biasing.  Methods to obtain such constraints have been developed 
and have already been applied to preliminary versions of the POTENT 
Mark III fields.
For example,
the divergence of the flow in underdense regions in our local cosmological
vicinity, and in particular in the Sculptor void at $\sim 40\hmpc$ from
the LG roughly towards the south Galactic pole, provides a straightforward
lower bound of $\Omega > 0.3$ at the $2.5\sigma$ level of confidence 
(Dekel \& Rees 1994).  The analogous constraint by POTENT
from the SFI data in a preliminary analysis seems to be even stronger 
(Eldar \etal 1999).
As another example,
the deviation of the present-day probability distribution function (PDF)
of $\divv$ from a Gaussian distribution, and in particular the skewness
of this distribution, both measurable from the POTENT output,
provide a similar lower bound on $\Omega$ (Bernardeau \etal 1995).
Finally, the deviation from a Gaussian distribution of the recovered 
initial PDF of 
density fluctuations, derived by following the fluctuations back in
time to the linear regime, enables a similar constraint with even higher
statistical significance (Nusser \& Dekel 1993).
These and related methods are being currently applied to the new POTENT
fields from the latest Mark III and SFI datasets.

To compare to the CMB, the POTENT potential field was translated to 
hypothetical maps of temperature fluctuations in the last-scattering
surface of the microwave background.  The earliest version of these maps, 
by Bertschinger, Gorski \& Dekel (1990) based on the Mark II data, 
was used to predict CMB fluctuations at the level of $\delta T/T \sim 10^{-5}$ 
under the assumption that the local cosmological neighborhood sampled
by peculiar velocities is roughly a ``fair sample" of the universe.
This prediction was made before the DMR instrument on board the COBE 
satellite and balloon-borne experiments actually detected these fluctuations.
A more recent reconstruction along similar lines, by Zaroubi \etal (1997a) 
from the Mark III data, treats more accurately the physical sources of 
temperature fluctuations and deals with different cosmological models.
The most robust conclusion from the comparison of the POTENT results and
the observed fluctuations in the CMB is their mutual consistency with
the growth rate predicted by GI. 

The POTENT density field has also been used to determine the power
spectrum of mass-density fluctuations (Kolatt \& Dekel 1997).
It is characterized by an amplitude of 
$P(k) f(\Omega)^2 =(4.6 \pm 1.4) \times 10^3 (\hmpc)^3$
at $k=0.1 (\hmpc)^{-1}$,
and a high value of rms fluctuations in a top-hat window of radius $8\hmpc$,
$\sigma_8 \,\Omega^{0.6} \simeq 0.7-0.8$.
%
The observed mass power spectrum from velocities
is then compared to the linear predictions of
a family of Inflation-motivated flat CDM models ($\omm+\oml=1$)
that are normalized by the COBE DMR measurements of CMB temperature
fluctuations on large angular scales.
Assuming no tensor fluctuations, 
the result is approximated by $\Omega h_{60}^{1.2} n^2 \approx 0.6 \pm 0.2$,
where $h_{60}=H_0/(60\kms{\rm Mpc}^{-1})$ 
and $n$ is the logarithmic slope of the power spectrum on large scales.
These results were confirmed and elaborated on by a POTENT-independent 
likelihood analysis of the Mark III velocities (Zaroubi \etal 1997b) 
and the SFI velocities (Freudling \etal 1998).
The implication is that slight modifications of the standard CDM model are 
needed for a good fit with these data: either an $\Omega$ value 
somewhat smaller than unity or a moderate tilt in the power spectrum 
on large scales (limited though to $n\gsim 0.9$ in order to
ensure a noticeable acoustic peak in the CMB angular power spectrum 
at $\sim 1^\circ$, Zaroubi \etal 1997a).
An alternative solution is provided by
mixing $\sim 20\%$ massive neutrinos with the dominant cold dark matter.

Much of the use of peculiar velocity data so far has been in comparing
to the galaxy density field extracted from whole-sky redshift surveys,
or to the peculiar velocities predicted from these redshift
data using linear GI.  Such comparisons can provide
degenerate constraints on $\Omega$ and on the biasing relation
between the density fluctuations of galaxies and mass. When this biasing
relation is parameterized by a linear proportionality factor, 
$b$, the comparison
leads to an estimate of the degenerate parameter $\beta\equiv\Omega^{0.6}/b$
(reviews: Dekel 1994; Strauss \& Willick 1995; Strauss 1998; Dekel 1998a).
The latest comparison of the {\it density} fields of POTENT Mark III and
the IRAS 1.2 Jy redshift survey within $40\hmpc$ yields 
$\betai= 0.89 \pm 0.12$ (Sigad \etal 1998). 
This is somewhat lower than
earlier estimates from Mark II velocities and IRAS 1.9 Jy redshifts
(Dekel \etal 1993) but is consistent with the result 
$\betao=0.8\pm 0.2$ from the more-clustered optical galaxies 
(Hudson \etal 1995).
Other methods of comparison of Mark III peculiar velocities and 
the {\it velocities} predicted from IRAS 1.2 Jy galaxies have led 
in general to somewhat lower estimates of the relevant $\beta$ 
parameters, near $0.5-0.6$ (\eg, ITF; VELMOD; VM2).
The span of $\beta$ estimates in the range $0.5-1.0$
may partly arise from underestimated systematic errors and
cosmic scatter between different volumes of sampling, 
it may partly reflect differences between comparisons of velocities
versus comparisons of density fields, and it may partly be
due to nontrivial features in the galaxy biasing scheme, including 
non-linearity, scale-dependence and stochasticity 
(\eg, Dekel \& Lahav 1998; Dekel 1998b; Somerville \etal 1998).
An important goal for future research is to better 
understand the sources of the observed span of values for $\beta$.
This can be achieved in the near future by a combination of  
observational and theoretical means.

\vskip -0.5truecm \subsection{Other Methods of Reconstruction}
\label{sec:disc_methods}


As mentioned already,
the main source of uncertainty in the POTENT reconstruction arises from the
random distance errors. Still within the family of methods based on 
inferred distances, the noise can be better dealt with
by implementing an alternative algorithm based on the Wiener Filter (WF)
plus constrained realizations (CR) (Zaroubi, Hoffman \& Dekel 1998).
The derived Wiener field is the most probable mean field given the noisy
data and a prior model for the power spectrum, but, like most other methods
except POTENT, it suffers
from a non-uniform effective smoothing which follows the spatial
variations in the errors.
The constrained realizations, based on an assumed power spectrum
that could be derived in a preceding stage from the same data, 
are of spatially uniform variance. Each of them could therefore equally
well be an approximation to the real structure in our cosmological
neighborhood. A working algorithm has been developed so far only in
the linear regime of gravitational instability theory.
Preliminary results indicate that
the Wiener mass-density field obtained from the SFI data 
is fairly similar to the one obtained from the
Mark III data (Zaroubi \etal 1999).
This method is complementary to POTENT in many respects.
In the nearby regions where the signal dominates over the noise,
the WF/CR fields are in general agreement with the POTENT fields,
but the WF/CR method allows an interesting extrapolation
into large distances where the noise dominates.

If the selection of the TF data does not explicitly depend on the
internal velocity $\eta$, the Malmquist bias
can be eliminated altogether by minimizing $\eta$ residuals in redshift space
without ever inferring actual distances to individual galaxies
(Schechter 1980).
The distance is replaced by $r=z-u_\alpha(z)$, where $u_\alpha$
is a parametric model for the radial peculiar velocity field,
\eg, an expansion in spherical harmonics and radial Bessel functions
(Davis, Nusser, \& Willick 1996) or a Fourier expansion 
(Blumenthal, Dekel \& Yahil 1999, MFPOT).

Several methods use the data from a whole-sky
redshift survey such as IRAS 1.2 Jy as an intrinsic part of the
reconstruction from peculiar velocities.
These methods are typically geared to estimating goodness of fit and 
certain parameters of the model (\eg, $\beta$)
without ever reconstructing statistically uniform maps of density 
or velocity fields.
The use of the ITF method so far is basically of this nature;
the recovered model velocity field has been compared mode by mode
to a similar expansion of the IRAS 1.2 Jy predicted velocity field
with the aim of determining $\beta$.
The alternative VELMOD method (Willick \etal 1997b)
compares the raw peculiar velocity data with a high-resolution
``model" velocity field
that is predicted from the IRAS 1.2 Jy redshift survey under the
assumption of linear biasing.
A key feature of VELMOD is that it explicitly allows for a
non-unique mapping between real space and redshift space.
Triple-valued zones in the redshift field as
well as non-negligible small-scale velocity ``temperature" are treated
in a unified way.
This method has been used to determine $\beta$ as well as the TF parameters 
and other parameters characterizing the velocity field and the errors.
The $\betai$ values obtained from these high-resolution
velocity-velocity comparisons are typically in the range 0.5 to 0.6,
with typical errors of $\pm 0.1$
(compared with the POTENT-IRAS density-density result of 
$\betai=0.89\pm0.12$ at G12 smoothing).

The dynamical fields, the TF parameters {\it and} the cosmological 
parameter $\beta$ can be recovered {\it simultaneously\,}
by a fit of a parametric model of the potential field
to the combined data of the observed radial peculiar velocities
and the distribution of galaxies in redshift space
(Nusser \& Dekel 1999, SIMPOT).
This procedure takes advantage of the complementary features of
the data in the recovery of the fields, it enforces the same effective
smoothing on the data without a preliminary reconstruction procedure such
as POTENT, and it obtains a more reliable best fit by
simultaneous rather than successive minimization. It has been implemented
so far to the forward TF data, but it can be generalized in principle to
minimize inverse TF residuals.

A natural extension of the POTENT procedure is to push the recovered
mildly non-linear fluctuation fields of the present epoch 
back in time into the linear regime.
Nusser \& Dekel (1992) have developed such a ``time machine"
by limiting the solution to the growing mode via an Eulerian version of
the Zel'dovich approximation (1970). This operation involves
solving a simple differential equation for the potential field,
termed the Zel'dovich-Bernoulli equation. It has been applied to an early
version of the Mark III catalog.
%
A more accurate solution to this mixed-boundary-condition problem
is provided by following trajectories of particles in time using 
the Least Action principle (Shaya \etal 1995).
This was used in combination with redshift data to determine $\beta$
for a local optical redshift survey, obtaining low values.
This original application was limited to a specific 
biasing scheme in which all
the mass is assigned to point-mass galaxies, ignoring the overlap
of dark-matter halos and thus overestimating the forces between galaxies.

The current bottom line is that the different estimates of $\betai$ are
in the range $0.5-1.0$.  Recall that this span of estimates was briefly 
discussed at the end of the previous subsection.

\vskip -0.5truecm \subsection{Other Results from Peculiar Velocities}
\label{sec:disc_sfi}

The early application of a simple version of POTENT 
(da Costa \etal 1996) to the complementary SFI data 
(Haynes \etal 1999; Wagner \etal 1999)
led to an apparent mass distribution in which the gross features roughly agree
with the reconstruction from Mark III, but the amplitude of the
perturbations is lower and other details also differ.
Most of these differences, which were partly due to ignoring the window bias,
seem to go away when the improved version of POTENT is used; 
in fact, the density fields out of Mark III and SFI are fairly consistent
within the errors (Eldar \etal 1999; Zaroubi \etal 1999, 
using POTENT and Wiener respectively).
A similar conclusion is obtained via maximum-likelihood determinations
of the mass power spectrum (not involving POTENT reconstruction)
from the Mark III velocities (Zaroubi \etal 1997b) and from the
SFI velocities (Freudling \etal 1998). The two power spectra are found to
be nearly identical and also very similar to the power spectrum derived 
from the POTENT density of Mark III (Kolatt \& Dekel 1997).
The main remaining difference in the reconstructions from Mark III and SFI  
is in the flows at large radii. The SFI bulk velocity in the CMB frame
in shells of radii beyond $50\hmpc$ about the LG is small and 
practically consistent with zero, while in Mark III it remains high, 
at the level of $(350 \pm 110)\kms$.
This difference may reflect errors in the differential calibration of the
zero points of the Tully-Fisher relations of the different datasets
that constitute these catalogs.  The VM2 calibration of Mark III
seems to bring the outer bulk velocity to better agreement with SFI,
but this does not necessarily indicate which calibration is more accurate.
The differences occur in the outer regions of the sampled volume 
where the errors are large and the bias corrections are particularly 
uncertain, so the apparent discrepancy is not entirely surprising.  
A comparison of the basic data for galaxies in common between the SFI
and Mark III datasets may prove useful for assessing the source for
this discrepancy.  It is encouraging though that this difference in 
the outer bulk flow, independent of its origin, hardly affects the 
density field, which is only a function of the derivatives of the 
velocity field.

\section{CONCLUSION}
\label{sec:conc}

The main purpose of this paper is to present the second-generation 
POTENT method for reconstructing the underlying dynamical fields of
velocity and mass density from observed peculiar velocities.
The main effort in the method development is directed at understanding
the systematic and random errors, adopting a compromise optimization
of the method such that the different errors are minimized simultaneously,
and then quantifying the remaining errors for subsequent analysis using
the POTENT output.
This is done using realistic mock data drawn from simulations
designed to mimic our local cosmological neighborhood,
and in particular the Mark III catalog of peculiar velocities.

The method is applied to the real Mark III data,
and the results are presented via maps of the fields of velocity and
mass density in our cosmological neighborhood, and via measurements
of the simplest statistic --- the bulk flow. The current reconstructions are
carried out with Gaussian smoothings of radii $10-12\hmpc$ inside a
sphere of radius $80\hmpc$, but the reliable results for quantitative
analyses are limited to effective radii of $\sim 40$ to $50\hmpc$.
Within $40\hmpc$, the rms systematic error in the local G12 
density fluctuation field $\delta$ is $\pm 0.13$, and the rms 
random error is $\pm 0.18$.
The typical weighted errors in this volume are smaller by $\sim 30\%$.

The recovered mass distribution with G12 smoothing
resembles in its gross features the recovery from other peculiar-velocity
data and the galaxy distribution extracted from redshift surveys. 
The robust structures in our cosmological neighborhood
are the two giant superstructures --- the Great
Attractor and Perseus Pisces --- both of mean mass density about twice
the average within regions of $\sim 50\hmpc$ in diameter,
and an extended underdense region stretching out between them.
There are differences in the fine details of the different reconstructions;
most of them can be attributed to known errors, some may reflect
a non-trivial biasing relation between galaxies and mass, and the
remaining differences may indicate additional errors not accounted for
in the error analysis above.

Both large-scale ($\sim\!100\hmpc$) and small-scale ($\sim\!10\hmpc$)
features in the recovered velocity field are of cosmological interest.
For example, the bulk velocity within spheres about the Local Group, \eg,
$V_{50}=370\pm110\kms$ inside $50\hmpc$,
reflects properties of the initial fluctuation power spectrum on scales
$\sim 100\hmpc$ and larger. On the other hand,
the small-scale variations of the velocity field, related to the 
density field, allow direct estimates of the value of the cosmological 
density parameter $\Omega$.
The statistically uniform POTENT fields and their associated errors
allow direct comparisons to theory and to other data in an effort to 
extract cosmological implications.

\acknowledgments{
Parts of this work are associated with the Ph.D. theses of TK (1994) 
and AE (in preparation) at the Hebrew University of Jerusalem.       
We thank our collaborators in related projects:
E. Bertschinger, G. Blumenthal, M. Davis, A. Dressler, G. Ganon, 
Y. Hoffman, M. Hudson, D. Mathewson, A. Nusser, Y. Sigad, M. Strauss, 
S. Zaroubi and I. Zehavi.
This research was supported by the US-Israel Binational Science
Foundation grants 89-00194, 92-00355, 95-00330,
by the Israel Science Foundation grants 462/92, 950/95,
by NASA ATP grant NAG 5-301 at UCSC,
by NASA grant NAS-5-1661 to the WFPC1 IDT,
by NSF grant AST-9617188 at Stanford,
and by NSF grant AST-9016930 at ASU. 
}

\def\re{\reference}
\def\jeru{in {\it Formation of Structure in the Universe},
     eds.~A. Dekel \& J.P. Ostriker (Cambridge Univ. Press)\ }


\end{document}


\begin{deluxetable}{lccccc}
\scriptsize
\tablewidth{0pt}
\tablecaption{Datasets in the Mark III Catalog \label{tab:data}}
\tablehead{
  Sample& Type ${}\tablenotemark{1}$ &$z$ cutoff
${}\tablenotemark{2}$ &
Angular boundaries & \# galaxies & \# galaxies   
\\
&&&& Willick \etal 97a & used here}
\startdata
HMCL ${}\tablenotemark{3}$ & C(S) & & & 428 & 329 \nl
A82  ${}\tablenotemark{4}$ & S &3000 &$ -6.5 \arcdeg < \delta $ & 359 & 206 \nl
MAT ${}\tablenotemark{5}$ & S &8500 & $\delta <  0\arcdeg $ & 1355 & 1198 \nl
W91PP ${}\tablenotemark{6}$ & S & & $21.5^\circ\!<\! \delta \!<\! 39.5^\circ $;
 $300^\circ \!<\! \alpha \!<\! 30^\circ $ & 326 & 347 \nl
WCF ${}\tablenotemark{7}$ & S & & $-2.5 \arcdeg < \delta $ & 321 & 212 \nl
7S ${}\tablenotemark{8}$ & E, C(E) &9000 & & 544 & 529 \nl
\enddata
\tablenotetext{1}{S = spirals; E = ellipticals/S0s;  C = clusters}
\tablenotetext{2}{in $\kms$}
\tablenotetext{3}{Han \& Mould (1990, 1992), Mould \etal 1991}
\tablenotetext{4}{Aaronson \etal (1982)}
\tablenotetext{5}{Mathewson \etal (1992)}
\tablenotetext{6}{Willick (1991)}
\tablenotetext{7}{Courteau \& Faber (1992)}
\tablenotetext{8}{Lynden-Bell \etal (1988)}
\end{deluxetable}

\begin{deluxetable}{lccccccccccccc}
 \scriptsize
\tablecolumns{16}
\tablewidth{0pt}
\tablecaption{Performance of POTENT with Mock Data: Unweighted 
\label{tab:eval}}

\tablehead{
Effect &
Input & Method & $\Rs$ & $\Re$ &
  $\sigt$  & $s$  & $c$ & $r$   & $\epssys$  & $\epsran$  & $\epstot$
& \S & Fig.
}
\startdata
$u(r)$ to $\delta (r) $ &
exact $u(r)$ & &12& 40&
0.21 & 1.00 & 0.003 & 0.99 & 0.12 &   &   &5.2&4
\nl
 &              & &10& 40&
0.28 & 1.00 & 0.002 & 0.99 & 0.11 &   &   &"&"
\nl
&&&&&&&&&\nl
+window &
uniform           &3 param& 12& 40&
0.21 & 0.56 & 0.021 & 0.81 & 0.60 &   &   &6.1&5,6
\nl
bias & true dist.              &9 param& 12& 40&
0.21 & 0.81 & -0.002 & 0.93 & 0.38 &   &   &" &"
\nl
 &                              &       & 10& 40&
0.28 & 0.90 & -0.006 & 0.95 & 0.30 &   &   &"&"
\nl
&&&&&&&&&\nl
+sampling&
nonuniform&$V_4$ &12&40&
0.22 & 0.90 & -0.008 & 0.92 & 0.39 & 0.23 & 0.45 &6.2&7-9,15c,16c
\nl
bias & true dist.                 & "    &10&40&
0.27 & 0.93 & -0.007 & 0.93 & 0.39 & 0.25 & 0.46 &"&"
\nl
&&&&&&&&&\nl
+random &
nonuniform &$V_4\sigma ^{-2}$, grp & 12&30&
0.24 & 0.97 &0.002 & 0.92 & 0.42 & 0.49 & 0.65 &6.3&10-14,15d,16d
\nl
errors & noisy                        & "             &{\bf 12} &{\bf 40}&
{\bf 0.21} & {\bf 0.89} & {\bf -0.007} & {\bf 0.82} &{\bf 0.62}&{\bf 0.84}&
{\bf 1.04} &" &"
\nl
 &                                 &    "                      & 12&50&
0.19 & 0.81 & -0.001 & 0.72 & 0.80 & 1.22 & 1.46 &"&"
\nl
&&&&&&&&&\nl
 &                                  &   "                       & 10&30&
0.32 & 0.95 &-0.006 & 0.92 & 0.42 & 0.49 & 0.65 &"&"
\nl
 &                                  &   "                       & 10&40&
0.27 & 0.94 &-0.043& 0.82 & 0.65 & 0.84 & 1.06 &"&"
\nl
 &                                  &   "                       & 10&50&
0.24 & 0.90 &-0.097& 0.68 & 1.05 & 1.27 & 1.65 &"&"
\nl
&&&&&&&&&\nl
 & & $V_4\sigma ^{-2}$, ung &{\bf 12}&{\bf 40}&
{\bf 0.22}&{\bf 0.96}&{\bf -0.038}&{\bf 0.83}&{\bf 0.65}&{\bf 0.83}&
{\bf 1.05}&"&"
\nl
 & & $V_4\sigma ^{-2}$, inv &{\bf 12}&{\bf 40}&
{\bf 0.21}&{\bf 0.95}&{\bf 0.008}&{\bf 0.85}&{\bf 0.60}&{\bf 0.89}&
{\bf 1.07}&"&"
\nl
\enddata
\end{deluxetable}

\begin{deluxetable}{lccccccccccccc}
\scriptsize
\tablecolumns{16}
\tablewidth{0pt}
\tablecaption{Performance of POTENT with Mock Data: Error Weighted 
	      \label{tab:evalw}}

\tablehead{
Effect & Input & Method & $\Rs$ & $\Re$ &
  $\sigtw$ & $\sw$ &$c_{\rm w}$ & $\rw$ & $\epssysw$ & $\epsranw$ & $\epstotw$
& \S &Fig.
}
\startdata
&&&&&&&&&\nl
+sampling &
nonuniform &$V_4$ &12&40&
0.25 & 0.81 &-0.014 & 0.94 & 0.34 & 0.13 & 0.37 & 6.2 & 7-9,15c,16c
\nl
bias  &  true dist.                   & "    &10&40&
..34 & 0.84 &-0.022 & 0.94 & 0.34 & 0.14 & 0.33 &"&"
\nl
&&&&&&&&&\nl
+random &
nonuniform &$V_4\sigma ^{-2}$, grp & 12&30&
0.26 & 0.97 & 0.003 & 0.95 & 0.31 & 0.34 & 0.47 & 6.3& 10-14,15d,16d
\nl
error & noisy                            &  "            &{\bf 12} &{\bf 40}&
{\bf 0.25} & {\bf 0.96} &{\bf -0.003} & {\bf 0.93} &{\bf 0.38}&{\bf 0.46}
&{\bf 0.60} &"&"
 \nl
  &                                 &       "                   & 12&50&
0.24 & 0.96 &-0.005& 0.90 & 0.46 & 0.61 & 0.76 &"&"
\nl
&&&&&&&&&\nl
   &                                &       "                   & 10&30&
0.36 & 0.93 &-0.012& 0.95 & 0.30 & 0.33 & 0.44 &"&"
\nl
    &                               &       "                   & 10&40&
0.34 & 0.94 &-0.043& 0.93 & 0.38 & 0.45 & 0.59 &"&"
\nl
     &                              &       "                   & 10&50&
0.33 & 0.97 &-0.051& 0.88 & 0.55 & 0.60 & 0.82 &"&"
\nl
&&&&&&&&&\nl
  & & $V_4\sigma ^{-2}$, ung &{\bf 12}&{\bf 40}&
{\bf 0.25}&{\bf 1.10}&{\bf -0.037} &{\bf 0.94}&{\bf 0.40}&{\bf 0.44}&
{\bf 0.59} &"&"
\nl
 & & $V_4\sigma ^{-2}$, inv &{\bf 12}&{\bf 40}&
{\bf 0.24}&{\bf 1.05}&{\bf 0.007}&{\bf 0.92}&{\bf 0.42}&{\bf 0.58}&{\bf 0.72} &"&"
\nl
\enddata
\end{deluxetable}


\end{document}